\documentclass[reqno,12pt,a4paper]{article}

\usepackage{amsfonts,amssymb,amsthm, 
amsmath,amscd, bbm, bm, mathabx, 
mathrsfs,delarray,subfigure}

\usepackage[dvipsnames]{xcolor}[svgnames,x11names]

\usepackage{hyperref}
\usepackage{cite}
\usepackage{xypic}
\usepackage{sectsty}

\usepackage{fancyhdr}
\fancyheadoffset[RE,LO]{0.\textwidth} 

\hypersetup{
  colorlinks,
  linkcolor=blue,
  linktoc=all,
  citecolor=blue,
   urlcolor=blue
}

\usepackage{sectsty}

\colorlet{sectitlecolor}{blue}
\colorlet{sectboxcolor}{white}
\colorlet{secnumcolor}{blue}

\sectionfont{\color{sectitlecolor}}

\makeatletter
\renewcommand\@seccntformat[1]{%
  \colorbox{sectboxcolor}{\textcolor{secnumcolor}{\csname the#1\endcsname}}%
  \quad
}
\makeatother

\usepackage{etoolbox}
\patchcmd{\thebibliography}{\section*{\refname}}{}{}{}

\DeclareSymbolFont{sfletters}{OML}{cmbrm}{m}{it}  

\DeclareMathSymbol{\sfGamma}{\mathord}{sfletters}{"00}
\DeclareMathSymbol{\sfDelta}{\mathord}{sfletters}{"01}
\DeclareMathSymbol{\sfTheta}{\mathord}{sfletters}{"02}
\DeclareMathSymbol{\sfLambda}{\mathord}{sfletters}{"03}
\DeclareMathSymbol{\sfXi}{\mathord}{sfletters}{"04}
\DeclareMathSymbol{\sfPi}{\mathord}{sfletters}{"05}
\DeclareMathSymbol{\sfSigma}{\mathord}{sfletters}{"06}
\DeclareMathSymbol{\sfUpsilon}{\mathord}{sfletters}{"07}
\DeclareMathSymbol{\sfPhi}{\mathord}{sfletters}{"08}
\DeclareMathSymbol{\sfPsi}{\mathord}{sfletters}{"09}
\DeclareMathSymbol{\sfOmega}{\mathord}{sfletters}{"0A}
\DeclareMathSymbol{\sfalpha}{\mathord}{sfletters}{"0B}
\DeclareMathSymbol{\sfbeta}{\mathord}{sfletters}{"0C}
\DeclareMathSymbol{\sfgamma}{\mathord}{sfletters}{"0D}
\DeclareMathSymbol{\sfdelta}{\mathord}{sfletters}{"0E}
\DeclareMathSymbol{\sfepsilon}{\mathord}{sfletters}{"0F}
\DeclareMathSymbol{\sfzeta}{\mathord}{sfletters}{"10}
\DeclareMathSymbol{\sfeta}{\mathord}{sfletters}{"11}
\DeclareMathSymbol{\sftheta}{\mathord}{sfletters}{"12}
\DeclareMathSymbol{\sfiota}{\mathord}{sfletters}{"13}
\DeclareMathSymbol{\sfkappa}{\mathord}{sfletters}{"14}
\DeclareMathSymbol{\sflambda}{\mathord}{sfletters}{"15}
\DeclareMathSymbol{\sfmu}{\mathord}{sfletters}{"16}
\DeclareMathSymbol{\sfnu}{\mathord}{sfletters}{"17}
\DeclareMathSymbol{\sfxi}{\mathord}{sfletters}{"18}
\DeclareMathSymbol{\sfpi}{\mathord}{sfletters}{"19}
\DeclareMathSymbol{\sfrho}{\mathord}{sfletters}{"1A}
\DeclareMathSymbol{\sfsigma}{\mathord}{sfletters}{"1B}
\DeclareMathSymbol{\sftau}{\mathord}{sfletters}{"1C}
\DeclareMathSymbol{\sfupsilon}{\mathord}{sfletters}{"1D}
\DeclareMathSymbol{\sfphi}{\mathord}{sfletters}{"1E}
\DeclareMathSymbol{\sfchi}{\mathord}{sfletters}{"1F}
\DeclareMathSymbol{\sfpsi}{\mathord}{sfletters}{"20}
\DeclareMathSymbol{\sfomega}{\mathord}{sfletters}{"21}
\DeclareMathSymbol{\sfvarepsilon}{\mathord}{sfletters}{"22}
\DeclareMathSymbol{\sfvartheta}{\mathord}{sfletters}{"23}
\DeclareMathSymbol{\sfvarpi}{\mathord}{sfletters}{"24}
\DeclareMathSymbol{\sfvarrho}{\mathord}{sfletters}{"25}
\DeclareMathSymbol{\sfvarsigma}{\mathord}{sfletters}{"26}
\DeclareMathSymbol{\sfvarphi}{\mathord}{sfletters}{"27}

\DeclareMathSymbol{\spartial}{\mathord}{sfletters}{"40}

\DeclareMathSymbol{\sfA}{\mathord}{sfletters}{"41}
\DeclareMathSymbol{\sfB}{\mathord}{sfletters}{"42}
\DeclareMathSymbol{\sfC}{\mathord}{sfletters}{"43}
\DeclareMathSymbol{\sfD}{\mathord}{sfletters}{"44}
\DeclareMathSymbol{\sfE}{\mathord}{sfletters}{"45}
\DeclareMathSymbol{\sfF}{\mathord}{sfletters}{"46}
\DeclareMathSymbol{\sfG}{\mathord}{sfletters}{"47}
\DeclareMathSymbol{\sfH}{\mathord}{sfletters}{"48}
\DeclareMathSymbol{\sfI}{\mathord}{sfletters}{"49}
\DeclareMathSymbol{\sfJ}{\mathord}{sfletters}{"4A}
\DeclareMathSymbol{\sfK}{\mathord}{sfletters}{"4B}
\DeclareMathSymbol{\sfL}{\mathord}{sfletters}{"4C}
\DeclareMathSymbol{\sfM}{\mathord}{sfletters}{"4D}
\DeclareMathSymbol{\sfN}{\mathord}{sfletters}{"4E}
\DeclareMathSymbol{\sfO}{\mathord}{sfletters}{"4F}
\DeclareMathSymbol{\sfP}{\mathord}{sfletters}{"50}
\DeclareMathSymbol{\sfQ}{\mathord}{sfletters}{"51}
\DeclareMathSymbol{\sfR}{\mathord}{sfletters}{"52}
\DeclareMathSymbol{\sfS}{\mathord}{sfletters}{"53}
\DeclareMathSymbol{\sfT}{\mathord}{sfletters}{"54}
\DeclareMathSymbol{\sfU}{\mathord}{sfletters}{"55}
\DeclareMathSymbol{\sfV}{\mathord}{sfletters}{"56}
\DeclareMathSymbol{\sfW}{\mathord}{sfletters}{"57}
\DeclareMathSymbol{\sfX}{\mathord}{sfletters}{"58}
\DeclareMathSymbol{\sfY}{\mathord}{sfletters}{"59}
\DeclareMathSymbol{\sfZ}{\mathord}{sfletters}{"5A}
\DeclareMathSymbol{\sfa}{\mathord}{sfletters}{"61}
\DeclareMathSymbol{\sfb}{\mathord}{sfletters}{"62}
\DeclareMathSymbol{\sfc}{\mathord}{sfletters}{"63}
\DeclareMathSymbol{\sfd}{\mathord}{sfletters}{"64}
\DeclareMathSymbol{\sfe}{\mathord}{sfletters}{"65}
\DeclareMathSymbol{\sff}{\mathord}{sfletters}{"66}
\DeclareMathSymbol{\sfg}{\mathord}{sfletters}{"67}
\DeclareMathSymbol{\sfh}{\mathord}{sfletters}{"68}
\DeclareMathSymbol{\sfi}{\mathord}{sfletters}{"69}
\DeclareMathSymbol{\sfj}{\mathord}{sfletters}{"6A}
\DeclareMathSymbol{\sfk}{\mathord}{sfletters}{"6B}
\DeclareMathSymbol{\sfl}{\mathord}{sfletters}{"6C}
\DeclareMathSymbol{\sfm}{\mathord}{sfletters}{"6D}
\DeclareMathSymbol{\sfn}{\mathord}{sfletters}{"6E}
\DeclareMathSymbol{\sfo}{\mathord}{sfletters}{"6F}
\DeclareMathSymbol{\sfp}{\mathord}{sfletters}{"70}
\DeclareMathSymbol{\sfq}{\mathord}{sfletters}{"71}
\DeclareMathSymbol{\sfr}{\mathord}{sfletters}{"72}
\DeclareMathSymbol{\sfs}{\mathord}{sfletters}{"73}
\DeclareMathSymbol{\sft}{\mathord}{sfletters}{"74}
\DeclareMathSymbol{\sfu}{\mathord}{sfletters}{"75}
\DeclareMathSymbol{\sfv}{\mathord}{sfletters}{"76}
\DeclareMathSymbol{\sfw}{\mathord}{sfletters}{"77}
\DeclareMathSymbol{\sfx}{\mathord}{sfletters}{"78}
\DeclareMathSymbol{\sfy}{\mathord}{sfletters}{"79}
\DeclareMathSymbol{\sfz}{\mathord}{sfletters}{"7A}

\usepackage[LGR,T1]{fontenc}
\usepackage{amsmath,etoolbox}

\newcommand{\declarebsfgreek}[2]{%
  \protected\csdef{bsf#1}{\mathord{\text{\bsfgreekfont#2}}}%
}
\newcommand{\bsfgreekfont}{\usefont{LGR}{cmss}{bx}{it}}

\declarebsfgreek{alpha}{a}
\declarebsfgreek{beta}{b}
\declarebsfgreek{gamma}{g}
\declarebsfgreek{delta}{d}
\declarebsfgreek{epsilon}{e}
\declarebsfgreek{zeta}{z}
\declarebsfgreek{eta}{h}
\declarebsfgreek{theta}{j}
\declarebsfgreek{iota}{i}
\declarebsfgreek{kappa}{k}
\declarebsfgreek{lambda}{l}
\declarebsfgreek{mu}{m}
\declarebsfgreek{nu}{n}
\declarebsfgreek{xi}{x}
\declarebsfgreek{omicron}{o}
\declarebsfgreek{pi}{p}
\declarebsfgreek{rho}{r}
\declarebsfgreek{sigma}{s}
\declarebsfgreek{tau}{t}
\declarebsfgreek{upsilon}{u}
\declarebsfgreek{phi}{f}
\declarebsfgreek{chi}{q}
\declarebsfgreek{psi}{y}
\declarebsfgreek{omega}{w}
\declarebsfgreek{varsigma}{c}

\declarebsfgreek{Gamma}{G}
\declarebsfgreek{Delta}{D}
\declarebsfgreek{Upsilon}{U}
\declarebsfgreek{Omega}{W}
\declarebsfgreek{Theta}{J}
\declarebsfgreek{Lambda}{L}
\declarebsfgreek{Xi}{X}
\declarebsfgreek{Pi}{P}
\declarebsfgreek{Sigma}{S}
\declarebsfgreek{Phi}{F}
\declarebsfgreek{Psi}{Y}

\newcommand{\declarebsfitalic}[2]{%
  \protected\csdef{bsf#1}{\mathord{\text{\bsfitalicfont#2}}}%
}
\newcommand{\bsfitalicfont}{\usefont{T1}{cmss}{bx}{it}}

\declarebsfitalic{a}{a}
\declarebsfitalic{b}{b}
\declarebsfitalic{c}{c}
\declarebsfitalic{d}{d}
\declarebsfitalic{e}{e}
\declarebsfitalic{f}{f}
\declarebsfitalic{g}{g}
\declarebsfitalic{h}{h}
\declarebsfitalic{i}{i}
\declarebsfitalic{j}{j}
\declarebsfitalic{k}{k}
\declarebsfitalic{l}{l}
\declarebsfitalic{m}{m}
\declarebsfitalic{n}{n}
\declarebsfitalic{o}{o}
\declarebsfitalic{p}{p}
\declarebsfitalic{q}{q}
\declarebsfitalic{r}{r}
\declarebsfitalic{s}{s}
\declarebsfitalic{t}{t}
\declarebsfitalic{u}{u}
\declarebsfitalic{v}{v}
\declarebsfitalic{x}{x}
\declarebsfitalic{y}{y}
\declarebsfitalic{z}{z}

\declarebsfitalic{A}{A}
\declarebsfitalic{B}{B}
\declarebsfitalic{C}{C}
\declarebsfitalic{D}{D}
\declarebsfitalic{E}{E}
\declarebsfitalic{F}{F}
\declarebsfitalic{G}{G}
\declarebsfitalic{H}{H}
\declarebsfitalic{I}{I}
\declarebsfitalic{J}{J}
\declarebsfitalic{K}{K}
\declarebsfitalic{L}{L}
\declarebsfitalic{M}{M}
\declarebsfitalic{N}{N}
\declarebsfitalic{O}{O}
\declarebsfitalic{P}{P}
\declarebsfitalic{Q}{Q}
\declarebsfitalic{R}{R}
\declarebsfitalic{S}{S}
\declarebsfitalic{T}{T}
\declarebsfitalic{U}{U}
\declarebsfitalic{V}{V}
\declarebsfitalic{X}{X}
\declarebsfitalic{Y}{Y}
\declarebsfitalic{Z}{Z}

\newcommand{\rmb}{{\mathrm{b}}}

\newcommand{\rme}{{\mathrm{e}}}
\newcommand{\rmf}{{\mathrm{f}}}

\newcommand{\rmA}{{\mathrm{A}}}
\newcommand{\rmB}{{\mathrm{B}}}
\newcommand{\rmC}{{\mathrm{C}}}

\newcommand{\rmF}{{\mathrm{F}}}
\newcommand{\rmG}{{\mathrm{G}}}
\newcommand{\rmH}{{\mathrm{H}}}

\newcommand{\rmJ}{{\mathrm{J}}}
\newcommand{\rmK}{{\mathrm{K}}}

\newcommand{\rmN}{{\mathrm{N}}}

\newcommand{\rmQ}{{\mathrm{Q}}}

\newcommand{\rmS}{{\mathrm{S}}}
\newcommand{\rmT}{{\mathrm{T}}}
\newcommand{\rmU}{{\mathrm{U}}}
\newcommand{\rmV}{{\mathrm{V}}}
\newcommand{\rmW}{{\mathrm{W}}}
\newcommand{\rmX}{{\mathrm{X}}}

\newcommand{\rmZ}{{\mathrm{Z}}}

\newcommand{\msr}{{\mathsans{r}}}

\newcommand{\msA}{{\mathsans{A}}}

\newcommand{\msC}{{\mathsans{C}}}
\newcommand{\msD}{{\mathsans{D}}}
\newcommand{\msE}{{\mathsans{E}}}

\newcommand{\msG}{{\mathsans{G}}}
\newcommand{\msH}{{\mathsans{H}}}

\newcommand{\msJ}{{\mathsans{J}}}

\newcommand{\msL}{{\mathsans{L}}}
\newcommand{\msM}{{\mathsans{M}}}
\newcommand{\msN}{{\mathsans{N}}}

\newcommand{\msQ}{{\mathsans{Q}}}

\newcommand{\msT}{{\mathsans{T}}}
\newcommand{\msU}{{\mathsans{U}}}
\newcommand{\msV}{{\mathsans{V}}}

\newcommand{\fke}{{\mathfrak{e}}}

\newcommand{\fkg}{{\mathfrak{g}}}
\newcommand{\fkh}{{\mathfrak{h}}}

\newcommand{\fkk}{{\mathfrak{k}}}
\newcommand{\fkl}{{\mathfrak{l}}}
\newcommand{\fkm}{{\mathfrak{m}}}

\newcommand{\fks}{{\mathfrak{s}}}
\newcommand{\fkt}{{\mathfrak{t}}}
\newcommand{\fku}{{\mathfrak{u}}}

\newcommand{\bbC}{{\mathbb{C}}}

\newcommand{\bbI}{{\mathbb{I}}}

\newcommand{\bbR}{{\mathbb{R}}}
\newcommand{\bbS}{{\mathbb{S}}}

\newcommand{\bbZ}{{\mathbb{Z}}}

\newcommand{\scD}{{\matheul{D}}}

\newcommand{\scI}{{\matheul{I}}}

\newcommand{\scL}{{\matheul{L}}}

\newcommand{\clC}{{\mathcal{C}}}
\newcommand{\clD}{{\mathcal{D}}}
\newcommand{\clE}{{\mathcal{E}}}
\newcommand{\clF}{{\mathcal{F}}}
\newcommand{\clG}{{\mathcal{G}}}
\newcommand{\clH}{{\mathcal{H}}}
\newcommand{\clI}{{\mathcal{I}}}

\newcommand{\clK}{{\mathcal{K}}}
\newcommand{\clL}{{\mathcal{L}}}

\newcommand{\clO}{{\mathcal{O}}}
\newcommand{\clP}{{\mathcal{P}}}

\newcommand{\clS}{{\mathcal{S}}}

\newcommand{\clV}{{\mathcal{V}}}

\usepackage{accents}

\sectionfont{\fontsize{12}{15}\selectfont}
\subsectionfont{\fontsize{12}{15}\selectfont}

\usepackage{enumitem}


\DeclareMathOperator{\ad}{ad}
\DeclareMathOperator{\Ad}{Ad}
\DeclareMathOperator{\vol}{vol}

\DeclareMathOperator{\id}{id}

\DeclareMathOperator{\Hom}{Hom}

\DeclareMathOperator{\Fun}{Fun}
\DeclareMathOperator{\iFun}{\textsc{Fun}} 

\DeclareMathOperator{\Map}{Map}
\DeclareMathOperator{\iMap}{\textsc{Map}}   
\DeclareMathOperator{\iDFnc}{\textsc{DFnc}}

\DeclareMathOperator{\Emb}{Emb}

\DeclareMathOperator{\Vect}{Vect}
\DeclareMathOperator{\iVect}{\textsc{Vect}}

\DeclareMathOperator{\ev}{ev}
\DeclareMathOperator{\tgr}{tgr}

\DeclareMathOperator{\INN}{\textsc{Inn}}    

\DeclareMathOperator{\tr}{tr}

\DeclareMathOperator{\ee}{e}

\DeclareMathOperator{\BB}{B}
\DeclareMathOperator{\W}{We\hspace{-1pt}}
\DeclareMathOperator{\iW}{\textsc{We}\hspace{-1pt}}
\DeclareMathOperator{\WO}{OWe\hspace{-1pt}}
\DeclareMathOperator{\iWO}{\textsc{OWe}\hspace{-1pt}}  
\DeclareMathOperator{\NN}{N\hspace{-1pt}}

\DeclareMathOperator{\WW}{W\hspace{-1pt}}

\DeclareMathOperator{\ZZ}{Z\hspace{-1pt}}
\DeclareMathOperator{\SD}{SD\hspace{-1pt}}
\DeclareMathOperator{\DD}{D\hspace{-1pt}}

\DeclareMathOperator{\iOOO}{\textsc{Op}\hspace{-1pt}}     

\numberwithin{equation}{subsection} 
\numberwithin{subsection}{section} 

\newcommand{\ceqref}[1]{{\textcolor{blue}{\eqref{#1}}}}
\newcommand{\cref}[1]{{\textcolor{blue}{\ref{#1}}}}
\newcommand{\ccite}[1]{{\textcolor{blue}{\!\cite{#1}}}}

\newcommand{\ddd}{{\hbox{\large $\bigoplus$}}}

\newcommand{\ul}[1]{{\underline{#1}}}

\newcommand{\sdot}{\hspace{.5pt}\dot{}\hspace{.5pt}}
\newcommand{\hfpt}{\hspace{.75pt}}
\newcommand{\mhfpt}{\hspace{-.75pt}}
\newcommand{\dd}{\text{d}}

\newcommand{\mathsans}[1]{{{\sf #1}}}

\font\euler=eusm10 at 12.8 truept
\font\scripteuler=eusm7
\font\scriptscripteuler=eusm5 
\def\eul{\fam=12}
\newcommand{\matheul}[1]{{{\eul #1}}}

\newtheorem{defi}{{\sf Definition}}[section]
\newtheorem{prop}{{\sf Proposition}}[section]

\newtheorem{lemma}{{\sf Lemma}}[section]

\DeclareMathSymbol{*}{\mathbin}{symbols}{"03} 

\begin{document}

\thispagestyle{empty} 

\vskip1.5cm
\begin{large}
{\flushleft\textcolor{blue}{\sffamily\bfseries Quantum field theoretic representation of Wilson surfaces:}}  
{\flushleft\textcolor{blue}{\sffamily\bfseries II higher topological coadjoint orbit model}}  
\end{large}
\vskip1.3cm
\hrule height 1.5pt
\vskip1.3cm
{\flushleft{\sffamily \bfseries Roberto Zucchini}\\
\it Department of Physics and Astronomy,\\
University of Bologna,\\
I.N.F.N., Bologna division,\\
viale Berti Pichat, 6/2\\
Bologna, Italy\\
Email: \textcolor{blue}{\tt \href{mailto:roberto.zucchini@unibo.it}{roberto.zucchini@unibo.it}}, 
\textcolor{blue}{\tt \href{mailto:zucchinir@bo.infn.it}{zucchinir@bo.infn.it}}}


\vskip.7cm
\vskip.6cm 
{\flushleft\sc
Abstract:} 
This is the second of a series of two papers devoted to the partition function realization
of Wilson surfaces in strict higher gauge theory. 
A higher 2--dimensional counterpart of the topological coadjoint orbit quantum mechanical model 
computing Wilson lines is presented based on the derived geometric
framework, which has shown its usefulness in 4--dimensional higher Chern--Simons theory.
 Its symmetries are described. Its quantization is analyzed in the functional integral framework. Strong
evidence is provided that the model does indeed underlie the partition function realization of Wilson
surfaces. The emergence of the vanishing fake curvature condition is explained 
and homotopy invariance for a flat higher gauge field is shown. 
The model's Hamiltonian formulation is further furnished highlighting
the model's close relationship to the derived Kirillov-Kostant-Souriau theory developed in the companion paper.

\vspace{2mm}
\par\noindent
MSC: 81T13 81T20 81T45  

\vfil\eject

{\color{blue}\tableofcontents}

\vfil\eject

\vfil\eject

\renewcommand{\sectionmark}[1]{\markright{\thesection\ ~~#1}}

\section{\textcolor{blue}{\sffamily Introduction}}\label{sec:intro}

\vspace{-1.5mm}

Wilson loops were introduced by Wilson in 1974 \ccite{Wilson:1974sk}
as a natural set of gauge invariant variables suitable for the description of the non
perturbative regime of quantum chromodynamics. 
Since then, they have found a wide range of applications in many branches of theoretical physics. 

In the loop formulation of gauge theory \ccite{Gambini:1980wm,Gambini:1986ew,Giles:1981ej},
Wilson loops constitute a basis of the Hilbert space
of gauge invariant wave functionals on the gauge field configuration space. 
Wilson loops are also the fundamental constitutive elements of loop quantum gravity, 
in particular of the spin network and foam approaches of this latter \ccite{Rovelli:1987df,Rovelli:1995ac}. 
Wilson loops are further relevant in condensed matter physics at low energy, specifically 
in the study of topologically ordered phases of matter described by topological quantum
field theories \ccite{Kitaev:2003fta,Levin:2004mi,Walker:2011mda}. 
Finally, Wilson loops can be employed to study knot and link topology using basic techniques of quantum field
theory in 3--dimensional Chern--Simons (CS) theory \ccite{Witten:1988hf}.

Higher gauge theory is a generalization of ordinary gauge theory where gauge fields
are higher degree forms \ccite{Baez:2010ya,Saemann:2016sis} that is relevant in string theory \ccite{Jurco:2019woz},  
spin foam theory \ccite{Baez:1999sr} and condensed matter physics \ccite{Zhu:2018kzd}. 
Wilson surfaces \ccite{Alvarez:1997ma,Chepelev:2001mg,Zucchini:2015wba,Zucchini:2015xba,Zucchini:2019mbz},
2--dimensional counterparts of Wilson loops, enter naturally in field theories with
higher gauge fields and are expected to be essential elements in the analysis of important aspects 
of them for reasons analogous to those for which Wilson loops are. 

In 4 spacetime dimensions, fractional braiding statistics is adequately described
through the correlation functions of Wilson loops and surfaces in BF type topological quantum
field theories \ccite{Balachandran:1992qg,Bergeron:1994ym,Szabo:1998ej}. 
Wilson surfaces also should be a basic element of any field theoretic approach to the study
of 2--dimensional knot topology \ccite{CottaRamusino:1994ez,Cattaneo:2002tk} through an appropriate
4--dimensional version of CS theory \ccite{Soncini:2014ara,Zucchini:2015ohw,Zucchini:2021bnn}.

The goal of the present two--part study is constructing a 2--dimensional topological 
sigma model whose quantum partition function computes a Wilson surface
in strict higher gauge theory on the same lines as the 1--dimensional topological 
sigma model furnishing a Wilson loop in ordinary gauge theory.


The idea of representing a given Wilson loop as the partition function
of a suitable quantum mechanical system 
can be traced back to the work of Balachandran {\it et al.}
\ccite{Balachandran:1977ub}. The approach was subsequently further 
developed by Alekseev {\it et al.} in \ccite{Alekseev:1988vx} 
and Diakonov and Petrov in \ccite{Diakonov:1989fc,Diakonov:1996zu}.
It was more recently applied to the canonical quantization of CS 
theory by Elitzur {\it et al.} in \ccite{Elitzur:1989nr}.
See \ccite{Witten:1999ams,Beasley:2009mb,Alexandrov:2011ab}
for readable reviews.

The quantum system underlying the partition function realization of a Wilson loop has an explicitly
Hamiltonian description. The underlying phase space is a coadjoint orbit.
As a symplectic manifold, it is described by Kirillov--Kostant--Souriau (KKS) theory \ccite{Kirillov:2004lom}
and is quantized using the methods of geometric quantization \ccite{Kostant:1970qur,Souriau:1970sds}. 
The resulting quantum theory can be understood at the light of the Borel--Weil theorem
\ccite{Bott:1957abc,Kirillov:1976etr}. It can also be reproduced as the functional integral
quantization of a 1--dimensional sigma model,
the topological coadjoint orbit (TCO) model. 


The problem of obtaining a partition function realization of a Wilson surface,
the main object of our study, has been tackled previously in the literature from
different perspectives \ccite{Alekseev:2015hda,Chekeres:2018kmh,Chekeres:2019xit}. 
Our approach to the topic is firmly framed in higher gauge theory. 
It aims to formulate a higher KKS theory
and to construct a higher version of the TCO model through the 
derived geometrical framework worked out in  refs.
\ccite{Zucchini:2019rpp,Zucchini:2019pbv}.


\subsection{\textcolor{blue}{\sffamily Plan of the endeavour}}\label{subsec:wsproject}

The present endeavour is naturally divided in two parts, which we refer to as I and II,
of which the present paper is the second.

In I, a higher version of the KKS theory of coadjoint orbits is elaborated based
on the derived geometric framework. An original definition of derived coadjoint orbit
is proposed. A theory of derived unitary line bundles and Poisson structures on regular
derived orbits is built. The proper derived counterpart of the Bohr–Sommerfeld
quantization condition is then identified. A version of derived prequantization is put
forward. The problems hindering a full quantization are discussed and a possible
solution is suggested. The theory worked out and the results derived, mostly of a
geometric nature, provide the grounding for the field theoretic constructions of II.

In II, the derived TCO sigma model is presented and studied in depth. Its symmetries
are described. Its quantization is analyzed in the functional integral set--up. Substantial 
evidence is provided that the model does indeed underpins the partition function realization of a
Wilson surface. It is shown how the vanishing fake curvature condition arises in this
context and homotopy invariance for flat derived  gauge field is proven.
The model's Hamiltonian formulation is further furnished highlighting
the model's close relationship to the derived KKS theory developed in I.

\vfil\eject

\renewcommand{\sectionmark}[1]{\markright{\thesection\ ~~#1}}

\section{\textcolor{blue}{\sffamily Part II: derived TCO model}}\label{sec:wsparttwo}

The present paper, which constitutes part II of our endeavour, is
devoted to the derived TCO model. 
In this section, we provide an introductory overview of the model 
and an outlook on future developments.

The derived TCO model is a 2--dimensional field theory, which under certain conditions
turns out to be a sigma model. The model is most naturally formulated in the derived geometrical
framework reviewed in great detail in sect. 3 of I. The derived 
set--up makes evident the structural affinity of the derived model to the ordinary one.

The derived TCO model is a higher extension of the ordinary TCO model just
as derived KKS theory is a higher analog of ordinary KKS theory. 
From a formal point of view, however, the formulations of the derived and
ordinary models are not as close as those of the derived and ordinary theory as 
presented in I are. The derived model has novel features and is richer than its ordinary
counterpart is several respects. There is nevertheless
a characteristic version of the derived model, whose analogy to the ordinary model
is particularly evident and which is closely related to the derived geometric orbit theory much
as the ordinary model is related to ordinary theory. 



\subsection{\textcolor{blue}{\sffamily Plan of part II}}\label{subsec:plantwo}

Paper II is organized in three sections with the content described below.
The main results are contained in the last section. 

In sect. \cref{sec:higau}, we present derived gauge theory, a formulation of higher
gauge theory based on the derived field framework of I which brings to light the formal
affinity of higher to ordinary gauge theory. 


In sect. \cref{sec:tcoreview}, we review the ordinary TCO model. 
The topics covered are the model's formulation as a classical field theory,
basic symmetries, sigma model interpretation, functional integral quantization
and canonical analysis. The presentation of this subject we provide
is intentionally structured in a way that directly suggests
the derived extension elaborated later.

In sect. \cref{sec:hafsmod}, we finally introduce the derived TCO model.
The construction is patterned on that of the ordinary TCO model expounded in sect.
\cref{sec:tcoreview}. The model's formulation as a classical
field theory, basic symmetries, interpretation as a sigma model, functional integral quantization
and canonical analysis are so studied in depth. Relevant original traits that distinguish
the derived model from the ordinary one, in particular the existence of gauge background
preserving gauge symmetry are highlighted. We present a number of arguments pointing to the
conclusion that the partition function of the derived model is to be identified with 
a Wilson surface depending on the model's data.
We show in particular how the vanishing fake curvature condition emerges and 
homotopy invariance for a flat higher gauge field is proven.
Finally, by means of a Hamiltonian formulation, we highlight the close relationship of the derived TCO
model to the derived KKS theory developed in I.
We also call attention to limitations of our analysis related to the non rigorous nature of the formal 
functional integral techniques used and the lack of a in--depth analysis of gauge fixing.

\subsection{\textcolor{blue}{\sffamily Overview of the derived TCO model}}\label{subsec:tcoproject}

As we anticipated above, the formulation of the derived TCO model hinges on the derived
geometrical set up introduced and described in I. 
As our remarks about this model have been merely qualitative up to this point,
in the rest of this subsection we provide a somewhat more formal introduction to it.
A complete more rigorous analysis of the material surveyed below is available in the main body
of the paper according to the plan stated in subsect. \cref{subsec:plantwo} above.
To justify and provide motivation for the derived approach followed by us, we shall
initially review briefly the derived CS theory worked out in ref. \ccite{Zucchini:2021bnn}.

Higher gauge symmetry is described by crossed modules. 
A Lie group crossed module $\msM$ consists of a source and target Lie group, $\msE$ and $\msG$, 
together with a target and an action structure map, $\tau$ and $\mu$,
relating them with certain properties \ccite{Baez5,Baez:2003fs}
In the derived set--up, with a Lie group crossed module $\msM$ there is associated
a derived Lie group $\DD\msM$ with derived Lie algebra $\DD\fkm$.

Derived gauge fields on a manifold $X$ are just $\DD\msM$-- or $\DD\fkm$--valued
fields. Integration is expressed by means of the Berezinian $\varrho_X$
of $X$. Upon equipping $\msM$ with an invariant pairing $\langle\cdot,\cdot\rangle$,
it is possible to construct a degree 1 graded symmetric pairing $(\cdot,\cdot)$ of derived fields.

A higher gauge theory with gauge 
crossed module $\msM$ features a higher gauge field $\Omega$
with 1-- and 2--form components $\omega$, $\varOmega$ valued respectively in the Lie algebras $\fkg$, $\fke$ of
$\msG$, $\msE$ and its higher gauge curvature $\Phi$ with 2--and 3--form components $\phi$, $\varPhi$ valued
in the same way and expressed in terms of $\omega$, $\varOmega$ in a definite way 
\ccite{Baez:2004in,Baez:2005qu}. A gauge transformation $\rmU$ comprises a $\msG$--valued component
$u$ and an $\fke$--valued 1--form component $U$ acting on the gauge field components $\omega$, $\varOmega$
according to a precise rule. 
In the derived formulation of higher gauge theory, a higher gauge field $\Omega$ is just
a $\DD\fkm[1]$--valued derived field. 
The curvature $\Phi$ of \linebreak $\Omega$ is the $\DD\fkm[2]$--valued derived field
given in terms of $\Omega$ by the usual gauge theoretic relation,
$\Phi=\dd\Omega+\frac{1}{2}[\Omega,\Omega]$. 
Similarly, a higher gauge transformation $\rmU$ is a $\DD\msM$--valued derived field acting on $\Omega$
and consequently $\Phi$ in the familiar manner, viz
$\Omega^\rmU=\Ad\rmU(\Omega)+\rmU^{-1}\dd\rmU$, $\Phi^\rmU=\Ad\rmU(\Phi)$.
The components of $\Omega$, $\Phi$ as well as $\rmU$ as derived fields correspond
precisely to the components of the fields $\Omega$, $\Phi$ and $\rmU$ stand for in higher gauge theory.
Further, the above relations, once expressed in components, take precisely
the same form as they do in higher gauge theory.  

The higher 4--dimensional CS theory worked out in ref. \ccite{Zucchini:2021bnn}
is most naturally formulated in the derived field framework and so  
it is referred to as derived CS theory. The derived CS model is characterized
by two basic data: a crossed mo\-dule $\msM$ encoding the model's symmetry
equipped with an invariant pairing and a base 4--fold $M$. The model's field
content consists of a derived gauge field $\Omega$.
The derived CS action reads as 
\begin{equation}
\sfC\hspace{-.75pt}\sfS(\Omega)
=\frac{k}{4\pi}\int_{T[1]M}\varrho_M\!\left(\Omega,\dd\Omega+\tfrac{1}{3}[\Omega,\Omega]\right)\!,
\label{phys14}
\end{equation}
where $k$ is a constant, the CS level. 
In spite of evident similarities, the derived CS model differs from the ordinary one
in important respects. It is 4--dimensional as claimed because of the 1 unit of degree provided by the
pairing $(\cdot,\cdot)$. Further, it is fully gauge invariant
if the 4–fold $M$ has no boundary. When it does, it is gauge variant, the gauge
variation of the action reducing to a boundary term. Level quantization can occur only when special boundary
conditions are obeyed by the derived gauge fields and gauge transformations.
Finally, the models' canonical formulation on a 3--fold with boundary exhibits 
a rich edge field mode structure. 

In the present paper, we proceed along the same lines as CS theory 
and employ all the main elements of the derived field set--up to formulate the higher
2--dimensional TCO model relevant for Wilson surface
partition function realization directly as a derived TCO model.

The derived TCO model we have in mind is therefore a 2--dimensional topological sigma model with
a derived homogeneous space as target space (cf. subsect. 5.2 of I). 
As derived CS theory has been built on the assumption of a formal correspondence to ordinary CS theory,
it is reasonable to hypothesize that derived TCO theory may be built by invoking 
a similar formal accordance to ordinary TCO theory. Proceeding in this way leads to the
derived field theory outlined next. The data of the model are a crossed module $\msM$
with invariant pairing, a base 2--fold $N$, a target manifold $M$ and an embedding $\varsigma$ of $N$ into $M$.
The field content consists of a $\DD\msM$--valued derived TCO field $\rmG$. The model's  action is 
\begin{equation}
\sfS(\rmG;\Omega)=\int_{T[1]N}\varrho_N(\rmK,\Ad\rmG^{-1}(\varsigma^*\Omega)+\rmG^{-1}\dd\rmG),
\label{phys16}
\end{equation}
where $\rmK$ is a fixed degree 0 $\DD\fkm$--valued level current such that $\dd\rmK=0$
and $\Omega$ is a background target space
derived gauge field. The structural congruity of $\sfS$ with the 1--dimensional
TCO action (1.2.8) of I is manifest. Derived TCO theory
is however 2--dimensional again because of the 1 unit of degree supplied 
by the pairing $(\cdot,\cdot)$. Further, when formulated through the components
of $\rmG$ in conventional terms, it features a $\msG$--valued field $g$ and a $\fke$--valued 1--form field $G$.


There are non trivial conditions which the level current $\rmK$ must obey in order the TCO model
to be a sigma model having a derived homogeneous space as target space as desired. 
In the absence of these restrictions, \ceqref{phys16} describes only a sigma model with $\DD\msM$ as target space.
A $\DD\msM$--valued map $\Upsilon$ is called a level preserving gauge transformation if $\Ad\Upsilon(\rmK)=\rmK$. 
Under right multiplication of the derived TCO fields $\rmG$ by any such gauge transformation $\Upsilon$,
the action $\sfS$ is invariant up to an additive term depending only on $\Upsilon$ and homotopically invariant. 
If the level current $\rmK$ satisfies an appropriate integrality condition, $\sfS$ is invariant mod $2\pi$.  
As a consequence, at the quantum level, the TCO model enjoys a level preserving gauge symmetry.
Under mild conditions, the level preserving gauge transformations $\Upsilon$ are precisely those which are
$\DD\msM_\rmK$--valued, where $\msM_\rmK$ is a certain crossed submodule of $\msM$ depending on $\rmK$. In this way, 
the model is a sigma model having the derived homogeneous space $\DD\msM/\DD\msM_\rmK$ as its target space.

In close analogy to the 
CS model,  the derived TCO model has a number of features with no analogue in the ordinary one. 
The field equations are integrable only if a certain integrability condition is satisfied,
which reduces to the fake flatness of $\Omega$ well--known in higher gauge theory. It has
further a novel gauge background preserving gauge symmetry associated with the special 
gauge transformations leaving the gauge background $\varsigma^*\Omega$ invariant. 

A detailed functional integral analysis indicates that the quantum partition function $\sfZ(\Omega)$
of the derived TCO model enjoys the properties which a Wilson surface $W_{R}(N)$ is supposed
to do. In fact, it is properly defined only when the background derived gauge field $\Omega$
is fake flat. As a functional of $\Omega$, $\sfZ(\Omega)$ is gauge invariant. Furthermore,
when $\Omega$ is flat, $\sfZ(\Omega)$ is also invariant under smooth variations of the embedding
$\varsigma$ by virtue of certain Schwinger--Dyson relations under mild assumptions on the symmetry
crossed module $\msM$. 
All this provides strong evidence that $\sfZ(\Omega)$ does indeed constitute a functional integral
realization of an underlying Wilson surface. 

For a special choice of the level current $\rmK$ specified by an element $\varLambda$
of the source Lie algebra $\fke$, the derived TCO sigma model takes a special form, which
we dub `characteristic'. The model's action, expressed explicitly in components, takes the form
\begin{multline}
\sfS(g, G;\omega,\varOmega)
=\int_{T[1]N}\varrho_N
\left\langle\dot\tau(\varLambda),\mu\sdot\left(g^{-1},
\varsigma^*\varOmega+\sdot\mu\sdot(\varsigma^*\omega, G)
+dG+\tfrac{1}{2}[G, G]\right)\right\rangle
\\
-\int_{T[1]\partial N}\varrho_{\partial N}
\left\langle\Ad g^{-1}\mhfpt\left(\varsigma^*\omega+dg g^{-1}
+\dot\tau(G)\right),\varLambda\right\rangle.
\label{}
\end{multline}
The characteristic model so features an ordinary TCO field, $g$, coupling to a kind of gauge field, $G$.
Under rather general assumptions on the crossed module $\msM$, its pairing $\langle\cdot,\cdot\rangle$
and the datum $\varLambda$, the model is in fact a sigma model. Its target space is the derived coadjoint
orbit $\clO_\varLambda$ of $\varLambda$, precisely the kind of derived homogeneous space studied
by the derived KKS theory elaborated in I.  The relationship of the model to this latter
is in this way established, showing convincingly that its quantization must definitely
proceed in the realm of 2--dimensional quantum field theory.



\subsection{\textcolor{blue}{\sffamily Outlook}}\label{subsec:outlooktwo}

To summarize, 
in this paper we work out a 2--dimensional topological quantum field theory, the derived TCO
model, whose partition function provides a candidate field theoretic expression of a corresponding
Wilson surface. We present mounting evidence for such an identification based on the matching of several
relevant properties. The data specifying the derived TCO model encode those defining the Wilson surface.
Equally significantly, the relationship of the characteristic version of model and derived KKS theory sheds light
on the quantization of this latter. 

Though we have made considerable progress toward the goal we set, a few basic issues remain unsolved. 
The derived TCO model has an extra gauge background preserving gauge symmetry with no analog in the ordinary model.
This symmetry has to be appropriately taken care of by fixing the gauge and introducing ghost
and antighost fields. Given the geometric nature of the model, this problem is likely to
be amenable by the quantization scheme of Batalin and Vilkovsky \ccite{BV1,BV2}
especially in the formulation of this elaborated by Alexandrov {\it et al.} \ccite{Alexandrov:1995kv}.
Finding a viable gauge fixing condition or equivalently a suitable gauge fermion
may however turn out to be a hard problem.

Detailed calculations on specific input data would be desirable to further test the model.
Further, following the path set long ago in refs. \ccite{CottaRamusino:1994ez,Cattaneo:2002tk},
one may try to use the model in 4--dimensional CS theory
to study 2--dimensional knot invariants with the basic techniques of quantum field theory
along the same lines as ordinary CS theory \ccite{Witten:1988hf}.

\vfill\eject

\vfil\eject

\renewcommand{\sectionmark}[1]{\markright{\thesection\ ~~#1}}

\section{\textcolor{blue}{\sffamily Higher gauge theory in the derived formulation}}\label{sec:higau}

In this section, we present derived gauge theory, a formulation of higher gauge theory based on the derived 
framework of sect. 3 of I. 
The derived field formalism has the virtue of bringing to light
the close relationship of higher to ordinary gau\-ge theory and allows so to import many ideas and techniques
of the latter to the former. The benefits of this approach, which showed themselves previously in the construction
of 4d CS theory in ref. \ccite{Zucchini:2021bnn}, will become evident in the elaboration 
of the derived TCO model in sect. \cref{sec:hafsmod} below. 

The topics covered include derived gauge fields and gauge transformations and special derived gauge
symmetry. The latter reflects a gauge for gauge symmetry of fake flat gauge fields in higher gauge theory
and underlies a basic gauge symmetry of the derived TCO model with no ordinary counterpart.


\subsection{\textcolor{blue}{\sffamily Derived gauge fields and transformations}}\label{subsec:higau}

In this subsection, we introduce derived gauge theory. Based on the derived field
formalism of I, this constitutes an equivalent reformulation of
higher gauge theory as ordinary gauge theory with exotic gauge group, the derived group
of the relevant symmetry crossed module.

In the derived field formalism, fields on a base manifold $X$ are valued either in the derived Lie group $\DD\msM$ of
a Lie group crossed module $\msM=(\msE,\msG,\tau,\mu)$
or in the derived Lie algebra $\DD\fkm$ of the associated Lie algebra crossed module
$\fkm=(\fke,\fkg,\dot\tau,\sdot\mu{}\sdot\hfpt)$. 
$\DD\msM$--valued derived fields are elements of the mapping space $\Map(T[1]X,\DD\msM)$.
If $\rmW\in\Map(T[1]X,\DD\msM)$ is one such field, then 
\begin{equation}
\rmW(\alpha)=\ee^{\alpha W}w 
\label{superfield1}
\end{equation}
with $\alpha\in\bbR[-1]$, 
where $w\in\Map(T[1]X,\msG)$, $W\in\Map(T[1]X,\fke[1])$. $w$, $W$
are called the components of $\rmW$. Similarly, 
$\DD\fkm$--valued derived fields are elements of the mapping space
$\Map(T[1]X,\DD\fkm[p])$ for some integer $p$. If $\Psi\in\Map(T[1]X,\DD\fkm[p])$
is a field of this kind, then \hphantom{xxxxxxxxxxx}
\begin{equation}
\Psi(\alpha)=\psi+(-1)^p\alpha\varPsi,  
\label{superfield3}
\end{equation}
where $\psi\in\Map(T[1]X,\fkg[p])$, $\varPsi\in\Map(T[1]X,\fke[p+1])$, the components of $\Psi$.
A more comprehensive review of the derived field formalism is provided in subsect. 3.3 of I.
Here, we employ the ordinary non internal version of the formalism.


In higher gauge theory, a Lie group crossed module $\msM$ 
is assigned and a higher principal $\msM$--bundle $P$ on a base manifold $X$ is given. 
Higher gauge fields and gauge transformations consist in collections of local
Lie valued map and form data organized respectively as non Abelian differential cocycles
and cocycle morphisms \ccite{Baez:2004in,Baez:2005qu}.
As the analysis provided below is of a local nature, this level of generality is not necessary.
We thus restrict ourselves to the case where
$P$ is the trivial $\msM$--bundle for which gauge fields and gauge transformations turn out to be 
maps and forms globally defined on $X$.

%

The basic field of higher gauge theory is the derived gauge field, 
that is a map
$\Omega\in\Map(T[1]X,\DD\fkm[1])$. In components, this reads as 
\begin{equation}
\Omega(\alpha)=\omega-\alpha\varOmega, 
\label{higau1}
\end{equation}
where $\omega\in\Map(T[1]X,\fkg[1])$, $\varOmega\in\Map(T[1]X,\fke[2])$
(cf. eq. (3.3.5) of I). $\omega$, $\varOmega$ are nothing but the familiar $1$-- and $2$--form
gauge fields of 
higher gauge theory.

The derived gauge field $\Omega$ is characterized by its curvature $\Phi$ defined by
\begin{equation}
\Phi=\dd\Omega+\tfrac{1}{2}[\Omega,\Omega], 
\label{higau2}
\end{equation}
where the Lie bracket $[\cdot,\cdot]$ and the differential $\dd$ are defined by (3.3.6) 
and (3.3.10) of I, respectively. The expression of $\Phi$ is otherwise formally identical to that
of the curvature of a gauge field in ordinary gauge theory. By construction,
$\Phi\in\Map(T[1]X,\DD\fkm[2])$. Expressed in components, $\Phi$ reads as 
\begin{equation}
\Phi(\alpha)=\phi+\alpha\varPhi,  
\label{higau3}
\end{equation}
where $\phi\in\Map(T[1]X,\fkg[2])$, $\varPhi\in\Map(T[1]X,\fke[3])$. $\phi$, $\varPhi$
are just the usual higher gauge theoretic $2$-- and $3$--form curvatures. They 
are expressible in terms of $\omega$, $\varOmega$ through the familiar relations 
\begin{align}
&\phi=d\omega+\tfrac{1}{2}[\omega,\omega]-\dot\tau(\varOmega),
\vphantom{\Big]}
\label{higau4}
\\
&\varPhi=d\varOmega+\sdot\mu\sdot(\omega,\varOmega). 
\vphantom{\Big]}
\label{higau5}
\end{align}

The derived curvature $\Phi$ satisfies the derived Bianchi identity
\begin{equation}
\dd\Phi+[\Omega,\Phi]=0,
\label{higau6}
\end{equation}
which follows from \ceqref{higau2} in the usual way. This turns into a pair of Bianchi identities for the
curvature components $\phi$, $\varPhi$, viz
\begin{align}
&d\phi+[\omega,\phi]+\dot\tau(\varPhi)=0,
\vphantom{\Big]}
\label{higau7}
\\
&d\varPhi+\sdot\mu\sdot(\omega,\varPhi)-\sdot\mu\sdot(\phi,\varOmega)=0.
\vphantom{\Big]}
\label{higau8}
\end{align}

A derived gauge transformation is codified in a derived Lie group valued map $\rmU\in\Map(T[1]X,\DD\msM)$. 
$\rmU$ acts on the derived gauge field $\Omega$ as
\begin{equation}
\Omega^{\rmU}=\Ad\rmU^{-1}(\Omega)+\rmU^{-1}\mathrm{dU},
\label{higau9}
\end{equation}
where the adjoint action and pulled--back Maurer--Cartan form of $U$ in the right hand side are defined 
in eqs. (3.3.8) and (3.3.14) of I, respectively. Again, in the derived formulation
the derived gauge transformation action is formally identical to that of ordinary gauge theory.
The derived curvature transforms as
\begin{equation}
\Phi^{\rmU}=\Ad\rmU^{-1}(\Phi), 
\label{higau10}
\end{equation}
as expected. The gauge transformation $\rmU$ can be expressed in components as
\begin{equation}
\rmU(\alpha)=\ee^{\alpha U}u
\label{higau11}
\end{equation}
with $u\in\Map(T[1]X,\msG)$, $U\in\Map(T[1]X,\fke[1])$  
according to (3.3.1) of I. In terms of these,
using systematically relations (3.3.8), (3.3.14) of I,
it is possible to write down the gauge transforms of the derived gauge field components
$\omega$, $\varOmega$,  
\begin{align}
&\omega^{u,U}=\Ad u^{-1}\!\left(\hspace{.5pt}\omega+duu^{-1}+\dot\tau(U)\right),
\vphantom{\Big]}
\label{higau12}
\\
&\varOmega^{u,U}=\mu\sdot\!\left(u^{-1},\varOmega+\sdot\mu\sdot(\omega,U)+dU+\tfrac{1}{2}[U,U]\right),
\vphantom{\Big]}
\label{higau13}
\end{align}
as well as those of the derived curvature components $\phi$, $\varPhi$,
\begin{align}
&\phi^{u,U}=\Ad u^{-1}\!\left(\hfpt\phi\right),
\vphantom{\Big]}
\label{higau14}
\\
&\varPhi^{u,U}=\mu\sdot\!\left(u^{-1},\varPhi+\sdot\mu\sdot(\phi,U)\right).
\vphantom{\Big]}
\label{higau15}
\end{align}
These relations are the well--known expressions of the gauge transforms of the 
1-- and 2--form gauge field and 2-- and 3--form curvature components of higher gauge theory.

An infinitesimal derived gauge transformation is a derived Lie algebra valued map
$\Theta\in\Map(T[1]X,\DD\fkm)$. The gauge variation of the derived gauge field $\Omega$ is
\begin{equation}
\delta_\Theta\Omega=\dd\Theta+[\Omega,\Theta],
\label{higau20}
\end{equation}
where as before the Lie bracket $[\cdot,\cdot]$ and the differential $\dd$ are given by (3.3.6) 
and (3.3.10) of I, respectively. In the derived formulation, the infinitesimal derived gauge transformation
action is again formally identical to that of ordinary gauge theory, in particular it is the linearized form
of its finite counterpart. As expected, so, the gauge variation of the derived curvature $\Phi$ reads as 
\begin{equation}
\delta_\Theta\Phi=[\Phi,\Theta]. 
\label{higau21}
\end{equation}
The gauge transformation $\Theta$ can be expressed in components as
\begin{equation}
\Theta(\alpha)=\theta+\alpha\varTheta,
\label{higau22}
\end{equation}
where $\theta\in\Map(T[1]X,\fkg)$, $\varTheta\in\Map(T[1]X,\fke[1])$  
according to (3.3.1) of I. In terms of these,
exploiting relations (3.3.8), (3.3.14) of I,
we can write down the gauge variations of the derived gauge field components
$\omega$, $\varOmega$,  
\begin{align}
&\delta_{\theta,\varTheta}\omega=d\theta+[\omega,\theta]+\dot\tau(\varTheta),
\vphantom{\Big]}
\label{higau23}
\\
&\delta_{\theta,\varTheta}\varOmega=d\varTheta+\sdot\mu\sdot(\omega,\varTheta)-\sdot\mu\sdot\left(\theta,\varOmega\right),
\vphantom{\Big]}
\label{higau24}
\end{align}
as well as those of the derived curvature components $\phi$, $\varPhi$, 
\begin{align}
&\delta_{\theta,\varTheta}\phi=[\phi,\theta],
\vphantom{\Big]}
\label{higau25}
\\
&\delta_{\theta,\varTheta}\varPhi=\sdot\mu\sdot\left(\phi,\varTheta\right)-\sdot\mu\sdot\left(\theta,\varPhi\right).
\vphantom{\Big]}
\label{higau26}
\end{align}
Once more, these are the well--known expressions of the gauge variations 
of the 1-- and 2--form gauge fields and the 2-- and 3--form curvatures in higher gauge theory.

\vfil\eject



A gauge transformation $\rmT$ is special if its components $t$, $T$ have the form 
\begin{align}
&t=\tau(A),
\vphantom{\Big]}
\label{higau16}
\\
&T=-dAA^{-1}-\sdot\mu(\omega,A), 
\vphantom{\Big]}
\label{higau17}
\end{align}
where $A\in\Map(T[1]X,\msE)$. Note the dependence on an underlying derived gauge field $\Omega$.
By \ceqref{higau12}, \ceqref{higau13}, 
its action on the gauge field components $\omega$, $\varOmega$ is 
\begin{align}
&\omega^{t,T}=\omega,
\vphantom{\Big]}
\label{higau18}
\\
&\varOmega^{t,T}=\varOmega+\sdot\mu(\phi,A^{-1}).
\vphantom{\Big]}
\label{higau19}
\end{align}
$\omega$ is so invariant. $\varOmega$ is not except for when the curvature component $\phi$ vanishes.
The requirement that $\phi=0$ is known in higher gauge theory as zero 
fake curvature condition.
A derived gauge field $\Omega$ with this property is called fake flat.
Fake flat gauge fields play an fundamental role in higher gauge theory. 


An infinitesimal special gauge transformation $\Xi$ has components 
\begin{align}
&\xi=\dot\tau(\varPi),
\vphantom{\Big]}\vphantom{\ul{\ul{\ul{g}}}}
\label{higau27}
\\
&\varXi=-d\varPi-\sdot\mu\sdot\left(\omega,\varPi\right),
\label{higau28}
\end{align}
where $\varPi\in\Map(T[1]X,\fke)$. In keeping with \ceqref{higau18}, \ceqref{higau19},
the corresponding variations of the derived gauge field components $\omega$, $\varOmega$ are 
\begin{align}
&\delta_{\xi,\varXi}\omega=0, 
\vphantom{\Big]^f}
\label{higau29}
\\
&\delta_{\xi,\varXi}\varOmega=-\sdot\mu\sdot\left(\phi,\varPi\right)\!,
\vphantom{\Big]_f}
\label{higau30}
\end{align}
with $\varOmega$ invariant if the fake flatness condition $\phi=0$ obtains. 

We conclude this subsection introducing some notation that will be used frequently in the following. 
The field space of pure higher gauge theory on the manifold $X$ is precisely 
the derived gauge field manifold
$\clC_{\msM}(X)=\Map(T[1]X,\DD\fkm[1])$. Finite derived gauge transformations are organized
in an infinite dimensional Lie group, the derived gauge transformation group $\clG_{\msM}(X)=\Map(T[1]X,\DD\msM)$,
acting on $\clC_{\msM}(X)$ according to \ceqref{higau9}. Infinitesimal derived gauge transformations 
are similarly structured in an infinite dimensional Lie algebra, the derived gauge transformation algebra
$\fkg_{\msM}(X)=\Map(T[1]X,\DD\fkm)$, acting variationally on $\clC_{\msM}(X)$ through \ceqref{higau20}.
Their operations are pointwise Lie group multiplication and inversion and bracketing, respectively. 
$\fkg_{\msM}(X)$ is further the virtual Lie algebra of $\clG_{\msM}(X)$. 

Special gauge transformations constitute in their finite and infinitesimal form
a Lie subgroup $\clG_{\msM,\Omega}(X)$
of $\clG_{\msM}(X)$ and a Lie subalgebra $\fkg_{\msM,\Omega}(X)$ of $\fkg_{\msM}(X)$
depending on an assigned gauge field $\Omega$. 


\subsection{\textcolor{blue}{\sffamily Distinguished features of special gauge symmetry}}\label{subsec:hisym}

Special gauge symmetry is a manifestation of a gauge for gauge symmetry
of higher gauge theory: special gauge transformations are non trivial gauge transformations acting trivially
on fake flat gauge fields. In the derived TCO model elaborated in sect. \cref{sec:hafsmod},
special gauge symmetry is associated with a basic gauge background preserving
gauge symmetry. Given the importance that this latter holds in the model,
we analyze its distinctive features in greater depth in this subsection. 

Relations \ceqref{higau16}, \ceqref{higau17} define a surjective map 
$\sfT_\Omega:\clH_\msM(X)\rightarrow\clG_{\msM,\Omega}(X)$ depending on $\Omega$, where $\clH_\msM(X)=\Map(T[1]X,\msE)$.
$\sfT_\Omega$ is a Lie group morphism, as is straightforward to demonstrate. 
The kernel of $\sfT_\Omega$, $\clK_{\msM,\Omega}(X)$, is a noteworthy \ subgroup of $\clH_\msM(X)$, as it 
consists of the elements $A\in\clH_\msM(X)$ whose associated special 
gauge transformations are trivial. Explicitly, $\clK_{\msM,\Omega}(X)$
is formed by the maps $A\in\Map(T[1]X,\ker\tau)$ obeying the equation 
\begin{equation}
dAA^{-1}+\sdot\mu(\omega,A)=0. 
\label{higau31}
\end{equation}
This equation is of a form analogous to that of the equation obeyed by the gauge transformations leaving a
given gauge field invariant. We thus expect that 
$\clK_{\msM,\Omega}(X)$, unlike $\clH_\msM(X)$, is generally finite dimensional. 
Infinitesimally, we have a surjective map 
$\dot\sfT_\Omega:\fkh_\msM(X)\rightarrow\fkg_{\msM,\Omega}(X)$ defined by relations \ceqref{higau27}, \ceqref{higau28},
where $\fkh_\msM(X)=\Map(T[1]X,\fke)$ is the Lie algebra of $\clH_\msM(X)$. $\dot\sfT_\Omega$ is a Lie 
algebra morphism whose kernel is the Lie subalgebra
$\fkk_{\msM,\Omega}(X)$ of $\fkh_\msM(X)$ formed by the elements $\varPi\in\fkh_\msM(X)$ with trivial 
associated infinitesimal special gauge transfor\-mations. $\fkk_{\msM,\Omega}(X)$
consists so of the maps $\varPi\in\Map(T[1]X,\ker\dot\tau)$ such that 
\begin{equation}
d\varPi+\sdot\mu\sdot(\omega,\varPi)=0
\label{higau32}
\end{equation}
and is generally finite dimensional.

Via $\sfT_\Omega$, the group $\clH_\msM(X)$ can be considered as a parameter space for the special gauge
transformation group $\clG_{\msM,\Omega}(X)$. The parametrization however is generally not one--to--one
because of the kernel $\clK_{\msM,\Omega}(X)$. Similarly, the algebra $\fkh_\msM(X)$ is a parameter space 
for the infinitesimal special gauge transformation algebra $\fkg_{\msM,\Omega}(X)$ which again is generally not
one--to--one because of the kernel $\fkk_{\msM,\Omega}(X)$.

For a fixed derived gauge field $\Omega\in\clC_\msM(X)$, denote by $\clG^*{}_{\msM,\Omega}(X)$ the set of all
gauge transformations $\rmT\in\clG_\msM(X)$ with the property that $\Omega^\rmT=\Omega$. $\clG^*{}_{\msM,\Omega}(X)$ 
is a subgroup of $\clG_\msM(X)$, the invariance subgroup of $\Omega$. 
When $\Omega$ is fake flat, as we shall assume, the structure of $\clG^*{}_{\msM,\Omega}(X)$ can be
analyzed precisely. In this case, as is not difficult to check 
from \ceqref{higau16}, \ceqref{higau17}, the special gauge transformation group $\clG_{\msM,\Omega}(X)$ is contained
in $\clG^*{}_{\msM,\Omega}(X)$ as a normal subgroup.
The essential invariance group of $\Omega$, the quotient group
\begin{equation}
\clI_{\msM,\Omega}(X)=\clG^*{}_{\msM,\Omega}(X)/\clG_{\msM,\Omega}(X)
\label{higau33/q}
\end{equation}
is then defined. 
Intuitively, $\clI_{\msM,\Omega}(X)$ describes the gauge transformations leaving $\Omega$ invariant
not reducible to `trivial' special gauge transformations.  

There are indications 
that $\clI_{\msM,\Omega}(X)$ is a finite dimensional Lie group. We illustrate this for the simplest choice
of $\Omega$, namely $\Omega=0$. In this case, a gauge transformation
$\rmT$ belongs to $\clG^*{}_{\msM,\Omega}(X)$ if and only if 
\begin{equation}
dtt^{-1}+\dot\tau(T)=0, \qquad dT+\tfrac{1}{2}[T,T]=0
\label{higau33}
\end{equation}
by \ceqref{higau12}, \ceqref{higau13}.
From \ceqref{higau27}, \ceqref{higau28}, further, two gauge transformations $\rmT,\rmT'\in\clG^*{}_{\msM,\Omega}(X)$
are equivalent mod $\clG_{\msM,\Omega}(X)$ if and only if 
\begin{equation}
t'=\tau(A)t,\qquad T'=\Ad A(T)-dAA^{-1}
\label{higau34}
\end{equation}
for some $A\in\clH_\msM(X)$. By \ceqref{higau33}, when $\rmT\in\clG^*{}_{\msM,\Omega}(X)$, $T$ is a flat
ordinary $\msE$--gauge field and $t$ is a $\msG$--valued function gauge trivializing the associated $\msG$--gauge
field $\dot\tau(T)$, so that $T$ has $\ker\tau$, hence central, holonomy. By \ceqref{higau34}, then,
$\clI_{\msM,\Omega}(X)$ classifies such pairs $(t,T)$ modulo $\msE$--gauge transformations. Under rather mild assumptions,
that is that $X$ is connected with torsion free first homology group, it is possible to show that
$\clI_{\msM,\Omega}(X)\simeq \msG\times H^1(X,\ker\tau)$, where the two factors correspond respectively
to the value of $t$ at a base point of $X$ and the holonomy of $T$.
$\clI_{\msM,\Omega}(X)$ is thus a finite dimensional Lie group. As 
$\Omega=0$ is a non generic gauge field with a large invariance subgroup, we expect 
that for a generic fake flat gauge field $\Omega$ with a smaller invariance group $\clG^*{}_{\msM,\Omega}(X)$
the Lie group $\clI_{\msM,\Omega}(X)$ is still 
finite dimensional. 
The infinite dimensional special gauge transformation group $\clG_{\msM,\Omega}(X)$ so essentially exhausts
$\clG^*{}_{\msM,\Omega}(X)$. 

The space of fake flat gauge fields, $\clC_{\rmf\hspace{.25pt}\rmf\msM}(X)$, gets naturally
subdivided into strata based on the isomorphism class of the essential invariance group of the
gauge fields: two fake flat gauge fields $\Omega,\Omega'\in\clC_{\rmf\hspace{.25pt}\rmf\msM}(X)$
belong to the same stratum if and only if $\clI_{\msM,\Omega'}(X)\simeq\clI_{\msM,\Omega}(X)$.
Each stratum is invariant under the gauge transformation action. 
Indeed, if $\rmU\in\clG_\msM(X)$ is a gauge transformation and $\Omega$ is a fake flat gauge field,
then $\Omega^\rmU$ also is by \ceqref{higau14}. Furthermore, we have that 
$\clG^*{}_{\msM,\Omega^\rmU}(X)=\rmU^{-1}\clG^*{}_{\msM,\Omega}(X)\rmU$
and by virtue of \ceqref{higau16}, \ceqref{higau17} also that
$\clG_{\msM,\Omega^\rmU}(X)=\rmU^{-1}\clG_{\msM,\Omega}(X)\rmU$, as is straightforward to demonstrate.
The conjugate group $\rmU^{-1}\clI_{\msM,\Omega}(X)\rmU$ of $\clI_{\msM,\Omega}(X)$ is hence defined and we have
in addition that $\clI_{\msM,\Omega^\rmU}(X)=\rmU^{-1}\clI_{\msM,\Omega}(X)\rmU$.
Thus, $\clI_{\msM,\Omega^\rmU}(X)\simeq\clI_{\msM,\Omega}(X)$ and so 
$\Omega$, $\Omega^\rmU$ belong to the same stratum. 
The stratum is made in general of several gauge orbits.

Each stratum $\clS\subset\clC_{\rmf\hspace{.25pt}\rmf\msM}(X)$ is thus characterized by a group 
$\clI_{\msM\clS}$ defined up to isomorphism by the property that $\clI_{\msM,\Omega}(X)\simeq\clI_{\msM\clS}$
for $\Omega\in\clS$. Strata $\clS$ with a higher dimensional essential invariance group $\clI_{\msM\clS}$
are higher codimensional, since they are constituted by more symmetric and thus more constrained 
gauge fields. In general, it is expected that there exist a leading 0--codimensional stratum $\clS_0$ whose associated
group $\clI_{\msM\clS_0}$ is trivial together with subleading higher codimensional strata $\clS$ with non trivial
groups $\clI_{\msM\clS}$.




\subsection{\textcolor{blue}{\sffamily Ordinary gauge theory from a derived perspective
}}\label{subsec:gaudeord} 

In subsect. 3.4 of I, we showed \pagebreak that the ordinary geometric framework
is in fact a special case of the derived geometric framework. This important 
property allows one to view ordinary gauge theory from a derived perspective
as a special case of derived gauge theory. We devote this final subsection to
a brief illustration of this point.

In subsect. 3.4 of I, we showed that a Lie group $\msG$ is fully codified by the unique Lie group
crossed module, also denoted as $\msG$, with trivial source group $1$ and target group $\msG$. Its Lie algebra
$\fkg$ is similarly fully codified by the unique Lie algebra crossed module denoted as $\fkg$ too
with trivial source algebra $0$ and target algebra $\fkg$. The derived Lie group
$\DD\msG$ of $\msG$ is just $\msG$ itself and similarly the derived Lie algebra $\DD\fkg$ of $\fkg$
is $\fkg$. Thus, the mapping spaces $\Map(T[1]X,\DD\msG)$ and $\Map(T[1]X,\msG)$ as well as 
the mapping spaces $\Map(T[1]X,\DD\fkg)$ and $\Map(T[1]X,\fkg)$ and their degree shifted versions
can be identified. 

Since the source Lie algebra is the trivial algebra $0$ for the crossed module $\fkg$,
a derived gauge field $\Omega$ of $\msG$ has a necessarily vanishing degree 2 component
$\varOmega$ and thus reduces to the degree 1 component $\omega$. Accordingly, the derived curvature 
$\Phi$ of $\Omega$ has the degree 2 component $\phi$ as its only non vanishing component, the degree
3 component $\varPhi$ vanishing identically by virtue of \ceqref{higau5}. Since $\varOmega=0$, $\phi$ is given
from \ceqref{higau4} by the familiar expression of ordinary gauge theory. Further, since $\varPhi=0$, 
$\phi$ satisfies by \ceqref{higau7} the usual Bianchi identity. 

A derived gauge transformation $\rmU$ of $\msG$ has similarly a necessarily vanishing degree 1 component
$U$ and thus reduces to the degree 0 component $u$. The component expression of the gauge transform of a
derived gauge field $\Omega$, eqs. \ceqref{higau12}, \ceqref{higau13}, reproduce the ordinary expression
for the degree 1 component $\omega$ and leave the degree 2 component $\varOmega$ vanishing as required.

The only special gauge transformations $\rmT$ of $\msG$ is the trivial one, as again
the source Lie group of the Lie group crossed module $\msG$ is the trivial group $1$.
For a derived gauge field $\Omega$, so, the invariance subgroup $\clG^*{}_{\msM,\Omega}(X)$
of $\Omega$ is non trivial if its component $\omega$
has an accidental symmetry. The essential invariance subgroup $\clI_{\msM,\Omega}(X)$ reduces then to 
$\clG^*{}_{\msM,\Omega}(X)$ itself.

\vfill\eject

\renewcommand{\sectionmark}[1]{\markright{\thesection\ ~~#1}}

\section{\textcolor{blue}{\sffamily Review of the ordinary TCO model}}\label{sec:tcoreview}


In this section, we shall review the ordinary TCO model. 
The topics covered are the model's formulation as a classical field theory,
basic symmetries, sigma model interpretation, functional integral quantization
and canonical analysis. 
Our exposition of this subject is purposefully patterned in a way that plainly alludes
to the derived extension presented in sect. \cref{sec:hafsmod}, even though the latter
has significant novel features. Alternative detailed exposition 
can be found in refs. \ccite{Beasley:2009mb,Alexandrov:2011ab}. 
We refer the reader to subsect. \cref{subsec:gaudeord} above for the 
gauge theoretic framework used. 


\subsection{\textcolor{blue}{\sffamily Ordinary TCO model}}\label{subsec:safsmod}

In this subsection, we shall illustrate the ordinary TCO model 
as a classical field theory. The model's basic elements are the following.
\begin{enumerate}

\item
The base manifold: an oriented compact connected 1--dimensional manifold $N$ perhaps
with boundary. 

\end{enumerate}
So, $N$ is either the circle $\bbS^1$ or the closed interval $\bbI$. 
\begin{enumerate}[resume]

\item  The ambient manifold: an oriented manifold $M$.

\end{enumerate}
The case where $M$ is a 3--fold is the most relevant for knot theoretic analyses. 
We however shall not impose such a restriction here.  
\begin{enumerate}[resume]
  
\item
The embedded curve: a base to ambient manifold embedding $\varsigma:N\rightarrow M$.

\end{enumerate}
The embedding provides a homeomorphic image of $N$ in $M$. 
\begin{enumerate}[resume]
\item
The symmetry structure: a Lie group $\msG$ and the associated Lie algebra 
$\fkg$. $\fkg$ is equipped with an invariant pairing $(\cdot,\cdot)$.

\end{enumerate}
$(\cdot,\cdot)$ allows one to systematically
identify $\fkg$ and its dual $\fkg^*$ as vector spaces.

\begin{enumerate}[resume]
\item
The ordinary TCO dynamical field space: $\clF_{\msG}(N):=\Map(T[1]N,\msG)$,
the space of ordinary Lie group valued fields $g$. 

\item
The ordinary TCO background field space: $\clC_{\msG}(M)$,  
the space of ordinary gauge fields $\omega$. 

\end{enumerate}

\noindent
The model is further characterized by a parameter: 
an element $\kappa\in\fkg$ which we shall refer to as the model's level. 


The action of the ordinary TCO model reads
\begin{equation}
\sfS(g;\omega)=\int_{T[1]N}\varrho_N(\kappa,\Ad g^{-1}(\varsigma^*\omega)+g^{-1}d g).
\label{safsmod1}
\end{equation}
Formally, so, the Lagrangian is the component along $\kappa$ of 
the gauge transform of $\omega$ by $g$. The action \ceqref{safsmod1}
defines a Schwarz type 1--dimensional topological field theory or topological quantum mechanics. 

Provided appropriate boundary conditions are imposed on $g$ when $\partial N\neq\emptyset$,
the field equations read 
\begin{equation}
[\Ad g^{-1}(\varsigma^*\omega)+g^{-1}d g,\kappa]=0.
\label{safsmod5}
\end{equation}
Since $\dim N=1$, \ceqref{safsmod5} is automatically integrable.
\ceqref{safsmod5} constrains the combination 
$\omega^g:=\Ad g^{-1}(\varsigma^*\omega)+g^{-1}d g$ to lie in  the centralizer of
the level $\kappa$ in $\fkg$. A more specific analysis can be made in the case where 
the Lie group $\msG$ is compact, as is often assumed.  
If $\kappa$ is a regular element of $\fkg$, then \ceqref{safsmod5} demands that 
$\omega^g$ belongs to $\Map(T[1]N,\fkt[1])$, where $\fkt$
is the maximal toral subalgebra of $\fkg$ in which $\kappa$ lies. If $\kappa$ is not regular, then 
\ceqref{safsmod5} allows $\omega^g$ to vary in $\Map(T[1]N,\fkh[1])$ for a larger subalgebra $\fkh$
of $\fkg$ containing $\fkt$.
In the extreme non regular case where $\kappa=0$, $\omega^g$ is completely
unconstrained by \ceqref{safsmod5}.


\subsection{\textcolor{blue}{\sffamily Symmetries of the ordinary TCO model}}\label{subsec:safssym}

In this subsection, we shall describe the symmetries of ordinary TCO theory:
the level preserving gauge symmetry and the accidental gauge background preserving symmetry,
the latter mainly for the importance its counterpart holds in the derived model. 
We shall also consider the model's background gauge symmetry.

The first distinguished symmetry of the TCO model is the level preserving gauge symmetry.\pagebreak 
The associated gauge transformation group, $\clG_{\msG,\kappa}(N)$, is the subgroup of the full
gauge transformation group $\clG_\msG(N)$ formed by the elements $\upsilon\in\clG_\msG(N)$
whose adjoint action leaves the model's level $\kappa$ invariant, 
\begin{equation}
\Ad\upsilon(\kappa)=\kappa.
\label{safssym1}
\end{equation}
$\clG_{\msG,\kappa}(N)$ is therefore just the gauge transformation
group $\clG_{\msG_{\kappa}}(N)$, where $\msG_\kappa$ denotes the invariance subgroup of $\kappa$
under the adjoint action of $\msG$. 
$\clG_{\msG,\kappa}(N)$ acts on the TCO field space $\clF_{\msG}(N)$; the action
is given by \hphantom{xxxxxxxxx}
\begin{equation}
g^\upsilon=g\upsilon 
\label{safssym2}
\end{equation}
with $\upsilon\in\clG_{\msG,\kappa}(N)$ and $g\in\clF_{\msG}(N)$. 
$\clG_{\msG,\kappa}(N)$ is instead inert on the TCO background gauge
field space $\clC_{\msG}(M)$. The TCO action $\sfS$ of eq. \ceqref{safsmod1}
is not level preserving gauge invariant.
The gauge variation it suffers under a gauge transformation $\upsilon$ 
however depends on 
$\upsilon$ but not on $g$; we have indeed \hphantom{xxxxxxxxxxxx}
\begin{equation}
\sfS(g^\upsilon;\omega)=\sfS(g;\omega)+\sfA(\upsilon),
\label{safssym4}
\end{equation}
where the (classical) gauge anomaly $\sfA$ is given by 
\begin{equation}
\sfA(\upsilon)=\int_{T[1]N}\varrho_N(\kappa,\upsilon^{-1}d\upsilon).
\label{safssym5}
\end{equation}
We shall discuss in the next subsection the reasons why the non invariance of $\sfS$ does not
compromise the level preserving gauge invariance of TCO theory both at the classical and quantum level.

The gauge background preserving symmetry is the second symmetry of the ordinary TCO model. It is 
rarely mentioned in the literature on the subject as it is merely accidental, depending as it does
on the gauge invariance properties of the pull-backed gauge field $\varsigma^*\omega$, 
which are generically trivial. It is a rigid symmetry.
The associated symmetry group, $\clG^*{}_{\msG,\varsigma^*\omega}(N)$,
is the is the subgroup of the full gauge transformation group $\clG_\msG(N)$ of the elements
$t\in\clG_\msG(N)$ such that 
\begin{equation}
\varsigma^*\omega^t=\varsigma^*\omega. 
\label{safssym12/1}
\end{equation}
$\clG^*{}_{\msG,\varsigma^*\omega}(N)$ is a finite dimensional group, which is non trivial 
only when $\varsigma^*\omega$ belongs to a finite codimensional subspace of \pagebreak 
the base manifold gauge field space $\clC_\msG(N)$.  
$\clG^*{}_{\msG,\varsigma^*\omega}(N)$ acts on the TCO field space $\clF_{\msG}(N)$:
the action reads as 
\begin{equation}
g^t=t^{-1}g.
\label{safssym12}
\end{equation}
for $t\in\clG^*{}_{\msG,\varsigma^*\omega}(N)$ and $g\in\clF_{\msG}(N)$. $\clG^*{}_{\msG,\varsigma^*\omega}(N)$ is
instead inert on the background gauge field space $\clC_{\msG}(M)$ consistently with
the invariance condition \ceqref{safssym12/1}. 
\ceqref{safssym12/1} itself immediately implies that \hphantom{xxxxxxxxx}
\begin{equation}
\sfS(g^t;\omega)=\sfS(g;\omega).
\label{safssym13}
\end{equation}
The background preserving gauge symmetry of the TCO model can be interpreted as the residual 
unbroken gauge symmetry left by the breaking of the full left $\msG$--gauge symmetry of TCO theory
acting 
according to \ceqref{safssym12} by the gauge background $\omega$.

The last symmetry of the ordinary TCO model is the background gauge
symmetry. The associated gauge transformation group is the ambient manifold 
gauge transformation group $\clG_{\msG}(M)$.
$\clG_{\msG}(M)$ acts on the TCO field space $\clF_{\msG}(N)$ as 
\begin{equation}
g^{\mathrm{u}}=\varsigma^*u^{-1}g, 
\label{safssym9}
\end{equation}
where $u\in\clG_{\msG}(M)$ and $g\in\clF_{\msG}(N)$. $\clG_{\msG}(M)$ acts similarly on the background gauge 
field space $\clC_{\msG}(M)$ by associating with any $\omega\in\clC_{\msG}(M)$ its gauge transform
$\omega^u$ as in \ceqref{higau12}. These transformations leave the action invariant,  
\begin{equation}
\sfS(g^u;\omega^u)=\sfS(g;\omega). \vphantom{\ul{\ul{g}}}
\label{safssym10}
\end{equation}


\subsection{\textcolor{blue}{\sffamily Ordinary TCO sigma model}}\label{subsec:safssigmod}

In this subsection, we explain why the ordinary TCO model is effectively 
a 1--dimensional sigma  model by virtue of the model's level preserving gauge symmetry
analyzed in subsect. \cref{subsec:safssym}

Upon modding out the level preserving gauge symmetry, the effective field space
of the ordinary TCO model is the $\clG_{\msG,\kappa}(N)$--orbit space 
\begin{equation}
\overline{\clF}_{\msG,\kappa}(N)=\clF_\msG(N)/\clG_{\msG,\kappa}(N). \vphantom{\ul{\ul{\ul{g}}}}
\label{safspath0}
\end{equation}
Since the level preserving gauge transformation group $\clG_{\msG,\kappa}(N)$ is just the group
$\clG_{\msG_{\kappa}}(N)$ of $\msG_\kappa$--valued gauge transformations,
we have 
\begin{equation}
\overline{\clF}_{\msG,\kappa}(N)=\Map(T[1]N,\msG/\msG_{\kappa}).
\label{safspath0/1}
\end{equation}
Thus, the TCO model is ultimately a sigma model over the homogeneous space $\msG/\msG_\kappa$.
Viewing it in this way sheds light also on the precise nature of the level preserving gauge symmetry.
An objection may be raised against the above conclusion however: by \ceqref{safssym4}, 
the action $\sfS$ is not level preserving gauge invariant. 
The question must be posed in the appropriate terms classically and quantically. 

In classical theory, it is the level preserving gauge covariance of the field equations 
\ceqref{safsmod5} that matters. In order this property to hold, the invariance of the action $\sfS$
up to a field independent additive term $\sfA$ is sufficient and this requirement is indeed
satisfied owing to \ceqref{safssym4}, \ceqref{safssym5}. 

In quantum theory, it is the invariance of the exponentiated action $\ee^{i\sfS}$
entering in the functional integral formulation that matters. 
From \ceqref{safssym5}, the variation of the anomaly $\sfA(\upsilon)$ with respect to $\upsilon$ is  
$\delta\sfA(\upsilon)=\int_{T[1]\partial N}\varrho_{\partial N}(\kappa,\upsilon^{-1}\delta\upsilon)$.
Hence, if either $\partial N=\emptyset$ or we allow only gauge transformations $\upsilon$ belonging 
to a subgroup of $\clG_{\msG,\kappa}(N)$ of gauge transformations obeying 
boundary conditions making the above boundary integral vanish, 
$\delta\sfA=0$ and $\sfA$ is so a discrete homotopy invariant.
Consequently, owing to \ceqref{safssym4}, 
the invariance of $\ee^{i\sfS}$ is generically ensured
provided that an appropriate quantization condition on the model's level $\kappa$
rendering the exponentiated anomaly $\ee^{i\sfA}$ trivial
is satisfied. 

TCO theory is in this fashion akin to and can be in fact viewed as
a 1--di\-mensional analog of 3--dimensional CS theory.
The strict non invariance of the action $\sfS$
does not by itself compromise the level preserving gauge invariance of the TCO model: 
both classically and quantically the action enjoys the appropriate form of level
preserving gauge invariance. 
The TCO model, so, can legitimately be regarded as a sigma model over $\msG/\msG_\kappa$. 
As such, the model still features a background gauge field preserving
symmetry and a background gauge symmetry,
as the $\clG^*{}_{\msG,\varsigma^*\omega}(N)$ and $\clG_\msG(M)$--actions \ceqref{safssym12}
and \ceqref{safssym9} commute with the $\clG_{\msG,\kappa}(N)$--action
and leave the action $\sfS$ invariant by \ceqref{safssym13} and \ceqref{safssym10}.



\subsection{\textcolor{blue}{\sffamily Functional integral quantization
of the ordinary TCO sigma model}}\label{subsec:safspath}

In this subsection, we shall consider the functional integral quantization of the ordinary TCO sigma model.
For conciseness, we shall be sketchy about the details of basic computations.

As noticed previously in subsect. \cref{subsec:safssigmod}, the functional integral quantization
of the TCO sigma model requires that the exponentiated action $\ee^{i\sfS}$ is level preserving
gauge invariant or equivalently, by \ceqref{safssym4}, that the exponentiated anomaly 
$\ee^{i\sfA}$ is trivial. On account of \ceqref{safssym5}, this leads to a quantization condition for
the TCO level $\kappa$. The condition can be stated rather explicitly in the
case when the Lie group $\msG$ is compact, based on the fact that 
the level preserving gauge transformations are just the $\msG_\kappa$--valued
gauge transformations. 
When $\kappa$ is regular and $\msG_\kappa$ is so a maximal torus $\msT$
of $\msG$, $\kappa$ is required to belong to the dual $\Lambda_\msG{}^*$ of the integral lattice
$\Lambda_\msG$ in $\fkt$. When conversely
$\kappa$ is not regular, $\kappa$ is subject to further restrictions. 

Whatever the form the quantization condition of the level $\kappa$ takes, when it is met the
exponentiated action $\ee^{i\sfS}$ constitutes a genuine functional on the TCO sigma model's field space
$\overline{\clF}_{\msG,\kappa}(N)$ (cf. subsect. \cref{subsec:safssigmod})
and the model's quantum partition function
can be expressed as a functional integral of $\ee^{i\sfS}$ on $\overline{\clF}_{\msG,\kappa}(N)$, 
\begin{equation}
\sfZ(\omega)=
\int_{\overline{\clF}_{\msG,\kappa}(N)} \scD g\,\ee^{i\sfS(g;\omega)}.
\label{safspath2}
\end{equation}
The functional measure $\scD g$ of $\overline{\clF}_{\msG,\kappa}(N)$ is assumed to be
invariant under left $\clG_\msG(N)$ multiplicative shifts.
In view of the expected relationship of $\sfZ(\omega)$
to the Wilson line of the embedded curve $\varsigma$, an analysis of the gauge and homotopy
invariance properties of $\sfZ(\omega)$ is in order. 


The left $\clG_\msG(N)$--invariance of the measure $\scD g$ ensures that the quantum TCO sigma model enjoys
all the symmetries acting 
left multiplicatively on sigma model field space $\overline{\clF}_{\msG,\kappa}(N)$
which the classical model does.
In particular, the background gauge symmetry of the classical model implies that 
the partition function $\sfZ(\omega)$ is gauge invariant as a functional of $\omega$.
Therefore, 
\begin{equation}
\sfZ(\omega^u)=\sfZ(\omega)
\label{safspath3}
\end{equation}
for $u\in\clG_{\msG}(M)$.
The gauge background preserving symmetry has instead no apparent implications for $\sfZ(\omega)$. 

We shall analyze next the dependence of the partition function $\sfZ(\omega)$ on the embedded curve 
$\varsigma$. Specifically, we shall 
compute the variation $\delta\sfZ(\omega)$ of $\sfZ(\omega)$ under a variation $\delta\varsigma$ of
$\varsigma$ leaving the image $\varsigma(\partial N)$ of the boundary $\partial N$ of $N$ fixed, i.e 
such that $\delta\varsigma|_{\partial N}=0$. Formally, expression \ceqref{safspath2} yields
\vspace{.85mm}
\begin{equation}
\delta\sfZ(\omega)=%
\int_{\overline{\clF}_{\msG,\kappa}(N)}\scD g\,\ee^{i\sfS(g;\omega)}i\delta\sfS(g;\omega).
\label{safspath8}
\end{equation} 
%
\vspace{.5mm}
\!To make the above  expression  more explicit, a suitable variational framework is required. 
The embeddings of $N$ into $M$ form the infinite dimensional functional manifold $\clE_{N,M}=\Emb(N,M)$.
The variational problem we are dealing with is therefore naturally framed in the complex consisting of the
graded functional algebra $\Fun(T[1]\clE_{N,M})$ and the appended variational differential $\delta$.
However, given that the embeddings always show up through elements of the function algebra
$\Fun(T[1]N)$, it is necessary to enlarge our formal framework to the augmented embedding manifold
$\clE_{N,M}\times N$ and the associated complex $\Fun(T[1](\clE_{N,M}\times N))$, $\delta+d$. 
As $T[1](\clE_{N,M}\times N)=T[1]\clE_{N,M}\boxplus T[1]N$
\footnote{$\vphantom{\dot{\dot{\dot{a}}}}$ \label{foot:boxplus} If
$E_1$, $E_2$ are vector bundles on distinct manifolds $M_1$, $M_2$, their ordinary direct sum cannot be defined 
whilst their external direct sum can. This is the vector bundle $E_1\,\boxplus\,E_2$ $=\pi_1{}^*E_1\oplus\pi_2{}^*E_2$
of base $M_1\times M_2$, where $\pi_1$, $\pi_2$ are the standard projection of $M_1\times M_2$ on its Cartesian factors
$M_1$, $M_2$. The tangent bundle of a product manifold $X_1\times X_2$ is the external direct sum
$TX_1\,\boxplus\, TX_2$.}, this latter is endowed with \linebreak
an inherent bigrading 
\footnote{$\vphantom{\dot{\dot{\dot{a}}}}$  \label{foot:bernstein}
There are two conventions for the sign produced by the commutation of two homogeneous elements
of a bigraded commutative algebra $\msA$. If $x,y\in\msA$ have 
bidegrees $(m,p)$, $(n,q)$, then $xy=(-1)^{(m+p)(n+q)}yx$ according to Bernstein--Leites
and $xy=(-1)^{mn+pq}yx$ according to Deligne. It makes no difference which convention is used:
any statement can be expressed in principle in any one of them. In this paper, we adhere to the Bernstein--Leites
rule as it turns out to be more natural in our construction, 
although the Deligne rule is more commonly used in the physical literature. \vspace{-2.5mm}
} with $\delta$ and $d$ being the bidegree $(1,0)$ and $(0,1)$ contributions of the differential respectively. 
In this way, the pull--back of an ordinary field $\psi\in\Map(T[1]M,\fkg[p])$ by the evaluation map
$\epsilon:\clE_{N,M}\times N\rightarrow M$ yields a map
$\epsilon^*\psi\in\Map(T[1](\clE_{N,M}\times N),\fkg[p])$. $\epsilon^*\psi$ can be decomposed
in terms of definite bidegree. In particular, the terms of bidegree $(0,p)$ 
constitute the component $\epsilon^*\psi_N$ of $\epsilon^*\psi$ along $N$ in $\clE_{N,M}\times N$. 

The computation of the variation $\delta\sfZ(\omega)$ of $\sfZ(\omega)$ involves first the
computation through \ceqref{safsmod1}
of the variation $\delta\sfS(g;\omega)$ of $\sfS(g;\omega)$ regarded as a degree
$0$ element of $\Fun(T[1]\clE_{N,M})$ depending on $g$ and $\omega$.
The expression so obtained is then inserted in \ceqref{safspath8}, yielding 
\begin{align}
\delta\sfZ(\omega)
&=-\int_{\clF_{\msG}(N)} \scD g\,\ee^{i\sfS(g;\omega)}i
\int_{T[1]N}\varrho_N
\vphantom{\Big]}
\label{safspath12}
\\
&\hspace{1cm}
\left[\left(\kappa,\Ad g^{-1}(\epsilon^*\phi)\right)
+\left([\Ad g^{-1}(\epsilon^*\omega_N)+g^{-1}dg,\kappa],\Ad g^{-1}(\epsilon^*\omega)\right)\right].
\vphantom{\Big]}
\nonumber
\end{align}
where $\phi=d\omega+\frac{1}{2}[\omega,\omega]$ is the curvature of $\omega$. 
The second term within square brackets gives a vanishing contribution to the functional integral
under rather general assumptions. Indeed, we have 
\begin{align}
&\int_{\clF_{\msG}(N)} \scD g\,\ee^{i\sfS(g;\omega)}i\bigg\{
\sft\sfr\left(\sfA\sfd(\Ad g^{-1}(\epsilon^*\omega))\right)
\vphantom{\Big]}
\label{safspath15}%
\\
&\hspace{2.5cm}-\int_{T[1]N}\varrho_N\left([\Ad g^{-1}(\epsilon^*\omega_N)+g^{-1}dg,\kappa],
\Ad g^{-1}(\epsilon^*\omega)\right)\bigg\}=0,
\vphantom{\Big]}
\nonumber
\end{align}
where $\sft\sfr$, $\sfA\sfd$ denote functional trace and adjoint respectively, as follows from a standard 
Schwinger--Dyson type argument. Above,
$\sft\sfr\left(\sfA\sfd(\Ad g^{-1}(\epsilon^*\omega))\right)$ contains a factor
$\delta_N(0)$, which must be regularized, and a factor that pointwise is of the form $\tr\ad O$
for some element $O\in\fkg[1]$. 
The latter vanishes if 
$\tr\ad x=0$ for $x\in\fkg$, i.e. when 
the Lie algebra $\fkg$ is unimodular. If this property holds,
$\sft\sfr\left(\sfA\sfd(\Ad g^{-1}(\epsilon^*\omega))\right)=0$.
From \ceqref{safspath12} and \ceqref{safspath15}, we find then that 
\begin{equation}
\delta\sfZ(\omega)
=-\int_{\clF_{\msG}(N)} \scD g\,\ee^{i\sfS(g;\omega)}
i\int_{T[1]N}\varrho_N\left(\kappa,\Ad g^{-1}(\epsilon^*\phi)\right). 
\label{safspath17}
\end{equation}
In this way, we have that
\begin{equation}
\delta\sfZ(\omega)=0 \qquad \text{if $\phi=0$}.
\label{safspath18}
\end{equation}
Thus, $\sfZ(\omega)$ is invariant under smooth variations of the embedding $\varsigma$ 
when the gauge field $\omega$ is flat. 

The background gauge invariance property \ceqref{safspath3} and the homotopy
invariance property \ceqref{safspath18} of the partition function $\sfZ(\omega)$ 
is a clear indication that $\sfZ(\omega)$ computes the trace in a representation $R_\kappa$
of the holonomy $F_\omega(\varsigma)$ of $\omega$ along the parametrized curve $\varsigma$, viz \hphantom{xxxxxxx}
\begin{equation}
\sfZ(\omega)=\tr_{R_\kappa}(F_\omega(\varsigma)).
\label{safspath20}
\end{equation}
This relation shows that the TCO sigma model underlies the partition realization of Wilson lines. 
It has been verified by explicit evaluation of $\sfZ(\omega)$ for specific choices of $\msG$ and $\omega$
in \ccite{Alekseev:1988vx,Diakonov:1989fc,Diakonov:1996zu} and by this reason is by now considered
an established result.






\subsection{\textcolor{blue}{\sffamily Canonical formulation of the ordinary TCO model}}\label{subsec:safscan}

In this subsection, we shall examine the canonical theory of the ordinary TCO model,
which will furnish us new insight into TCO theory. 

For a canonical formulation of the TCO model, the base manifold $N$ is chosen to be $\bbR^1$ to be interpreted
as the time axis. The action \ceqref{safsmod1} reads then as 
\begin{equation}
\sfS(g;\omega)=\int_{T[1]\bbR^1}\varrho_{\bbR^1}\!
\left(\kappa,\Ad g^{-1}(\varsigma^*{}_t\omega)+g^{-1}d_tg\right)\!,
\label{nsafscan4}
\end{equation}
where $d_t$ denotes the ordinary de Rham differential of $\bbR^1$. 

An inspection of the kinetic term of the TCO action $\sfS$ given above, $(\kappa,g^{-1}d_tg)$, 
reveals that the ambient phase space of the TCO model is the group manifold $\msG$.
The appropriate form of the presymplectic potential 1--form $\varpi$ of phase space  is also indicated
by that of the kinetic term, viz
\begin{equation}
\varpi=-(\kappa,g^{-1}d g),
\label{safscan1}
\end{equation}
where $g$ is to be regarded here as a $\msG$--valued phase space variable.  
The associated presymplectic 2-form therefore is 
\begin{equation}
\psi=d\varpi=\frac{1}{2}(\kappa,[g^{-1}d g,g^{-1}d g]).
\label{safscan2}
\end{equation}
$\psi$ so depends on the model's level $\kappa$. 

In subsect. \cref{subsec:safssigmod}, we found that the ordinary TCO model actually is a 1--di\-mensional
sigma model over the homogeneous space $\msG/\msG_\kappa$.
A basic property such as this should emerge also in the canonical formulation. Indeed
it does, as we are going to verify next. 
The level preserving gauge transformation group
appears in the canonical formulation as the subgroup of the group $\msG$ formed by the elements $\upsilon\in\msG$
leaving $\kappa$ invariant in conformity with \ceqref{safssym1} 
\begin{equation}
\Ad\upsilon(\kappa)=\kappa.
\label{safscan3}
\end{equation}
It is therefore just the invariance subgroup of $\kappa$ in $\msG$, $\msG_\kappa$.
The gauge transformation action on the model's field space $\clF_{\msG}(N)$
translates in this way as the $\msG_\kappa$--action on the ambient phase space $\msG$ given by 
\begin{equation}
g^\upsilon=g\upsilon  
\label{safscan4}
\end{equation}
with $g\in\msG$ and $\upsilon\in\msG_\kappa$, in keeping with \ceqref{safssym2}.
From \ceqref{safscan3}, it follows that the 
infinitesimal level preserving 
gauge transformation algebra is the Lie algebra $\fkg_\kappa$ of $\msG_\kappa$ and that
the action \ceqref{safscan4} is implemented infinitesimally for any $z\in\fkg_\kappa$
by the vector field $X_z\in\Vect(\msG)$ such that 
\begin{equation}
j_{\hfpt X_z}(g^{-1}dg)=z,
\label{safscan5}
\end{equation}
where $j_V$ denotes contraction with respect to a vector field $V\in\Vect(\msG)$. 
From \ceqref{safscan2} and \ceqref{safscan5}, it is immediately checked that \hphantom{xxxxxxxxxxxx}
\begin{equation}
j_{\hfpt X_z}\psi=0.
\label{safscan7}
\end{equation}
Likewise, any vector field $X\in\Vect(\msG)$ such that
$j_{\hfpt X}\psi=0$ is of the form $X=X_z$ for some $z\in\fkg_\kappa$ pointwise in $\msG$.
The degeneracy of $\psi$ just exhibited and \ceqref{safscan4} together 
indicate that the phase space of the TCO model is precisely the 
$\msG_\kappa$--orbit space of the ambient phase space $\msG$, 
that is the homogeneous space $\msG/\msG_\kappa$.  
They imply further that $\psi$ induces a symplectic 2--form $\overline\psi$ on $\msG/\msG_\kappa$ 
and associated with this a Poisson bracket structure $\{\cdot,\cdot\}$ on the function algebra
$\Fun(\msG/\msG_\kappa)$ of $\msG/\msG_\kappa$.

We recall next how the Poisson bracket of a pair of functions of $\Fun(\msG/\msG_\kappa)$ is computed.
Though this prescription can be ascribed to ordinary KKS theory proper with no reference to the
TCO model we are examining, we review it because an infinite dimensional version of it will
be employed in the canonical formulation of the derived TCO model next section.
In practice, it is awkward to operate with orbits as such, while it is
relatively more straightforward to work with orbit representatives.
With this perspective in mind, we consider the following isomorphisms. The first isomorphism,
$\Fun(\msG/\msG_\kappa)\simeq\Fun(\msG)^{\msG_\kappa}$,
equates the function algebra $\Fun(\msG/\msG_\kappa)$ of $\msG/\msG_\kappa$
and the subalgebra $\Fun(\msG)^{\msG_\kappa}$
of $\Fun(\msG)$ of the functions invariant under the $\msG_\kappa$--action \ceqref{safscan4}.
The second isomorphism, $\Vect(\msG/\msG_\kappa)\simeq\WW\Vect_\kappa(\msG)$, 
identifies the vector field Lie algebra $\Vect(\msG/\msG_\kappa)$ of $\msG/\msG_\kappa$ and
the Weyl Lie algebra $\WW\Vect_\kappa(\msG)$ 
\footnote{$\vphantom{\dot{\dot{\dot{a}}}}$  \label{foot:weyl}
Recall that the Weyl Lie algebra of a Lie subalgebra $\fks$
of a Lie algebra $\fkl$ is defined as $\WW\fks=\NN\fks/\fks$, 
where $\NN\fks$ denotes the normalizer of $\fks$ in $\fkl$. The normalizer of $\NN\fks$ of $\fks$ is the largest
subalgebra of $\fkl$ containing $\fks$ as an ideal.}
of the Lie subalgebra $\Vect_\kappa(\msG)$ of $\Vect(\msG)$ of vector fields $V\in\Vect(\msG)$
of the form $V=X_z$ for some $z\in\fkg_\kappa$ pointwise in $\msG$.
So, a function $f\in\Fun(\msG/\msG_\kappa)$ is to be thought of as
a function $f\in\Fun(\msG)$ such that $f(g^\upsilon)=f(g)$ for $\upsilon\in\msG_\kappa$ and,
similarly, a vector field $V\in\Vect(\msG/\msG_\kappa)$ as a vector field 
$V\in\Vect(\msG)$ defined mod  vector fields $V'\in\Vect_\kappa(\msG)$ and
such that $[V,V']\in\Vect_\kappa(\msG)$ for any vector field $V'\in\Vect_\kappa(\msG)$. 
The computation of the Poisson bracket involving
a function $f\in\Fun(\msG/\msG_\kappa)$ requires the Hamiltonian vector field
$V_f\in\Vect(\msG/\msG_\kappa)$ of $f$, which is defined 
by the property that 
\begin{equation}
df+j_{V_f}\psi=0.
\label{safscan17}
\end{equation}
The Poisson bracket of a pair of functions $f,h\in\Fun(\msG/\msG_\kappa)$ 
is then the function $\{f,h\}\in\Fun(\msG/\msG_\kappa)$ given by the standard relation 
\begin{equation}
\{f,h\}=j_{V_f}dh=-j_{V_h}df. 
\label{safscan18}
\end{equation}  

As recalled in sect. 4 of I, the homogeneous space $\msG/\msG_\kappa$ is 
the coadjoint orbit $\clO_\kappa$ of $\kappa$ and the symplectic 2--form
$\overline{\psi}$ of $\msG/\msG_\kappa$ induced by \ceqref{safscan2} 
is the KKS one.
The canonical formulation of the TCO model in this way reproduces the KKS theory of $\clO_\kappa$.


In the quantum theory expounded in subsect. \cref{subsec:safspath}, the TCO model's level
$\kappa$ is quantized in a certain way. 
For such discrete
values of $\kappa$, the symplectic form $\overline{\psi}$  obeys the Bohr--Sommerfeld quantization condition
rendering the geometric quantization of $\clO_\kappa$ possible.


\vfil\eject

\vfil\eject

\renewcommand{\sectionmark}[1]{\markright{\thesection\ ~~#1}}

\section{\textcolor{blue}{\sffamily Derived TCO model}}\label{sec:hafsmod}

In this section, which is the central one of the present paper, we illustrate the derived
TCO model. The construction is patterned on that of the ordinary TCO model of sect.
\cref{sec:tcoreview} using systematically the derived geometrical set--up of
sect. 3 of I. The topics covered therefore are the model's formulation as a classical
field theory, basic symmetries, interpretation as a sigma model, functional integral quantization
and canonical analysis. 
However, from a formal standpoint, the formulations of the derived and ordinary TCO models are not as close as those
of the derived and ordinary KKS theory presented in part I are.
The derived TCO model is in many respects richer than its ordinary counterpart.

The derived counterpart of the ordinary TCO level is a closed derived level current.
The derived TCO model is a sigma model only if certain restrictions are imposed
on the level current and the symmetry crossed module.
Further, the relationship of the derived TCO model to derived KKS theory emerges only
for a special choice of the level current, which yields the so--called characteristic model. 


The derived TCO classical field equations are not automatically integrable as in the ordinary 
case, but they are only if the gauge background is fake flat. Further, the accidental gauge background preserving
symmetry of the ordinary model gets enhanced to a full gauge symmetry in the derived model.
These novel features impinge upon the model's quantization and canonical structure.


\subsection{\textcolor{blue}{\sffamily Derived TCO model}}\label{subsec:hafsmod}

In this subsection, we shall introduce the derived TCO model and study its most salient properties
as a classical field theory. Our formulations parallels to a important extent that of the ordinary model
reviewed in subsect. \cref{subsec:safsmod}.

The basic elements of the model are the following.

\begin{enumerate}

\item
The base manifold: an oriented compact connected 2--dimensional manifold $N$, possibly
with boundary.

\end{enumerate}
The topologies of this kind are fully classified and form 
a denumerable gamut. 
\begin{enumerate}[resume]

\item  
The ambient manifold: an oriented manifold $M$.

\end{enumerate} 
The case where $M$ is a 4--fold is the only relevant for the study of 2--knot topology. 
We however shall leave the dimension of $M$ undetermined in what follows. 
\begin{enumerate}[resume]

\item
The embedded surface: a base to ambient manifold embedding $\varsigma:N\rightarrow M$.

\end{enumerate}
The embedding provides a homeomorphic image of $N$ in $M$. 
\begin{enumerate}[resume]

\item
The symmetry structure: a Lie group crossed module $\msM=(\msE,\msG)$ 
equipped and the associated Lie algebra crossed module $\fkm=(\fke,\fkg)$. $\fkm$ is equipped 
with an invariant pairing $\langle\cdot,\cdot\rangle$. 


\end{enumerate}
$\msM$ is thus balanced as 
$\langle\cdot,\cdot\rangle$ yields an isomorphism 
of the vector spaces $\fke$ and $\fkg^*$.

\begin{enumerate}[resume]

\item
The derived TCO dynamical field space: $\clF_{\msM}(N):=\Map(T[1]N,\DD\msM)$,
the space of derived Lie group valued fields $\rmG$. 

\item
The derived TCO background field space: $\clC_{\msM}(M)$,  
the space of derived gauge fields $\Omega$ (cf. subsect. \cref{subsec:higau})

\end{enumerate}

\noindent
The derived model is moreover characterized by a parameter: a fixed level current
$\rmK\in\Map'_c(T[1]N,\DD\fkm[0])$ with compact support in the interior of $N$ obeying
\begin{equation}
\dd\rmK=0. \vphantom{\bigg]}
\label{hafsmod0}
\end{equation}
A level datum of this kind allows for a broader instantiation than that of the ordinary setting.
We notice here that, owing to (3.3.10) of I, condition \ceqref{hafsmod0} does not reduces
to the customary closedness of $\rmK$, but it is more involved. 


The action of the derived TCO model reads as
\begin{equation}
\sfS(\rmG;\Omega)=\int_{T[1]N}\varrho_N(\rmK,\Ad\rmG^{-1}(\varsigma^*\Omega)+\rmG^{-1}\dd\rmG) \vphantom{\Bigg]}
\label{hafsmod1}
\end{equation}
in the derived field formal framework expounded in subsect. 3.3 of I. 
The Lagrangian, formally consisting in the component along $\rmK$ of the gauge transform
of $\Omega$ by $\rmG$, is thus totally analogous to the one of the
usual TCO model (cf. eq. \ceqref{safsmod1}) with derived quantities replacing the ordinary ones. %

The variation of $\sfS$ with respect to $\rmG$ takes the form
\vspace{-.66mm}
\begin{multline}
\delta\sfS(\rmG;\Omega)=-\int_{T[1]N}\varrho_N(\rmG^{-1}\delta\rmG, 
[\Ad\rmG^{-1}(\varsigma^*\Omega)+\rmG^{-1}\dd\rmG,\rmK])
\\
+\int_{T[1]\partial N}\varrho_{\partial N}(\rmK,\rmG^{-1}\delta\rmG).
\label{hafsmod2}
\end{multline}
If a suitable boundary condition is imposed on $\rmG$
which makes the boundary term in \ceqref{hafsmod2} vanish identically,
$\sfS$ is 
differentiable in the sense of refs. \ccite{Regge:1974zd,Benguria:1976in}.
The field equations then take the form
\begin{equation}
[\Ad\rmG^{-1}(\varsigma^*\Omega)+\rmG^{-1}\dd\rmG,\rmK]=0.
\label{hafsmod5}
\end{equation}
As expected by design, the field equations \ceqref{hafsmod5} are analogous in form to those
of the ordinary TCO model (cf. eq. \ceqref{safsmod5}).
They are so, among other things, also by virtue of the level condition \ceqref{hafsmod0}. 

The discussion of the choice of the appropriate boundary condition is far
more involved in the derived TCO model than it is in the ordinary one, as the boundary $\partial N$
of the base manifold $N$ is 1-- instead than 0--dimensional. 
An in--depth analysis of this matter is beyond the scope of the present work and will not be tackled here.
In some generality, anyway, 
the boundary condition takes the following form. Consider the space 
$\clF_\msM(\partial N)=\Map(T[1]\partial N,\DD\msM)$ of boundary derived TCO fields 
$\rmG_\partial$ and the associated variational complex $\Fun(T[1]\clF_\msM(\partial N))$, $\delta_\partial$. 
The complex contains a special degree 1 element corresponding to the boundary contribution
to the action's variation $\delta\sfS(\rmG;\Omega)$ in \ceqref{hafsmod2}, viz 
\begin{equation}
\sfPi_\partial=-\int_{T[1]\partial N}\varrho_{\partial N}(\rmK,\rmG_\partial{}^{-1}\delta_\partial\rmG_\partial).
\label{}
\end{equation}
The boundary condition has then the basic form
\begin{equation}
i_\partial{}^*\rmG\in\clL,
\label{}
\end{equation}
where $i_\partial:\partial N\rightarrow N$  is the canonical injection of
$\partial N\subset N$ and $\clL$ is a functional submanifold of $\clF_\msM(\partial N)$ such that 
$\sfkappa_\clL{}^*\sfPi_\partial=0$, $\sfkappa_\clL:\clL\rightarrow\clF_\msM(\partial N)$
being the canonical injection of $\clL\subset\clF_\msM(\partial N)$.
There is very little more specific that can be said about the boundary condition without
breaking $\rmG$ into its components $g$, $G$.

The integrability of the field equations \ceqref{hafsmod5} requires that 
\begin{equation}
[\Ad\rmG^{-1}(\varsigma^*\Phi),\rmK]\approx 0,
\label{hafsmod6}
\end{equation}
where $\Phi$ is the curvature of the derived gauge field $\Omega$ (cf. eq. \ceqref{higau2})
and $\approx$ denotes equality on shell.
Since the base manifold $N$ of the TCO model is 2--dimensional, the degree 3 component
$\varsigma^*\varPhi$ of $\varsigma^*\Phi$ vanishes identically. The degree 2 component
$\varsigma^*\phi$ of $\varsigma^*\Phi$, conversely, may be non vanishing. Therefore, 
a broadly general sufficient condition for integrability to obtain is that
the pull--back $\varsigma^*\Omega$ of the gauge field $\Omega$ by the embedding $\varsigma$ is fake flat 
\begin{equation}
\varsigma^*\phi=0
\label{hafsmod7}
\end{equation}
(cf. subsect. \cref{subsec:higau}). The occurrence of a non trivial integrability condition
of the field equations is a novel feature of the derived TCO theory with no analogue
in the ordinary one, whose field equations are always integrable
(cf. subsect. \cref{subsec:safsmod}). 



\subsection{\textcolor{blue}{\sffamily Symmetries of the derived TCO model}}\label{subsec:hafssym}

In this subsection, we shall study the symmetries of derived TCO theory.
The derived model's main symmetries are essentially the same as those of the ordinary one:
the level preserving gauge symmetry, the gauge background preserving symmetry
and the background gauge symmetry. The main difference concerns the second, which is
an accidental rigid symmetry in the ordinary case while it gets promoted to a gauge symmetry
in the derived one with far--reaching implications.

The first symmetry of the derived TCO model is the level preserving gauge symmetry.
Its nature and action are totally analogous to those of its counterpart of the ordinary model
(cf. subsect. \cref{subsec:safssym}). The associated gauge transformation group, $\clG_{\msM,\rmK}(N)$,
is the subgroup of the full derived gauge transformation group $\clG_\msM(N)$ formed by the transformations 
$\Upsilon\in\clG_\msM(N)$ whose adjoint action leaves the model's level current $\rmK$ invariant,
\hphantom{xxxxxxxxxxxxx}
\begin{equation}
\Ad\Upsilon(\rmK)=\rmK
\label{hafssym1}
\end{equation}
(cf. eq. \ceqref{safssym1}). \pagebreak 
Unlike in the ordinary case, in the derived case there generally does not exist anything like an
invariance crossed submodule $\msM_\rmK$ of $\rmK$, 
by means of which $\clG_{\msM,\rmK}(N)$ can be equated to the derived gauge transformation group
$\clG_{\msM_\rmK}(N)$. (This feature of the model will be further
discussed in subsect. \cref{subsec:hafssigmod}.) 
$\clG_{\msM,\rmK}(N)$ acts on the derived TCO field space $\clF_{\msM}(N)$; the action reads as
\begin{equation}
\rmG^\Upsilon=\rmG\Upsilon 
\label{hafssym2}
\end{equation}
for $\Upsilon\in\clG_{\msM,\rmK}(N)$ and $\rmG\in\clF_{\msM}(N)$ (cf. eq. \ceqref{safssym2}).
$\clG_{\msM,\rmK}(N)$ is instead inert on the TCO derived background gauge field space $\clC_{\msM}(M)$. 
As in the ordinary case, the derived TCO action $\sfS$ is not level preserving gauge invariant. 
The gauge variation it undergoes by effect of a gauge transformation 
$\Upsilon$ has again a simple form depending only on $\Upsilon$ 
but not on $\rmG$, 
\begin{equation}
\sfS(\rmG^\Upsilon;\Omega)=\sfS(\rmG;\Omega)+\sfA(\Upsilon),
\label{hafssym4}
\end{equation}
where the (classical) gauge anomaly $\sfA$ is given by 
\begin{equation}
\sfA(\Upsilon)=\int_{T[1]N}\varrho_N(\rmK,\Upsilon^{-1}\dd\Upsilon).
\label{hafssym5}
\end{equation}
(cf. eqs. \ceqref{safssym4}, \ceqref{safssym5}). Later, we shall explain the reasons
why the non invariance of $\sfS$ does not mar the level preserving gauge invariance of TCO theory.


The derived TCO model is also characterized by a second gauge background preserving symmetry,
like the ordinary model (cf. subsect. \cref{subsec:safssym}).
The associated symmetry group, $\clG^*{}_{\msM,\varsigma^*\Omega}(N)$,
is the invariance subgroup of the pull--backed derived gauge field $\varsigma^*\Omega$, 
that is the subgroup of the full derived gauge transformation group
$\clG_\msG(N)$ formed by the elements $\rmT\in\clG_{\msM}(N)$ obeying 
\begin{equation}
\varsigma^*\Omega^\rmT=\varsigma^*\Omega
\label{hafssym12/1}
\end{equation}
(cf. eq. \ceqref{safssym12/1}).
In derived gauge theory, $\clG^*{}_{\msM,\varsigma^*\Omega}(N)$ is an infinite dimensional group
(cf. subsect. \cref{subsec:hisym}).
Indeed, the symmetry is no longer an accidental
rigid symmetry, but a full gauge symmetry, as discussed later below.  
$\clG^*{}_{\msM,\varsigma^*\Omega}(N)$ acts on the TCO field
space $\clF_{\msM}(N)$: for $\rmT\in\clG^*{}_{\msM,\varsigma^*\Omega}(N)$
and $\rmG\in\clF_{\msM}(N)$, we have 
\begin{equation}
\rmG^\rmT=\rmT^{-1}\rmG
\label{hafssym12}
\end{equation}
(cf. eq. \ceqref{safssym12}).
$\clG^*{}_{\msM,\varsigma^*\Omega}(N)$ is instead inert by design on the background gauge field space $\clC_{\msM}(M)$. 
The invariance condition \ceqref{hafssym12/1} implies that 
\begin{equation}
\sfS(\rmG^\rmT;\Omega)=\sfS(\rmG;\Omega)
\label{hafssym13}
\end{equation}
(cf. eq. \ceqref{safssym13}). Analogously to the ordinary model,
this symmetry property can be interpreted to the effect 
that the TCO field space $\clF_{\msM}(N)$ possess a $\clG_\msM(N)$--gauge symmetry acting 
according to \ceqref{hafssym12} that the gauge background $\Omega$ breaks down
to $\clG^*{}_{\msM,\varsigma^*\Omega}(N)$. We shall have more to say about this below. 


The third symmetry of the derived TCO model is the background gauge
symmetry. Its nature and action are also totally analogous to those of
its counterpart of the ordinary model (cf. subsect. \cref{subsec:safssym}).
The associated gauge transformation group is the ambient space 
gauge transformation group $\clG_{\msM}(M)$. 
$\clG_{\msM}(M)$ acts on the TCO field space $\clF_{\msM}(N)$ as 
\begin{equation}
\rmG^\rmU=\varsigma^*\rmU^{-1}\rmG
\label{hafssym9}
\end{equation}
for $\rmU\in\clG_{\msM}(M)$ and $\rmG\in\clF_{\msM}(N)$ (cf. eq. \ceqref{safssym9}). 
$\clG_{\msM}(M)$ acts further on the derived background gauge 
field space $\clC_{\msM}(M)$ by associating with any $\Omega\in\clC_{\msM}(M)$ its gauge transform
$\Omega^\rmU$ given by \ceqref{higau9}. %
The action is invariant,  
\begin{equation}
\sfS(\rmG^\rmU;\Omega^\rmU)=\sfS(\rmG;\Omega)
\label{hafssym10}
\end{equation}
(cf. eq. \ceqref{safssym10}).

If $\rmU\in\clG^*{}_{\msM,\Omega}(M)$ is a background gauge transformation 
of the invariance subgroup of $\Omega$ in $\clG_{\msM}(M)$,
then by \ceqref{hafssym10}
\begin{equation}
\sfS(\rmG^\rmU;\Omega)=\sfS(\rmG;\Omega), 
\label{hafssym11}
\end{equation}
since $\Omega^\rmU=\Omega$ in this case. 
A comparison of \ceqref{hafssym11} and \ceqref{hafssym13} 
indicates that the gauge background preserving 
gauge symmetry extends at the base manifold level
the gauge background preserving background gauge symmetry existing at the ambient manifold level,
because $\varsigma^*\clG^*{}_{\msM,\Omega}(M)\subseteq\clG^*{}_{\msM,\varsigma^*\Omega}(N)$.


The invariance properties of the derived TCO action $\sfS$ under the level preserving
and gauge background preserving gauge symmetries 
hold with no need to impose boundary conditions on either the TCO
field $\rmG\in\clF_\msM(N)$ or the gauge transformations $\Upsilon\in\clG_{\msM,\rmK}(N)$, 
$\rmT\in\clG^*{}_{\msM,\varsigma^*\Omega}(N)$ involved. If however
a boundary condition is applied to $\rmG$ to make the variational problem well-defined,
the gauge transformations $\Upsilon$, $\rmT$ generally modify such boundary condition, since
$\Upsilon$, $\rmT$ have boundary restrictions $i_\partial{}^*\Upsilon$, $i_\partial{}^*\rmT$ 
acting on the boundary TCO field $\rmG_\partial$ congruently
with eqs. \ceqref{hafssym2}, \ceqref{hafssym12} and this boundary action does not generally
preserve the functional submanifold $\clL\subset\clF_\msM(\partial N)$ that specifies
the boundary condition (cf. subsect. \cref{subsec:hafsmod}).

While the content of the level preserving gauge group $\clG_{\msM,\rmK}(N)$
is as a rule straightforward to find out, the content of the
gauge background preserving gauge group $\clG^*{}_{\msM,\varsigma^*\Omega}(N)$
is hard to determine for a generic gauge field $\Omega$.
Fortunately, more about it is known for a gauge field $\Omega$ obeying
the fake flatness condition \ceqref{hafsmod7} guaranteeing the fulfillment of the integrability
condition \ceqref{hafsmod6} of the classical field equations. 
In this important case, as explained in subsect. \cref{subsec:hisym}, 
$\clG^*{}_{\msM,\varsigma^*\Omega}(N)$ reduces up to a finite dimensional quotient
group $\clI_{\msM,\varsigma^*\Omega}(N)$ (cf. eq, \ceqref{higau33/q})
to its normal special gauge transformation subgroup $\clG_{\msM,\varsigma^*\Omega}(N)$,
whose explicit description was provided 
in subsect. \cref{subsec:higau}.
We submit that only the subgroup $\clG_{\msM,\varsigma^*\Omega}(N)$ codifies a genuine gauge
symmetry while the group $\clI_{\msM,\varsigma^*\Omega}(N)$, 
if non trivial, reflects an accidental residual rigid symmetry. It is this latter that is
the proper counterpart of the gauge background preserving symmetry of the ordinary TCO model
(cf. subsect. \cref{subsec:safssym}). 
For this reason, in what follows we shall generally concentrate on the special subgroup
$\clG_{\msM,\varsigma^*\Omega}(N)$ of $\clG^*{}_{\msM,\varsigma^*\Omega}(N)$.


For a fake flat gauge field $\Omega$, 
the existence of a special gauge background dependent surjective Lie group morphism
$\sfT_{\varsigma^*\Omega}:\clH_\msM(N)\rightarrow\clG_{\msM,\varsigma^*\Omega}(N)$,
shown in subsect. \cref{subsec:hisym}, allows us to regard the 
gauge transformation action of $\clG_{\msM,\varsigma^*\Omega}(N)$ as one of $\clH_\msM(N)$:
by way of relations \ceqref{higau16}, \ceqref{higau17},
$\sfT_{\varsigma^*\Omega}$ provides the gauge transformations of $\clG_{\msM,\varsigma^*\Omega}(N)$
through which $\clH_\msM(N)$ acts on the TCO sigma model 
field space $\overline{\clF}_{\msM,\rmK}(N)$.
Explicitly, for $A\in\clH_\msM(N)$, we have 
\begin{equation}
\rmG^A=\rmG^{\sfT_{\varsigma^*\Omega}(A)},
\label{hafssym14}
\end{equation}
where $\rmG^\rmT$ with $\rmT\in\clG_{\msM,\varsigma^*\Omega}(N)$
is given by \ceqref{hafssym12}. The $\clG_{\msM,\varsigma^*\Omega}(N)$ and $\clH_\msM(N)$
forms of the symmetry, which we shall call dressed and bare respectively,$\vphantom{\ul{\ul{\ul{g}}}}$
differ in several respects. On one hand, $\clG_{\msM,\varsigma^*\Omega}(N)$
depends on $\varsigma^*\Omega$ while $\clH_\msM(N)$ does not. On the other, the gauge transformation action
of $\clG_{\msM,\varsigma^*\Omega}(N)$ does not depend on $\varsigma^*\Omega$ while that of $\clH_\msM(N)$ does.
In addition, the $\clG_{\msM,\varsigma^*\Omega}(N)$--action is effective, while the $\clH_\msM(N)$
is not because of the non trivial kernel $\clK_{\msM,\varsigma^*\Omega}(N)$ of $\sfT_{\varsigma^*\Omega}$. 
As long as we are concerned 
with the analysis of the gauge symmetry at the classical level, it makes no
difference whether either the dressed or the bare standpoint is adopted. 
At the quantum level, as we shall explain in subsect.
\cref{subsec:hafspath} below, it does: the bare standpoint is the appropriate one. 

The special gauge symmetry superficially seems to be
of a nature analogous to that of the level preserving gauge symmetry, since
by virtue of \ceqref{hafssym2}, \ceqref{hafssym12} both symmetries appear to instantiate 
a certain form of multiplicative redundancy of the derived TCO field $\rmG$. 
A component analysis of their gauge transformation action, however,
brings to light their inherent difference, as we now show.
From \ceqref{hafssym2},
for $\Upsilon\in\clG_{\msM,\rmK}(N)$ the components of the gauge transform $\rmG^\Upsilon$ of $\rmG$ are  
$g^{\upsilon,\varUpsilon}=g\upsilon$, $G^{\upsilon,\varUpsilon}=G+\mu\sdot(g,\varUpsilon)$.
Similarly, from \ceqref{hafssym12}, the components of the bare form gauge transform $\rmG^A$
of $\rmG$ with $A\in\clH_\msM(N)$ read as 
$g^A=\tau(A)^{-1}g$, $G^A=\Ad A^{-1}(G+dAA^{-1}+\sdot\mu(\varsigma^*\omega,A))$. 
While the action of the component $g$ is indeed multiplicative in both cases, 
that on the component $G$ reduces to a shift in the former and an ordinary gauge transformation
of $G$ regarded as a kind of gauge field in the latter. The special gauge symmetry,
therefore, when properly analyzed through its bare expression,
turns out to be akin to an ordinary gauge theoretic symmetry in contrast to
the level preserving gauge symmetry. As such it should be treated in the quantum theory of the model,
as we shall see in greater detail again in subsect. \cref{subsec:hafspath}.


\subsection{\textcolor{blue}{\sffamily Derived TCO sigma model}}\label{subsec:hafssigmod}

Unlike the ordinary model, the derived TCO sigma model is not automatically
a sigma model. In this subsection, we discuss the conditions under which the derived model is
effectively a 2--dimensional sigma model in virtue of the model's level preserving gauge symmetry
studied in subsect. \cref{subsec:hafssym}. 

Upon modding out the level preserving gauge symmetry, the derived TCO model can
be interpreted as an effective theory of the $\clG_{\msM,\rmK}(N)$ orbit space 
\begin{equation}
\overline{\clF}_{\msM,\rmK}(N)=\clF_\msM(N)/\clG_{\msM,\rmK}(N) 
\label{hafspath0}
\end{equation}
just as in the ordinary model (cf. eq. \ceqref{safspath0}). In the derived model, however, 
$\overline{\clF}_{\msM,\rmK}(N)$ is not a genuine field space in general.
It is one provided there exists a crossed submodule $\msM_\rmK$ of the symmetry crossed module
$\msM$ such that the level preserving gauge transformation group $\clG_{\msM,\rmK}(N)$ is just the group
$\clG_{\msM_{\rmK}}(N)$ of $\DD\msM_\rmK$--valued gauge transformations.
In such a case, we have indeed that 
\begin{equation}
\overline{\clF}_{\msM,\rmK}(N)=\Map(T[1]N,\DD\msM/\DD\msM_{\rmK}) 
\label{hafspath0/1}
\end{equation}
and the model can be conceived as a sigma model over the derived
homogeneous space $\DD\msM/\DD\msM_{\rmK}$ (cf. eq. \ceqref{safspath0/1}).
Accordingly, a field $\rmG\in\overline{\clF}_{\msM,\rmK}(N)$ is to be
regarded as a field $\rmG\in\clF_\msM(N)$ defined up to
right multiplication by a field $\rmG'\in\clF_{\msM_\rmK}(N)$. 
We do not know general conditions on the crossed module $\msM$ and the level current $\rmK$
ensuring the existence of the submodule $\msM_{\rmK}$,
but we shall examine in depth in subsect. \cref{subsec:hafsspx} below
a special kind of TCO model, the characteristic one, where $\msM_{\rmK}$ 
can be shown to exist and explicitly described. 
In the rest of this subsection, we assume unless otherwise stated 
that $\msM_\rmK$ does exist and call the resulting model derived TCO sigma model. 

As in the ordinary set--up, an apparent problem with the sigma model reinterpretation
of the TCO model presented in the previous paragraph
is that the action $\sfS$ does not really enjoy the level preserving gauge symmetry
by \ceqref{hafssym4}. The issue is solved here essentially in the same manner  as in the ordinary theory
(cf. subsect. \cref{subsec:safssigmod}). 

Classically, the basic requirement to be met is the level preserving gauge
co\-variance of the field equations \ceqref{hafsmod5}. The invariance of the action $\sfS$
up to a field independent additive term $\sfA$ attested by \ceqref{hafssym4}, \ceqref{hafssym5}
is enough for that property to hold.  

Quantically, the provision to be satisfied is the level preserving gauge
invariance of the exponentiated action $\ee^{i\sfS}$ \pagebreak entering in the functional
integral formulation. This is secured provided that the exponentiated anomaly $\ee^{i\sfA(\Upsilon)}=1$
for any gauge transformation $\Upsilon\in\clG_{\msM,\rmK}(N)$ or equivalently that
\begin{equation}
\sfA(\Upsilon)\in 2\pi\bbZ.
\label{hafspath1}
\end{equation}
Inspection of expression \ceqref{hafssym5} shows that the range of values which $\sfA(\Upsilon)$ can take
depends on the symmetry crossed module $\msM$, the invariant pairing $\langle\cdot,\cdot\rangle$
of $\msM$ defining the pairing $(\cdot,\cdot)$ (cf. eq. (3.3.15) of I) and the level current $\rmK$.
We know no general conditions on such data ensuring that the integrality condition
\ceqref{hafspath1} is satisfied by $\sfA(\Upsilon)$. Standard arguments however indicate that this
is a clear possibility. 
As in the ordinary model, the variation of the anomaly $\sfA(\Upsilon)$ with respect to $\Upsilon$, given by  
$\delta\sfA(\Upsilon)=\int_{T[1]\partial N}\varrho_{\partial N}(\rmK,\Upsilon^{-1}\delta\Upsilon)$ from
\ceqref{hafssym5}, vanishes if either $\partial N=\emptyset$ or the gauge transformations $\Upsilon$
is restricted to belong to a subgroup of $\clG_{\msM,\rmK}(N)$ of gauge transformations obeying 
boundary conditions making the boundary integral vanish. 
Further, the values taken by $\sfA(\Upsilon)$ form a group, as by \ceqref{hafssym1} and \ceqref{hafssym5}
$\sfA(\Upsilon)+\sfA(\Upsilon')=\sfA(\Upsilon\Upsilon')$ and $-\sfA(\Upsilon)=\sfA(\Upsilon^{-1})$
for $\Upsilon,\Upsilon'\in\clG_{\msM,\rmK}(N)$.
Hence, $\sfA$ is likely a discrete homotopy invariant valued in a lattice of $\bbR$.
\ceqref{hafspath1} can so generically be satisfied if 
the level current $\rmK$ is suitably quantized. 
Admittedly, all this is conjectural. The characteristic model studied in subsect. \cref{subsec:hafsspx}
will furnish however a concrete realization of this scenario. 

Derived TCO theory is hence akin to 4--dimensional CS theory \ccite{Zucchini:2021bnn}
and indeed a 2--dimensional counterpart of this, analogously to ordinary TCO
theory and in agreement with the general expectations of categorification.  
The rigorous non invariance of the action $\sfS$
does not impinge upon  the level preserving gauge invariance of the derived TCO model, since  
the action enjoys the appropriate form of level preserving gauge invariance both classically and
quantically. In this way, when the crossed submodule $\msM_\rmK$ introduced above does actually exist,
the derived TCO model can be deemed as a sigma model over $\DD\msM/\DD\msM_\rmK$. 
The sigma model keeps featuring a gauge background preserving symmetry and a background
gauge symmetry, as the $\clG^*{}_{\msM,\varsigma^*\Omega}(N)$-- and $\clG_\msM(M)$--actions \ceqref{hafssym12}
and \ceqref{hafssym9} commute with the $\clG_{\msM,\rmK}(N)$--action
and leave the action $\sfS$ invariant by \ceqref{hafssym13} and \ceqref{hafssym10}.
$\vphantom{\ul{\ul{\ul{g}}}}$ 

\vfil\eject

\subsection{\textcolor{blue}{\sffamily Functional integral quantization
of the derived TCO sigma model}}\label{subsec:hafspath}


In this subsection, we shall study the functional integral quantization of the derived TCO sigma model.
Our considerations will lead to the identification of the model's partition function and a Wilson surface depending
on the models data.
The analysis we carry out employs formal functional integral techniques and hinges on certain assumptions
concerning the fixing of the special gauge symmetry stated in detail in the text. For these reasons, strictly speaking,
it provides only a strong, theoretically well--grounded validation of the identification
without being a conclusive proof of it.

The quantum functional integral formulation of the derived model is broadly patterned on that
of the ordinary model. The gauge nature of the gauge background preserving symmetry in the derived case
requires however a special treatment that is not necessary in the ordinary one. 

We posit again that there exists a crossed submodule $\msM_\rmK$ of the symmetry crossed module
$\msM$ such that the level preserving gauge transformation group $\clG_{\msM,\rmK}(N)$ equals the group
$\clG_{\msM_{\rmK}}(N)$ of $\DD\msM_\rmK$--valued gauge transformations. As argued in subsect.
\cref{subsec:hafssigmod}, this renders the derived TCO model a sigma model over the derived
homogeneous space $\DD\msM/\DD\msM_{\rmK}$. On condition that the level current $\rmK$ satisfies a suitable
quantization condition trivializing the anomaly $\ee^{i\sfA}$, 
the exponentiated action $\ee^{i\sfS}$ is then a genuine functional on
the sigma model's field space $\overline{\clF}_{\msM,\rmK}(N)$. 
In the quantum theory, therefore, functional integration 
can be performed directly on $\overline{\clF}_{\msM,\rmK}(N)$. 

We assume in what follows that the functional measure $\scD\rmG$ of
$\overline{\clF}_{\msM,\rmK}(N)$ is invariant under left $\clG_\msM(N)$ multiplicative shifts.
As the exponential action $\ee^{i\sfS}$ enjoys the gauge background preserving and background gauge symmetries
as a functional on $\overline{\clF}_{\msM,\rmK}(N)$, as highlighted in subsect.
\cref{subsec:hafssigmod}, all the symmetries of the classical sigma model then extend to the quantum one. 

There are convincing indications representing that at quantum level derived TCO field theory is 
unsound unless the background gauge field $\Omega$ satisfies the fake flatness requirement \ceqref{hafsmod7}. 
In the classical theory, as we saw, \ceqref{hafsmod7} is a general sufficient condition for the integrability
of the field equations. 
In the quantum theory, the same condition ensures the existence of a semiclassical
regime and, with this, the possibility of establishing a perturbative expansion around relevant classical
field configurations, a property that any sound quantum field theory presumably should enjoy. 
For this reason, we assume henceforth that $\Omega$ is fake flat.


Since the derived background gauge field $\Omega$ is fake flat, the gauge 
background preserving gauge transformation group $\clG^*{}_{\msM,\varsigma^*\Omega}(N)$
reduces to its normal special gauge transformation subgroup $\clG_{\msM,\varsigma^*\Omega}(N)$
up to a finite dimensional quotient group $\clI_{\msM,\varsigma^*\Omega}(N)$
(cf. subsect. \cref{subsec:hisym}). 
As we have argued in subsect. \cref{subsec:hafssym}, 
only the subgroup $\clG_{\msM,\varsigma^*\Omega}(N)$ codifies a genuine gauge
symmetry of the derived TCO sigma model, while the group $\clI_{\msM,\varsigma^*\Omega}(N)$, 
if non trivial, reflects an accidental residual rigid symmetry 
answering to the gauge background preserving symmetry of the ordinary TCO model
(cf. subsect. \cref{subsec:safssym}). 

From the above considerations, it follows that the quantum partition function 
of the derived TCO sigma model is given by 
\begin{equation}
\sfZ(\Omega)=\frac{1}{\vol\clG_{\msM,\varsigma^*\Omega}(N)}
\int_{\overline{\clF}_{\msM,\rmK}(N)} \scD\rmG\,\ee^{i\sfS(\rmG;\Omega)}.
\label{hafspath2}
\end{equation}
In the right hand side, the volume $\vol\clG_{\msM,\varsigma^*\Omega}(N)$ of the special 
gauge transformation group $\clG_{\msM,\varsigma^*\Omega}(N)$ has been divided out
to turn the integration on the sigma model field space
$\overline{\clF}_{\msM,\rmK}(N)$ into an effective one on the associated
$\clG_{\msM,\varsigma^*\Omega}(N)$--orbit space
$\overline{\clF}_{\msM,\rmK}(N)/\clG_{\msM,\varsigma^*\Omega}(N)$, as required by the 
nature of the special gauge symmetry. The above formal expression actually requires
further refinement, about which we shall say momentarily. 

The background gauge symmetry of the exponentiated action $\ee^{i\sfS}$ and the left  
multiplicative shift invariance of the functional measure $\scD\rmG$
imply that the partition function $\sfZ(\Omega)$ is gauge invariant as a functional of $\Omega$.
Consequently, 
\begin{equation}
\sfZ(\Omega^\rmU)=\sfZ(\Omega)
\label{hafspath3}
\end{equation}
for any background gauge transformation
$\rmU\in\clG_{\msM}(M)$, similarly to ordinary mo\-del (cf. eq. \ceqref{safspath3}).
The background gauge invariance of $\sfZ(\Omega)$, albeit quite simple, 
is a salient property of $\sfZ(\Omega)$. 

The partition function $\sfZ(\Omega)$ depends on the base to ambient space embedding $\varsigma$. 
In view of a possible relationship with Wilson surfaces, it is interesting to study the invariance
properties of $\sfZ(\Omega)$ under continuous deformations of $\varsigma$, analogously to what we did
for the ordinary model in subsect. \cref{subsec:safspath}. 
An adequate analysis of this issue, 
which is considerably complicated by the special gauge symmetry of the derived model,
requires the elaboration of apposite formal tools, as we shall do next. 


As a preliminary step in this direction, we necessitate a more accurate form of the
functional integral expression of $\sfZ(\Omega)$ in \ceqref{hafspath2}. As anticipated in subsect.
\cref{subsec:hafssym}, the existence of a gauge background dependent
surjective Lie group morphism $\sfT_{\varsigma^*\Omega}:\clH_\msM(N)\rightarrow\clG_{\msM,\varsigma^*\Omega}(N)$
allows one to convert the dressed special 
gauge transformation action of $\clG_{\msM,\varsigma^*\Omega}(N)$ for the bare one of $\clH_\msM(N)$.
If we adopt the second viewpoint, which we claim to be the appropriate one
for the model's quantization, one should properly divide by the volume
$\vol\clH_\msM(N)$ of $\clH_\msM(N)$ rather than the volume $\vol\clG_{\msM,\varsigma^*\Omega}(N)$ of
$\clG_{\msM,\varsigma^*\Omega}(N)$ in \ceqref{hafspath2}. 
\ceqref{hafspath2} is therefore to be replaced by the more precise expression 
\begin{equation}
\sfZ(\Omega)=\frac{1}{\vol\clH_\msM(N)}
\int_{\overline{\clF}_{\msM,\rmK}(N)} \scD\rmG\,\ee^{i\sfS(\rmG;\Omega)}.
\label{hafspath4}
\end{equation}
Unlike $\clG_{\msM,\varsigma^*\Omega}(N)$, the gauge transformation group $\clH_\msM(N)$ is independent
from the pull--back $\varsigma^*\Omega$ of the background gauge field $\Omega$. Such dependence has been
turned over to the action of $\clH_\msM(N)$ on the field space $\overline{\clF}_{\msM,\rmK}(N)$ 
encoded in the group morphism $\sfT_{\varsigma^*\Omega}$, rendering its
analysis more straightforward at least in principle. \raggedbottom 

We are now going to study the dependence of the partition function $\sfZ(\Omega)$ on the embedding $\varsigma$
by computing the variation $\delta\sfZ(\Omega)$ of $\sfZ(\Omega)$ under a variation $\delta\varsigma$ of
$\varsigma$ leaving the image $\varsigma(\partial N)$ of the boundary $\partial N$ of $N$ fixed, i.e 
such that $\delta\varsigma|_{\partial N}=0$. Formally, from \ceqref{hafspath4} we have 
\begin{equation}
\delta\sfZ(\Omega)=\frac{1}{\vol\clH_\msM(N)}
\int_{\overline{\clF}_{\msM,\rmK}(N)}\scD\rmG\,\ee^{i\sfS(\rmG;\Omega)}i\delta\sfS(\rmG;\Omega).
\label{hafspath8}
\end{equation}
This naive approach suffers however a potential problem: even if the action $\sfS$ is
gauge invariant, its variation $\delta\sfS$ may fail to be so because of the
dependence of the $\clH_\msM(N)$ gauge transformation action on $\varsigma^*\Omega$.
While in a gauge theory the formal quotient by the volume of the gauge transformation group
can be carried out via the Faddeev--Popov (FP) procedure when the integrand of 
the functional integral is gauge invariant such as in \ceqref{hafspath4}, 
it is not immediately evident whether the same can be done when the integrand is
not such as in \ceqref{hafspath8}. 
Before proceeding any further, it is therefore necessary to clarify this point.

In TCO theory, the FP approach is an algorithm employed for computing the formal quotient by the volume
$\vol\clH_\msM(N)$ of the gauge transformation group $\clH_\msM(N)$ in an expression of the form
\begin{equation}
\sfZ_\sfW(\Omega)=\frac{1}{\vol\clH_\msM(N)}
\int_{\overline{\clF}_{\msM,\rmK}(N)} \scD\rmG\,\ee^{i\sfS(\rmG;\Omega)}\sfW(\rmG;\Omega),
\label{hafspath5}
\end{equation}
where $\sfW$ is some gauge invariant functional. In outline, the method works as follows. 

The FP algorithm is based on the basic FP identity defining the FP determinant. It involves the choice 
of a suitable gauge fixing prescription. Here, we shall not provide an explicit one,
something in general very hard to do as well--known.
We only shall assume as a working hypothesis that one does exist. 
The gauge fixing condition is specified by means of
a suitable functional $\sfF:\overline{\clF}_{\msM,\rmK}(N)\rightarrow\clV$, where $\clV$ is some functional
vector space: the gauge fixing amounts to impose the condition $\sfF(\rmG)=0$ with $\rmG\in\overline{\clF}_{\msM,\rmK}(N)$
defining the gauge slice.
$\sfF$ should satisfy certain basic requirements: first, each gauge orbit should intersect the gauge slice and,
second, it should do so just once. In other words, Gribov type issues should not occur or be
somewhat harmless for the particular gauge choice made. 
We presently do not have a general proof of the existence of a gauge fixing with the above properties,
though it should be possible to get one for the characteristic TCO model of
subsect. \cref{subsec:hafsspx}, which is a more conventional kind of gauge theory. Though 
a simple Lorenz like gauge fixing prescription might work in that case, a BV formulation \ccite{BV1,BV2}
of the model in the AKSZ framework \ccite{Alexandrov:1995kv} may be required. At any rate, this is not a matter that
can be solved in the present paper and is left for future work. It is for this reason that,
as stated at the beginning of this subsection, our analysis is still to some measure conjectural.

In the TCO sigma model, the FP identity takes the form 
\begin{align}
&\det\big((\sfF\circ\sfB_{\varsigma^*\Omega|\rmG_\sfF})_*(1_N)|_{N_{1_N}\clK_{\msM,\varsigma^*\Omega}(N)}\big)
\vphantom{\Big]}
\nonumber
\\
&\hspace{4cm}\frac{1}{\vol\clK_{\msM,\varsigma^*\Omega}(N)}
\int_{\clH_\msM(N)}\scD\mhfpt A\,\delta_{\clV}\left(\sfF\circ\sfB_{\varsigma^*\Omega|\rmG}(A)\right)=1.
\label{hafspath6}
\end{align}
In the above relation,  $\sfB_{\varsigma^*\Omega|\rmG}:\clH_\msM(N)\rightarrow\overline{\clF}_{\msM,\rmK}(N)$ is
the $\clH_\msM(N)$--orbit map of an element $\rmG\in\overline{\clF}_{\msM,\rmK}(N)$ defined by 
$\sfB_{\varsigma^*\Omega|\rmG}(A)=\rmG^A$ with $A\in\clH_\msM(N)$, 
where the $A$-transform $\rmG^A$ of $\rmG$ is defined in \ceqref{hafssym14} through 
special gauge symmetry dressed to bare action conversion map 
$\sfT_{\varsigma^*\Omega}:\clH_\msM(N)\rightarrow\clG_{\msM,\varsigma^*\Omega}(N)$
met earlier. 
The kernel $\clK_{\msM,\varsigma^*\Omega}(N)$ of $\sfT_{\varsigma^*\Omega}$ is here the
subgroup of $\clH_\msM(N)$ of the elements $A\in\clH_\msM(N)$ 
such that $\sfB_{\varsigma^*\Omega|\rmG}(A)=\rmG$. 
$(\sfF\circ\sfB_{\varsigma^*\Omega|\rmG})_*(1_N):T_{1_N}\clH_\msM(N)
\rightarrow T_{\sfF(\rmG)}\clV$ is the tangent map of the map 
$\sfF\circ\sfB_{\varsigma^*\Omega|\rmG}:\clH_\msM(N)\rightarrow\clV$ 
at the neutral element $1_N\in\clH_\msM(N)$. 
$N\clK_{\msM,\varsigma^*\Omega}(N)$ is the normal 
subbundle of $\clK_{\msM,\varsigma^*\Omega}(N)$ in $\clH_\msM(N)$ and $N_{1_N}\clK_{\msM,\varsigma^*\Omega}(N)$ 
is the fiber of $N\clK_{\msM,\varsigma^*\Omega}(N)$ at
$1_N \!$\footnote{$\vphantom{\dot{\dot{\dot{a}}}}$  The formal definition of the functional measure $\clD\mhfpt A$
of $\clH_\msM(N)$ requires endowing $\clH_\msM(N)$ with a Riemannian metric.
The normal bundle $N\clK_{\msM,\varsigma^*\Omega}(N)$ of $\clK_{\msM,\varsigma^*\Omega}(N)$
is the orthogonal complement of the tangent bundle $T\clK_{\msM,\varsigma^*\Omega}(N)$
of $\clK_{\msM,\varsigma^*\Omega}(N)$ with respect to this metric.
More abstractly, we have
$N\clK_{\msM,\varsigma^*\Omega}(N)=T\clH_\msM(N)\big|_{\clK_{\msM,\varsigma^*\Omega}(N)}/T\clK_{\msM,\varsigma^*\Omega}(N)$.}.  
Lastly, $\rmG_\sfF\in\overline{\clF}_{\msM,\rmK}(N)$ is the unique element of the 
$\clH_\msM(N)$--orbit of $\rmG$ obeying $\sfF(\rmG_\sfF)=0$. 
As is well--known, assuming that the measure $\scD A$ is invariant under both left and right multiplicative shifts
of $A$, insertion of the left hand side of relation \ceqref{hafspath6} into the functional integral 
in \ceqref{hafspath5} allows one to factorize the volume $\vol\clH_\msM(N)$ of the gauge transformation group
$\clH_\msM(N)$ out of the integral on one hand and restrict
the integration to the gauge slice defined by $\sfF$ on the other, furnishing \raggedbottom 
\begin{align}
&\sfZ_\sfW(\Omega)=
\frac{1}{\vol\clK_{\msM,\varsigma^*\Omega}(N)}\int_{\overline{\clF}_{\msM,\rmK}(N)} \scD\rmG\,
\ee^{i\sfS(\rmG;\Omega)}\sfW(\rmG;\Omega)
\nonumber
\vphantom{\Big]}
\\
&\hspace{4.4cm}
\det\big((\sfF\circ\sfB_{\varsigma^*\Omega|\rmG})_*(1_N)|_{\NN_{1_N}\clK_{\msM,\varsigma^*\Omega}(N)}\big)
\delta_{\clV}\left(\sfF(\rmG)\right).
\label{hafspath7}
\end{align}
Making the above expression fully explicit in Lagrangian quantum field theory 
requires the introduction of ghost, antighost and Nakanishi--Lautrup fields.

At the formal level, one can give a meaning to expression \ceqref{hafspath5} even when $\sfW$ is not gauge
invariant again by inserting the left hand side of relation \ceqref{hafspath6} into the functional integral. 
The FP gauge fixed expression obtained in this way is of the form \ceqref{hafspath7} with 
$\sfW$ replaced by its $\clH_\msM(N)$--average 
\begin{equation}
\langle\hfpt\sfW\rangle_{\clH_\msM(N)}(\rmG;\Omega)=\frac{1}{\vol\clH_\msM(N)}
\int_{\clH_\msM(N)} \scD A\,\sfW(\sfB_{\varsigma^*\Omega|\rmG}(A),\Omega).
\label{hafspath9}
\end{equation}
Unlike $\sfW$, $\langle\hfpt\sfW\rangle_{\clH_\msM(N)}$ is always gauge invariant by the multiplicative
shift invariance of the measure $\scD A$. Further, when
$\sfW$ is gauge invariant, $\langle\hfpt\sfW\rangle_{\clH_\msM(N)}$ reduces to $\sfW$. 
A less formal definition of $\langle\hfpt\sfW\rangle_{\clH_\msM(N)}$ would however be desirable.
However, we can get information useful for our analysis even without one.

The variation $\delta\sfZ(\Omega)$ of $\sfZ(\Omega)$ under a variation $\delta\varsigma$ of $\varsigma$
given by eq. \ceqref{hafspath8} can be written in the form \ceqref{hafspath5} as 
\begin{equation}
\delta\sfZ(\Omega)=\sfZ_{i\delta\sfS}(\Omega).
\label{hafspath10}
\end{equation}
It can therefore be cast and analyzed through the FP framework that we have detailed above.  

As in the corresponding problem of the ordinary model (cf. subsect. \cref{subsec:safspath}), 
the variational analysis of the embedding dependence of the partition function $\sfZ(\Omega)$
requires a suitable variational framework. This is totally analogous to
that used for the ordinary model. The variational problem is thus naturally framed in the complex
$\Fun(T[1]\clE_{N,M})$, $\delta$, where $\clE_{N,M}=\Emb(N,M)$
is the infinite dimensional functional manifold of the embeddings of $N$ into $M$ and $\delta$
is the variational differential of $\clE_{N,M}$ and more broadly in the augmented complex 
$\Fun(T[1](\clE_{N,M}\times N))$, $\delta+d$, $d$ being the de Rham differential of $N$. 
Recall that the augmented complex has a natural
bigrading induced by the external direct sum decomposition
$T[1](\clE_{N,M}\times N)=T[1]\clE_{N,M}\boxplus T[1]N$ (cf. fnn. \cref{foot:boxplus}, \cref{foot:bernstein}),
with $\delta$, $d$ the bidegree $(1,0)$, $(0,1)$ terms of $\delta+d$. 
Recall moreover that the pull--back 
of a derived field $\Psi\in\Map(T[1]M,\DD\fkm[p])$ by the evaluation map
$\epsilon:\clE_{N,M}\times N\rightarrow M$, a derived field 
$\epsilon^*\Psi\in\Map(T[1](\clE_{N,M}\times N),\DD\fkm[p])$, 
can be decomposed in terms with definite bidegree, the terms of bidegree $(0,p)$, $(0,p+1)$ 
forming the component $\epsilon^*\Psi_N$ of $\epsilon^*\Psi$ along $N$.



The action $\sfS(\rmG;\Omega)$ can be regarded as a degree $0$ element of $\Fun(T[1]\clE_{N,M})$
depending on $\rmG$ and $\Omega$. \pagebreak The variation $\delta\sfS(\rmG;\Omega)$ can be straightforwardly
computed from  \ceqref{hafsmod1}. No boundary contributions occur owing to the boundary
condition $\delta\varsigma|_{\partial N}=0$. 
Inserting the resulting expression in \ceqref{hafspath8}, we obtain 
\vspace{.5mm}
\begin{align}
\delta\sfZ(\Omega)
&=\frac{1}{\vol\clH_\msM(N)}\int_{\clF_{\msM}(N)} \scD\rmG\,\ee^{i\sfS(\rmG;\Omega)}i
\int_{T[1]N}\varrho_N
\vphantom{\Big]}
\label{hafspath12}
\\
&\hspace{1cm}
\left[\left(\rmK,\Ad\rmG^{-1}(\epsilon^*\Phi)\right)
+\left([\Ad\rmG^{-1}(\epsilon^*\Omega_N)+\rmG^{-1}\dd\rmG,\rmK],\Ad\rmG^{-1}(\epsilon^*\Omega)\right)\right].
\vphantom{\Big]}
\nonumber
\end{align}
\noindent
We shall now show 
that under rather general assumptions the second term within square brackets
in the right hand side gives a vanishing contribution. 

Let $\rmJ,\rmH\in\Map_c(T[1]N,\DD\fkm[0])$. Under a variation of the field $\rmG$ such that
$\rmG^{-1}\delta_\rmH\rmG=\rmH$, by the left shift invariance of the functional measure $\scD\rmG$,
we have
\begin{equation}
\frac{1}{\vol\clH_\msM(N)}\int_{\clF_{\msM}(N)} \scD\rmG\,\delta_\rmH\left(\ee^{i\sfS(\rmG;\Omega)}
\int_{T[1]N}\varrho_N(\Ad\rmG^{-1}(\epsilon^*\Omega),\rmJ)\right)=0.
\label{hafspath13}
\end{equation}
By explicit calculation of the variation of the expression within round brackets,
we obtain the Schwinger--Dyson type identity  
\begin{align}
&\frac{1}{\vol\clH_\msM(N)}\int_{\clF_{\msM}(N)} \scD\rmG\,\ee^{i\sfS(\rmG;\Omega)}i\bigg\{
\int_{T[1]N}\varrho_N(\Ad\rmG^{-1}(\epsilon^*\Omega),[\rmH,\rmJ])
\vphantom{\Big]}
\label{hafspath14}%
\\
&-\int_{T[1]N}\varrho_N\left([\Ad\rmG^{-1}(\epsilon^*\Omega_N)+\rmG^{-1}\dd\rmG,\rmK],\rmH\right)
\int_{T[1]N}\varrho_N(\Ad\rmG^{-1}(\epsilon^*\Omega),\rmJ)\bigg\}=0. 
\vphantom{\Big]}
\nonumber
\end{align}
From this relation, by the arbitrariness of $\rmJ$, $\rmH$, it follows that 
\begin{align}
&\frac{1}{\vol\clH_\msM(N)}\int_{\clF_{\msM}(N)} \scD\rmG\,\ee^{i\sfS(\rmG;\Omega)}i\bigg\{
\sft\sfr\left(\sfA\sfd(\Ad\rmG^{-1}(\epsilon^*\Omega))\right)
\vphantom{\Big]}
\label{hafspath15}%
\\
&\hspace{2.5cm}-\int_{T[1]N}\varrho_N\left([\Ad\rmG^{-1}(\epsilon^*\Omega_N)+\rmG^{-1}\dd\rmG,\rmK],
\Ad\rmG^{-1}(\epsilon^*\Omega)\right)\bigg\}=0, 
\vphantom{\Big]}
\nonumber
\end{align}
where $\sft\sfr$, $\sfA\sfd$ denote respectively functional trace and adjoint respectively. Above,
$\sft\sfr\left(\sfA\sfd(\Ad\rmG^{-1}(\epsilon^*\Omega))\right)$ contains a factor
$\delta_N(0)$, which must be regularized, and a factor that pointwise is of the form $\tr\ad O$
for some $O\in\DD\fkm[1]$. 
On account of (3.3.4) of I, one has that 
$\tr\ad X=0$ for every $X\in\DD\fkm$
if and only if $\tr\ad x=0$ and $\tr\sdot\mu\sdot(x,\cdot)=0$ for $x\in\fkg$. We call
a Lie algebra crossed module $\fkm$ with such a property unimodular. If this property holds,
$\sft\sfr\left(\sfA\sfd(\Ad\rmG^{-1}(\epsilon^*\Omega))\right)=0$ and so 
\begin{align}
&-\frac{1}{\vol\clH_\msM(N)}\int_{\clF_{\msM}(N)} \scD\rmG\,\ee^{i\sfS(\rmG;\Omega)}
\hspace{5cm}
\vphantom{\Big]}
\label{hafspath16}
\end{align}
\begin{align}
&\hspace{2.5cm}\int_{T[1]N}\varrho_N\left([\Ad\rmG^{-1}(\epsilon^*\Omega_N)+\rmG^{-1}\dd\rmG,\rmK],
\Ad\rmG^{-1}(\epsilon^*\Omega)\right)=0,
\vphantom{\Big]}
\nonumber
\end{align}
as anticipated. From \ceqref{hafspath12} and \ceqref{hafspath16}, we find then that 
\begin{equation}
\delta\sfZ(\Omega)
=\frac{1}{\vol\clH_\msM(N)}\int_{\clF_{\msM}(N)} \scD\rmG\,\ee^{i\sfS(\rmG;\Omega)}
\int_{T[1]N}\varrho_N\left(\rmK,\Ad\rmG^{-1}(\epsilon^*\Phi)\right). 
\label{hafspath17}
\end{equation}
In this way, we have that
\begin{equation}
\delta\sfZ(\Omega)=0 \qquad \text{if $\Phi=0$}.
\label{hafspath18}
\end{equation}
We find in this way that $\sfZ(\Omega)$ is invariant under suitably
boundary restricted variations of the embedding $\varsigma$ 
when the derived gauge field $\Omega$ is flat. In this respect, the
derived TCO model is completely akin to  the ordinary one (cf. eq. \ceqref{safspath18}). 

In refs. \ccite{Zucchini:2015wba,Zucchini:2015xba} a general definition of
surface knot holonomy is provided. A closed base 2--fold $N$ of genus $\ell$ is said marked
when it is endowed with a choice of a point $p_N$ and $2\ell$ closed curves $C_{Ni}$
intersecting at $p_N$ only representing the homology classes of the standard $a$- and $b$--cycles. 
An ambient manifold $M$ is said marked when it is endowed with a choice of a point $p_M$
and $2\ell$ closed curves $C_{Mi}$ intersecting at $p_M$ only. A based surface knot of genus $\ell$
of the marked manifold $M$ is an embedding $\varsigma:N\rightarrow M$ that is  marking preserving,
i.e. such that $\varsigma(p_N)=p_M$ and $\varsigma(C_{Ni})=C_{Mi}$. 
In a higher gauge theory with gauge crossed module $\msM$, 
one can associate with any flat derived gauge field $\Omega$ and based surface knot $\varsigma$ 
the surface holonomy $F_\Omega(\varsigma)\in \msE$. $F_\Omega(\varsigma)$ is not invariant
under the gauge transformations of $\Omega$ of the type \ceqref{higau9} 
and depends also on the marking of $M$ for a fixed marking of
$N$. It can be shown however that gauge transforming $\Omega$ and smoothly deforming
the marking and the surface knot $\varsigma$ through ambient
isotopy affects $F_\Omega(\varsigma)$ only by a $\mu$--conjugation, 
\begin{equation}
F_\Omega(\varsigma)\rightarrow \mu(a)(F_\Omega(\varsigma))
\vphantom{\Big]}
\label{hafspath19}
\end{equation}
for some marking dependent element $a\in\msG$. 
Therefore, for given $\Omega$ and $\varsigma$, only the $\mu$--conjugation class of 
$F_\Omega(\varsigma)$ is 
uniquely determined in a gauge independent and ambient isotopy invariant fashion. 
If we wish to extract numerical invariants out of the holonomy of a surface knot $\varsigma$, we need 
therefore a $\mu$--trace over $\msE$, which we define as a mapping 
$\tr_\mu:\msE\rightarrow \bbC$ invariant under $\mu$--conjugation, viz $\tr_\mu(\mu(a)(A))=\tr_\mu(A)$
for $a\in\msG$ and $A\in\msE$. Our claim is that the partition function $\sfZ(\Omega)$ possibly computes
a $\mu$--trace of the holonomy of $\varsigma$
\begin{equation}
\sfZ(\Omega)=\tr_\mu(F_\Omega(\varsigma)).
\label{hafspath20}
\end{equation}
This conjecture is supported by the background gauge invariance and homotopy invariance
properties \ceqref{hafspath3} and \ceqref{hafspath18} and
the analogy to the ordinary case, see eq. \ceqref{safspath20}, where a similar identification
can be established for line knot holonomies. However, the arguments expounded above 
employ formal functional integral techniques and are based on certain assumptions
concerning the fixing of the special gauge symmetry, which albeit reasonable, remain
to be verified. For these reasons, our analysis does not provided strictly speaking a full proof of
the identification but only an argument in favour of it, albeit one with a sound theoretically grounding.
To conclusively establish the identification, it would be necessary to settle the gauge fixing issue
and check it by carrying out calculations of the partition function for simple choices of the model's data,
in particular for the characteristic TCO model of subsect. \cref{subsec:hafsspx}
whose relation to the derived KKS theory developed in I can be shown. 

While the derived TCO model is a topological quantum field theory, the ordinary one is just a
topological quantum mechanics. It is to be noticed, moreover, that the gauge fixing issue is
specific of the former, which is a gauge theory, and has no analog in the latter, which is not.
This explains why the functional integral analysis of the ordinary model reviewed in
sect. \cref{sec:tcoreview} culminating with relation \ceqref{safspath20} can be considered a proof
of the identification of the model's partition function and a Wilson loop 
while the corresponding analysis of the derived model leading to relation \ceqref{hafspath20},
in spite of many formal similarities, can be considered a proof
of the identity of the partition function and a Wilson surface somewhat less so. 
We add to that that \ceqref{safspath20} is supported by the calculations presented in refs.
\ccite{Alekseev:1988vx,Diakonov:1989fc,Diakonov:1996zu}, while similar calculations
checking \ceqref{hafspath20} are not available at present.







\vfil\eject

\subsection{\textcolor{blue}{\sffamily Canonical formulation of the derived TCO model}}\label{subsec:hafscan}

In this subsection, we shall construct the canonical theory of the derived TCO model.
The different perspective allowed by the canonical formulation will furnish us new insight into the model.

The canonical formulation of the derived TCO model broadly follows the pattern of that of the
ordinary model (cf. subsect. \cref{subsec:safscan}). In the derived case, however, the canonical theory
is far 
richer because of the appearance of first class constraints corresponding
to the special gauge symmetry.

To set up the canonical framework, we assume that the model's base manifold $N$
is $\widetilde L=\bbR^1\times L$, where $L$ is a compact connected 
1--fold, so either $\bbS^1$ or $\bbI^1$, the circle and the interval.
The factor $\bbR^1$ is just the time axis. 
The resulting external direct
sum decomposition $T[1]\widetilde L=T[1]\bbR^1\boxplus T[1]L$ (cf. fn. \cref{foot:boxplus})
of the degree shifted tangent bundle of $\widetilde L$ entails that 
the function algebra $\Fun(T[1]\widetilde L)$ is not only graded but also bigraded in compatible fashion 
(cf. fn. \cref{foot:bernstein}).
A  generic field $\widetilde\lambda\in\Map(T[1]\widetilde L,\bbR[p])$ can therefore be expressed as 
$\widetilde\lambda=\lambda_t+\lambda$, where $\lambda_t,\lambda\in\Map(T[1]\widetilde L,\bbR[p])$ 
have 
bidegrees $(1,p-1)$, $(0,p)$, respectively
\footnote{$\vphantom{\dot{\dot{\dot{a}}}}$  For dimensional reasons, $\widetilde\lambda$ is non vanishing
only for $p=0,1,2$. Furthermore, for $p=0$, $\lambda_t=0$; for $p=2$, $\lambda=0$.}.
Similarly, the de Rham differential
$\widetilde d$ of $\widetilde L$ splits as $\widetilde d=d_t+d$, where the bidegree $(1,0)$, $(0,1)$
$d_t$, $d$ terms are accordingly the de Rham differentials of $\bbR^1$, $L$. 

In the derived framework, fields are amenable to a similar bidegree analysis. 
A field $\widetilde\rmU\in\Map(T[1]\widetilde L,\DD\msM)$ factorizes as 
\begin{equation}
\widetilde\rmU=\rmU_t\rmU 
\label{nhafscan1}
\end{equation}
with $\rmU_t,\rmU\in\Map(T[1]\widetilde L,\DD\msM)$. The multiplicative field constituents 
$\rmU_t$, $\rmU$ are
uniquely determined by certain requirements on the their components (cf. eq. (3.3.1) of I),
viz that $u_t=1_\msG$ and that $U_t$, $U$ have bidegrees $(1,0)$, $(0,1)$, respectively. In similar fashion, 
a field $\widetilde\Psi\in\Map(T[1]\widetilde L,\DD\fkm[p])$ decomposes as
\begin{equation}
\widetilde\Psi=\Psi_t+\Psi, 
\label{nhafscan2}
\end{equation}
\eject\noindent
where $\Psi_t,\Psi\in\Map(T[1]\widetilde L,\DD\fkm[p])$. The additive field constituents 
$\Psi_t$, $\Psi$ are again uniquely determined by conditions on their components, namely that 
$\psi_t$, $\psi$, $\varPsi_t$, $\varPsi$ have bidegrees $(1,p-1)$, $(0,p)$, $(1,p)$, $(0,p+1)$,
respectively. Accordingly, the derived differential $\widetilde\dd$ of $\widetilde L$ can be expressed as 
\begin{equation}
\widetilde\dd=d_t+\dd,
\label{nhafscan3}
\end{equation}
where the bidegree $(1,0)$, $(0,1)$ components $d_t$, $\dd$ correspond to 
the ordinary de Rham differential of $\bbR^1$ and the derived differential of $L$. 

Relying the above geometrical set--up, we can write the derived TCO model's action
$\sfS$ on $N=\widetilde L$ in the form required for canonical analysis.
The TCO field factorizes as $\widetilde\rmG=\rmG_t\rmG$ in accordance with \eqref{nhafscan1}.
The base manifold pull--back of the background derived gauge field splits as
$\widetilde\varsigma^*\Omega=\varsigma^*{}_t\Omega+\varsigma^*\Omega$ as in \eqref{nhafscan2}.
The level current similarly reads as $\widetilde\rmK=\rmK_t+\rmK$. By explicit computation, we find that
$\sfS$ gets expressed in terms of the 
field constituents $\rmG_t$, $\rmG$, $\varsigma^*{}_t\Omega$, $\varsigma^*\Omega$ $\rmK_t$, $\rmK$ as 
\begin{align}
\sfS(\widetilde\rmG;\,&\Omega)=\int_{T[1]\widetilde L}\varrho_{\widetilde L}
\big[-\big(\Ad\rmG([\Ad\rmG^{-1}(\varsigma^*\Omega)+\rmG^{-1}\dd\rmG,\rmK]),\widehat\rmG_t\big)
\label{nhafscan4}
\\
+&\left(\rmK_t,\Ad\rmG^{-1}(\varsigma^*\Omega)+\rmG^{-1}\dd\rmG\right)
+\left(\rmK,\Ad\rmG^{-1}(\varsigma^*{}_t\Omega)+\rmG^{-1}d_t\rmG+\rmG^{-1}\dd\rmG\right)\big],
\vphantom{\Big]}
\nonumber
\end{align}
where $\widehat\rmG_t\in\Map(T[1]\widetilde L,\DD\fkm[0])$ is a derived field of components $\widehat g_t=0$,
$\widehat G_t=G_t$. \linebreak Inspection of relation \ceqref{nhafscan4} reveals that $\widehat\rmG_t$
is a non dynamical Lagrange multiplier field enforcing the constraints 
\begin{equation}
[\rmK,\Ad\rmG^{-1}(\varsigma^*\Omega)+\rmG^{-1}\dd\rmG]\approx 0.
\label{nhafscan5}
\end{equation}
As we shall show in due course,
when the background gauge field $\Omega$ obeys the fake flatness condition \eqref{hafsmod7},
these constraints reflect the special gauge symmetry
of the derived TCO model (cf. subsect. \cref{subsec:hafssym}).
In canonical derived TCO theory, the gauge symmetry is partially 
fixed by imposing the condition $\widehat\rmG_t=0$, that is, 
\begin{equation}
\rmG_t=1_{\DD\msM}. 
\label{nhafscan6}
\end{equation}
\ceqref{nhafscan6} leaves the special gauge transformations$\vphantom{\ul{\ul{\ul{g}}}}$
\pagebreak
$\widetilde\rmT\in\clG_{\msM,\varsigma^*\Omega}(\widetilde L)$ such
that $\rmT_t=1_{\DD\msM}$ with regard to the factorization \ceqref{nhafscan1} as the only allowed ones.


In canonical theory, we replace the derived TCO field 
$\rmG$ with a time independent field, also denoted as $\rmG$, viewed as a point 
of an ambient functional phase space $\clF_{\msM}(L):=\Map(T[1]L,\DD\msM)$. 
The study of the symplectic structure \linebreak 
of $\clF_{\msM}(L)$ is naturally carried out using the variational Cartan calculus of $\clF_{\msM}(L)$. 
This features a collection of variational derivations of the functional algebra $\Fun(T[1]\clF_{\msM}(L))$ 
comprising the degree $-1$ contractions $\iota_\sfV$ and degree $0$ Lie derivatives $\lambda_\sfV$
along the vector fields $\sfV\in\Vect(\clF_{\msM}(L))$ and the degree $1$ differential $\delta$
and satisfying graded commutation relations of the form (A.3.1)--(A.3.4) of I.
Since the fields appearing in our analysis can be reduced to fields belonging to the graded tensor product algebra
$\Fun(T[1]\clF_{\msM}(L))\hfpt\otimes\hfpt\Fun(T[1]L)$, they are naturally bigraded 
(cf. fn. \cref{foot:bernstein}). 

As in the canonical formulation of the ordinary model (cf. subsect. \cref{subsec:safscan}), 
the appropriate form of the presymplectic potential 1--form $\sfPi$ is indicated by that
of the kinetic term $(\rmK,\rmG^{-1}d_t\rmG)$ appearing in the expression of the 
TCO action $\sfS$ given in eq. \ceqref{nhafscan4}. It therefore depends on the level current term
$\rmK$. However, the requirement that $\sfPi$ and the deriving presymplectic 2--form $\sfPsi=\delta\sfPi$
be time independent necessitates that certain restrictions are imposed on $\widetilde\rmK$. The most obvious
and natural one is that \hphantom{xxxxxx}
\begin{equation}
\rmK_t=0.
\label{khafscan1}
\end{equation}
\ceqref{khafscan1} together with condition \ceqref{hafsmod0}, here reading as $\widetilde\dd\widetilde\rmK=0$,
then yield 
\begin{equation}
d_t\rmK=0\quad \text{and}\quad\dd\rmK=0.
\label{khafscan2}
\end{equation}
The first relation simply states the time independence of $\rmK$. The second
avoids problematic terms in our analysis. 
The presymplectic potential 1--form
\begin{equation}
\sfPi=\int_{T[1]L}\varrho_L (\rmK,\rmG^{-1}\delta\rmG)
\label{hafscan1}
\end{equation}
and the associated presymplectic 2-form 
\begin{equation}
\sfPsi=\delta\sfPi=\frac{1}{2}\int_{T[1]L}\varrho_L\mhfpt
\left(\rmK,[\rmG^{-1}\delta\rmG,\rmG^{-1}\delta\rmG]\right) \pagebreak 
\label{hafscan2}
\end{equation}
have in this way the required time independence properties.
The form of $\sfPi$ and $\sfPsi$ clearly mirrors that of the corresponding
objects of the ordinary model's canonical theory (cf. eqs. \ceqref{safscan1}, \ceqref{safscan2})

In subsect. \cref{subsec:hafssym}, we found that the derived TCO model is an effective theory
of the level preserving $\clG_{\msM,\rmK}(N)$-orbit space $\overline{\clF}_{\msM,\rmK}(N)$
(cf. eq. \ceqref{hafspath0}), leading under the assumption of the existence of the 
crossed submodule $\msM_\rmK$ of $\msM$ to the reinterpretation of the model as a sigma model.
Such primary property of the model is expected to turn up also in the model's canonical theory. 
In the canonical formulation, the level preserving gauge transformation group $\clG_{\msM,\rmK}(L)$
is the subgroup of $\clG_{\msM}(L)$ formed by the elements $\Upsilon\in\clG_{\msM}(L)$
obeying the condition
\begin{equation}
\Ad\Upsilon(\rmK)=\rmK,
\label{hafscan3}
\end{equation}
consistently with \ceqref{hafssym1}, and acting on the fields $\rmG\in\clF_{\msM}(L)$ as 
\begin{equation}
\rmG^\Upsilon=\rmG\Upsilon, 
\label{hafscan4}
\end{equation}
accordingly with \ceqref{hafssym2}. By \ceqref{hafscan3} and \ceqref{hafscan4}, the infinitesimal
level preserving gauge transformation algebra is the Lie algebra $\fkg_{\msM,\rmK}(L)$
of $\clG_{\msM,\rmK}(L)$ and the action \ceqref{hafscan4} is implemented infinitesimally
for any $\rmZ\in\fkg_{\msM,\rmK}(L)$ by the vector field $\sfX_\rmZ\in\Vect(\clF_{\msM}(L))$ acting as 
\begin{equation}
\iota_{\hfpt\sfX_\rmZ}(\rmG^{-1}\delta\rmG)=\rmZ.
\label{hafscan5}
\end{equation}
By \ceqref{hafscan2} and \ceqref{hafscan5}, we have 
\begin{equation}
\iota_{\hfpt\sfX_\rmZ}\sfPsi=0.
\label{hafscan7}
\end{equation}
Conversely, any vector field $\sfX\in\Vect(\clF_{\msM}(L))$ such that
$\iota_{\hfpt\sfX}\sfPsi=0$ is of the form $\sfX=\sfX_\rmZ$ for some $\rmZ\in\fkg_{\msM,\rmK}(L)$
pointwise in $\clF_{\msM}(L)$.
The degeneracy of $\sfPsi$ highlighted by these properties and its form
reveal that the phase space of the model is to be properly identified with the 
$\clG_{\msM,\rmK}(L)$--orbit space of $\clF_{\msM}(L)$, \hphantom{xxxxxxxxx}
\begin{equation}
\overline{\clF}_{\msM,\rmK}(L):=\clF_{\msM}(L)/\clG_{\msM,\rmK}(L).
\label{hafscan8}
\end{equation}
$\sfPsi$ indeed induces \pagebreak a symplectic 2--form $\overline\sfPsi$
on $\overline{\clF}_{\msM,\rmK}(L)$ together with
the  associated Poisson bracket structure $\{\cdot,\cdot\}$ on the functional algebra
$\Fun(\overline{\clF}_{\msM,\rmK}(L))$ of $\overline{\clF}_{\msM,\rmK}(L)$. These findings are
in line with the facts recalled at the start of this paragraph.
The above analysis closely parallel and is indeed the obvious derived extension of
the corresponding analysis for the ordinary model in subsect. \cref{subsec:safscan}
(cf. eqs. \ceqref{safscan3}--\ceqref{safscan7}). 

The canonical set--up of the derived TCO model is now fully in place. 
Expressing the Poisson bracket structure of $\overline{\clF}_{\msM,\rmK}(L)$
is hardly doable if one works with orbits, while it is apparently handier to
do if one relies on orbit representative, much as for the ordinary model in subsect. \cref{subsec:safscan}. 
We thus compute the Poisson bracket of functionals of $\Fun(\overline{\clF}_{\msM,\rmK}(L))$ 
relying on two basic isomorphisms. The first isomorphism,
$\Fun(\overline{\clF}_{\msM,\rmK}(L))\simeq\Fun(\clF_{\msM}(L))^{\clG_{\msM,\rmK}(L)}$,
identifies the functional algebra $\Fun(\overline{\clF}_{\msM,\rmK}(L))$ of $\overline{\clF}_{\msM,\rmK}(L)$
and the subalgebra $\Fun(\clF_{\msM}(L))^{\clG_{\msM,\rmK}(L)}$
of $\Fun(\clF_{\msM}(L))$ of the functionals invariant under the $\clG_{\msM,\rmK}(L)$--action \ceqref{hafscan4}.
The second isomorphism, $\Vect(\overline{\clF}_{\msM,\rmK}(L))\simeq\WW\Vect_\rmK(\clF_\msM(L))$, 
equates the vector field Lie algebra $\Vect(\overline{\clF}_{\msM,\rmK}(L))$ of $\overline{\clF}_{\msM,\rmK}(L)$ and
the Weyl Lie algebra $\WW\Vect_\rmK(\clF_\msM(L))$ 
of the Lie subalgebra $\Vect_\rmK(\clF_\msM(L))$ of $\Vect(\clF_\msM(L))$ of vector fields $\sfV\in\Vect(\clF_\msM(L))$
of the form $\sfV=\sfX_\rmZ$ for some $\rmZ\in\fkg_{\msM,\rmK}(L)$ pointwise in $\clF_{\msM}(L)$ 
(cf. fn. \cref{foot:weyl}).
In this way, a functional $\sfF\in\Fun(\overline{\clF}_{\msM,\rmK}(L))$ will be thought of as
a functional $\sfF\in\Fun(\clF_{\msM}(L))$ such that $\sfF(\rmG^\Upsilon)=\sfF(\rmG)$ for $\Upsilon\in\clG_{\msM,\rmK}(L)$.
Similarly, a vector field $\sfV\in\Vect(\overline{\clF}_{\msM,\rmK}(L))$ will be regarded as a vector field 
$\sfV\in\Vect(\clF_\msM(L))$ defined mod  vector fields $\sfV'\in\Vect_\rmK(\clF_\msM(L))$ and
such that $[\sfV,\sfV']\in\Vect_\rmK(\clF_\msM(L))$ for any vector field $\sfV'\in\Vect_\rmK(\clF_\msM(L))$. 
The expression of the Poisson bracket including
a functional $\sfF\in\Fun(\overline{\clF}_{\msM,\rmK}(L))$ involves the Hamiltonian vector field
$\sfV_\sfF\in\Vect(\overline{\clF}_{\msM,\rmK}(L))$ of $\sfF$ defined by the relation 
\begin{equation}
\delta\sfF+\sfiota_{\sfV_\sfF}\sfPsi=0
\label{hafscan17}
\end{equation}
(cf. eq. \ceqref{safscan17}). 
The Poisson bracket of a pair of functionals $\sfF,\sfG\in\Fun(\overline{\clF}_{\msM,\rmK}(L))$ 
is then the functional $\{\sfF,\sfG\}\in\Fun(\overline{\clF}_{\msM,\rmK}(L))$ given by the expression 
\begin{equation}
\{\sfF,\sfG\}=\sfiota_{\sfV_\sfF}\delta\sfG=-\sfiota_{\sfV_\sfG}\delta\sfF 
\label{hafscan18}
\end{equation}
(cf. eq. \ceqref{safscan18}). 

We can now tackle the issue of the canonical characterization of the constraints \ceqref{nhafscan5}.
A preliminary observation: 
since the restriction of the embedding $\widetilde\varsigma$ of $\widetilde L$
into $M$ to a time slice of $\{t\}\times L$ of $\widetilde L$ is an embedding $\{t\}\times L$ into $M$,
the canonical theoretic counterpart of $\widetilde\varsigma$ is an embedding $\varsigma:L\rightarrow M$.
Reflecting the structure of \ceqref{nhafscan5}, we now define the functional 
\begin{equation}
\sfN(\rmW)(\rmG;\Omega)=\int_{T[1]L}\varrho_L
\left([\rmK,\Ad\rmG^{-1}(\varsigma^*\Omega)+\rmG^{-1}\dd\rmG],\Ad\rmG^{-1}(\rmW)\right),
\label{nhafscan7}
\end{equation}
where $\rmW\in\clP_\msM(L)$ with $\clP_\msM(L)=\Map(T[1]L,\DD\fkm[-1])$. 
Recalling that $\dd\rmK=0$, one verifies 
that $\sfN(\rmW)$ is $\clG_{\msM,\rmK}(L)$--invariant, so that 
$\sfN(\rmW)\in\Fun(\overline{\clF}_{\msM,\rmK}(L))$.
The Hamiltonian vector field $\sfV_\rmW:=\sfV_{\sfN(\rmW)}\in\Vect(\clF_\msM(L))$
of $\sfN(\rmW)$ is readily expressed through its action on $\clF_\msM(L)$, which reads as 
\begin{equation}
\sfiota_{\sfV_\rmW}(\delta\rmG\rmG^{-1})=-\dd\sfV_\rmW-[\varsigma^*\Omega,\sfV_\rmW]
\label{nhafscan8}
\end{equation}
for $\rmG\in\clF_\msM(L)$. 
Using this relation, it is straightforward to compute the Poisson bracket of
$\sfN(\rmW)$, $\sfN(\rmW')$ for $\rmW,\rmW'\in\clP_\msM(L)$. We find, 
\begin{equation}
\{\sfN(\rmW),\sfN(\rmW')\}=\sfN([\rmW,\rmW']_*),
\label{nhafscan13}
\end{equation}
where $[\rmW,\rmW']_*\in\clP_\msM(L)$ is given by the expression 
\begin{equation}
  [\rmW,\rmW']_*
  =\frac{1}{2}\left([\rmW,\dd\rmW'+[\varsigma^*\Omega,\rmW']]-[\rmW',\dd\rmW+[\varsigma^*\Omega,\rmW]]\right).
\label{nhafscan14}
\end{equation}
The bracket $[\cdot,\cdot]_*$ is antisymmetric but does not obey the Jacobi identity.
The failure $[\cdot,\cdot]_*$ to be a Lie bracket remains however compatible with the Jacobi property
of the Poisson bracket $\{\cdot,\cdot\}$, as is not difficult to verify.

The emergence of the constraints \ceqref{nhafscan5} in the canonical analysis of the derived TCO model 
indicates that the constraints 
\begin{equation}
\sfN(\rmW)\approx 0
\label{nhafscan15}
\end{equation}
with $\rmW\in\clP_\msM(L)$ should be imposed. 
Relation \ceqref{nhafscan13} shows that the functionals $\sfN(\rmW)$ 
are first class. Therefore, these constraints are associated with a gauge symmetry.
We are now going to see that this is precisely the special symmetry
studied in subsect. \cref{subsec:hafssym}.

In subsect. \cref{subsec:hafssym}, we saw that the special gauge transformation action
can ta\-ke either the dressed $\clG_{\msM,\varsigma^*\Omega}(N)$ or the bare $\clH_\msM(N)$ form.
In subsect. \cref{subsec:hafspath},
we concluded that the second option is the most natural one for functional integral quantization. Based on these
findings, we suppose that $\clH_\msM(L)$ is the appropriate gauge transformation group
in the canonical formulation. The transform of a field $\rmG\in\clF_{\msM}(L)$
by a gauge transformation $A\in\clH_\msM(L)$ is
\begin{equation}
\rmG^A=\rmG^{\sfT_{\varsigma^*\Omega}(A)} 
\label{nhafscan16}
\end{equation}
in conformity with 
\ceqref{hafssym14}, where the morphism
$\sfT_{\varsigma^*\Omega}:\clH_\msM(L)\rightarrow\clG_{\msM,\varsigma^*\Omega}(L)$ is
defined in subsect. \cref{subsec:hisym}. Correspondingly, 
the action of an infinitesimal gauge transformation
$\varPi\in\fkh_\msM(L)$ on a field $\rmG\in\clF_{\msM}(L)$ reads as 
\begin{equation}
\delta_\varPi\rmG\rmG^{-1}=-\dot\sfT_{\varsigma^*\Omega}(\varPi).
\label{nhafscan17}
\end{equation}

The component $w$ of $\rmW\in\clP_\msM(L)$ vanishes identically,
since $w$  has degree $-1$ and $T[1]L$ is a non negatively graded manifold.
The component $W$ of $\rmW$, which has degree $0$, can conversely be non zero.
Therefore, there is an element $\varPi\in\Map(T[1]L,\fke)=\fkh_\msM(L)$  such that $\rmW=\rmW_\varPi$, where
$\rmW_\varPi\in\clP_\msM(L)$ with
\footnote{$\vphantom{\dot{\dot{\dot{a}}}}$  Note that under the sign convention (3.3.5) of I, 
one has $W_\varPi=-\varPi$.
}
\begin{equation}
\rmW_\varPi(\alpha)=\alpha\varPi. 
\label{nhafscan9}
\end{equation}
Set $\sfV_\varPi=\sfV_{\rmW_\varPi}$ for brevity. Inserting \ceqref{nhafscan9} into \ceqref{nhafscan8},
we get
\begin{equation}
\sfiota_{\sfV_\varPi}(\delta\rmG\rmG^{-1})(\alpha)
=-\dot\tau(\varPi)+\alpha\left(d\varPi+\sdot\mu\sdot(\omega,\varPi)\right).
\label{nhafscan10}
\end{equation}
Expressing alternatively $\dot\sfT_{\varsigma^*\Omega}(\varPi)$ in \ceqref{nhafscan17}
in terms of $\varPi$ by utilizing 
\ceqref{higau27}, \ceqref{higau28}, it emerges that 
\begin{equation}
\sfiota_{\sfV_\varPi}(\delta\rmG\rmG^{-1})=\delta_\varPi\rmG\rmG^{-1}.
\label{nhafscan18}
\end{equation}
$\sfV_\varPi$ is thus the vector field enacting the infinitesimal 
transformation $\varPi\in\fkh_\msM(L)$. The identification of the gauge symmetry associated with the first class
constraints \ceqref{nhafscan15} with the special gauge symmetry 
is thereby ascertained.

From the above discussion, it follows that 
\begin{equation}
\sfN(\varPi)=\sfN(\rmW_\varPi)
\label{nhafscan11}
\end{equation}
are the Hamiltonian functionals of the infinitesimal special 
gauge transformations $\varPi\in\fkh_\msM(L)$. From relations \ceqref{nhafscan13}, \ceqref{nhafscan14},
taking \ceqref{nhafscan9} into account, we find further that for $\varPi,\varPi'\in\fkh_\msM(L)$ 
\begin{equation}
\{\sfN(\varPi),\sfN(\varPi')\}=\sfN([\varPi,\varPi']).
\label{nhafscan12}
\end{equation}
The appearance in the right hand side of this relation of the Lie bracket of $\fkh_\msM(L)$ supports 
by its naturality our assumption that $\clH_\msM(L)$ is the appropriate realization
of the special gauge transformation group.

The above analysis indicates that the physical phase space
$\overline{\clF}_{\msM,\rmK\,{\rm phys}}(L)$ of the derived TCO sigma model is
the subspace of the ambient phase space $\overline{\clF}_{\msM,\rmK}(L)$ defined through the constraints 
\begin{equation}
\sfN(\varPi)\approx 0
\label{nhafscan19}
\end{equation}
for $\varPi\in\fkh_\msM(L)$. That $\overline{\clF}_{\msM,\rmK\,{\rm phys}}(L)$ is indeed a subspace of 
$\overline{\clF}_{\msM,\rmK}(L)$ follows from the $\clG_{\msM,\rmK}(L)$ invariance of the functionals
$\sfN(\varPi)$. 
The reduced physical phase space $\overline{\clF}_{\msM,\rmK\,{\rm red\,phys}}(L)$ is the 
quotient of $\overline{\clF}_{\msM,\rmK\,{\rm phys}}(L)$ by the special gauge symmetry 
\begin{equation}
\overline{\clF}_{\msM,\rmK\,{\rm red\,phys}}(L):=\overline{\clF}_{\msM,\rmK\,{\rm phys}}(L)/\clH_\msM(L).
\vphantom{\Big]}
\label{nhafscan20}
\end{equation}
Both $\overline{\clF}_{\msM,\rmK\,{\rm phys}}(L)$ and $\overline{\clF}_{\msM,\rmK\,{\rm red\,phys}}(L)$
depend secretly on the background gauge field $\Omega$, because the Hamiltonian functionals $\sfN(\varPi)$ and the
action of $\clH_\msM(L)$ on $\overline{\clF}_{\msM,\rmK\,{\rm phys}}(L)$, implemented by 
the dressed to bare action conversion map $\sfT_{\varsigma^*\Omega}$, do. The Poisson bracket structure of
$\overline{\clF}_{\msM,\rmK\,{\rm red\,phys}}(L)$ has a standard description.
Let $\scI_\sfN$ be the ideal of the algebra $\Fun(\overline{\clF}_{\msM,\rmK}(L))$ generated 
by the functionals $\sfN(\varPi)$ with $\varPi\in\fkh_\msM(L)$. By virtue of the first classness of
the $\sfN(\varPi)$, $\scI_\sfN$ is a Poisson subalgebra of $\Fun(\overline{\clF}_{\msM,\rmK}(L))$. 
The isomorphism $\Fun(\overline{\clF}_{\msM,\rmK\,{\rm red\,phys}}(L))\simeq\WW\scI_\sfN$ then holds,
where $\WW\scI_\sfN$ is the Weyl Poisson Lie algebra of $\scI_\sfN$ (cf. fn. \cref{foot:weyl}).
The isomorphism can be understood as follows.
$\scI_\sfN$ is constituted by the functionals  $\sfF\in\Fun(\overline{\clF}_{\msM,\rmK}(L))$
with $\sfF|_{\overline{\clF}_{\msM,\rmK\,{\rm phys}}(L)}=0$. 
As any functional of $\Fun(\overline{\clF}_{\msM,\rmK\,{\rm phys}}(L))$
can be extended non uniquely to some functional of $\Fun(\overline{\clF}_{\msM,\rmK}(L))$,
we have an isomorphism $\Fun(\overline{\clF}_{\msM,\rmK\,{\rm phys}}(L))\simeq\Fun(\overline{\clF}_{\msM,\rmK}(L))/\scI_\sfN$.  
From the first classness of the Hamiltonian generators $\sfN(\varPi)$
again, the infinitesimal special gauge transformation action on $\Fun(\overline{\clF}_{\msM,\rmK}(L))$
descends onto $\Fun(\overline{\clF}_{\msM,\rmK\,{\rm phys}}(L))$ with $\delta_\varPi(\sfF+\scI_\sfN)
=\{\sfN(\varPi),\sfF\}+\scI_\sfN$ for $\sfF\in\Fun(\overline{\clF}_{\msM,\rmK}(L))$ and $\varPi\in\fkh_\msM(L)$. 
A functional $\sfF+\scI_\sfN\in\Fun(\overline{\clF}_{\msM,\rmK\,{\rm red\,phys}}(L))$ is therefore one
such that $\{\sfF',\sfF\}\in\scI_\sfN$ for $\sfF'\in\scI_\sfN$, i.e. that $\sfF\in\NN\scI_\sfN$, the Poisson normalizer
of $\scI_\sfN$. As a consequence, 
$\Fun(\overline{\clF}_{\msM,\rmK\,{\rm red\,phys}}(L))\simeq\NN\scI_\sfN/\scI_\sfN=\WW\scI_\sfN$ as claimed.
Concretely, the upshot of this analysis is that  
a functional $\sfF+\scI_\sfN\in\Fun(\overline{\clF}_{\msM,\rmK\,{\rm red\,phys}}(L))$ can be viewed as 
a functional $\sfF\in\Fun(\overline{\clF}_{\msM,\rmK}(L))$ defined up to the addition of functionals  
$\sfF'\in\Fun(\overline{\clF}_{\msM,\rmK}(L))$ with $\sfF'\approx 0$ and with the property that
$\{\sfF',\sfF\}\approx 0$ for all $\sfF'\in\Fun(\overline{\clF}_{\msM,\rmK}(L))$  with $\sfF'\approx 0$. 
The Poisson bracket of two functionals
$\sfF+\scI_\sfN,\sfG+\scI_\sfN\in\overline{\clF}_{\msM,\rmK\,{\rm red\,phys}}(L)$ is 
given now by the natural expression 
\begin{equation}
\{\sfF+\scI_\sfN\,\sfG+\scI_\sfN\}=\{\sfF,\sfG\}+\scI_\sfN. \vphantom{\Big]_g^f}
\label{nhafscan22}
\end{equation}
It can be 
checked that this Poisson bracket structure is well--defined,
in particular that it has $\Fun(\overline{\clF}_{\msM,\rmK\,{\rm red\,phys}}(L))$ as its range,  
and has all the required properties.

The derived TCO phase space $\overline{\clF}_{\msM,\rmK}(L)$ is characterized by a wider phase space
gauge symmetry subsuming the special gauge symmetry. The associated
gauge transformation group is the full derived gauge transformation group $\clG_\msM(L)$.
The action $\clG_\msM(L)$ on $\overline{\clF}_{\msM,\rmK}(L)$
reads as 
\begin{equation}
\rmG^\rmT=\rmT^{-1}\rmG
\label{hafscan9}
\end{equation}
with $\rmT\in\clG_\msM(L)$ and $\rmG\in\overline{\clF}_{\msM,\rmK}(L)$.
The action of an infinitesimal gauge transformation $\rmS\in\fkg_\msM(L)$
is enacted by a vector field $\sfU_\rmS\in\Vect(\overline{\clF}_{\msM,\rmK}(L))$ such that
\begin{equation}
\iota_{\hfpt\sfU_\rmS}(\delta\rmG\rmG^{-1}) =-\rmS.
\label{hafscan10}
\end{equation}
This symmetry is Hamiltonian with respect to the symplectic structure
of $\overline{\clF}_{\msM,\rmK}(L)$: 
there is a functional $\sfF(\rmS)\in\Fun(\overline{\clF}_{\msM,\rmK}(L))$
obeying 
\begin{equation}
\delta\sfF(\rmS)+\iota_{\hfpt\sfU_\rmS}\sfPsi=0. 
\label{hafscan11}
\end{equation}
$\sfF(\rmS)$ is given by the expression 
\begin{equation}
\sfF(\rmS)(\rmG)=\int_{T[1]L}\varrho_L \left(\rmK,\Ad\rmG^{-1}(\rmS)\right).
\label{hafscan12}
\end{equation}
$\sfF(\rmS)$ is $\clG_{\msM,\rmK}(L)$--invariant and so
$\sfF(\rmS)\in\Fun(\overline{\clF}_{\msM,\rmK}(L))$, as is straightforward to verify.

The functionals $\sfF(\rmS)$ with $\rmS\in\fkg_\msM(L)$ constitute a first class set of
functionals. Indeed, they satisfy the Poisson bracket algebra
\begin{equation}
\{\sfF(\rmS),\sfF(\rmS')\}=\sfF([\rmS,\rmS'])
\label{hafscan13}
\end{equation}
with $\rmS,\rmS'\in\fkg_{\msM}(L)$. The Hamiltonians $\sfN(\rmW)$,
$\rmW\in\clP_\msM(L)$, of the special 
gauge symmetry are simply related to the $\sfF(\rmS)$,  
\begin{equation}
\sfN(\rmW)=\sfF(\rmS)|_{\rmS=\dd\rmW+[\varsigma^*\Omega,\rmW]}.
\label{hafscan14}
\end{equation}
This finding reveals how the special gauge symmetry arises from the 
wider phase space gauge symmetry. The former is the gauge symmetry left over after the breaking
of the latter due the TCO model's coupling to the background gauge field $\Omega$. 
However, while the special symmetry is dynamical,
the phase space symmetry is merely kinematical.



\subsection{\textcolor{blue}{\sffamily Characteristic derived TCO sigma model}}\label{subsec:hafsspx}

The characteristic model is a specialization of the derived TCO model
elaborated in the preceding subsections based on a specific choice of the level current.
Its distinguished features highlight the TCO model's connection to the derived KKS theory
worked out in sect. 5 of I and by this very reason provide the main motivation for studying
the model in the first place.


In the characteristic derived TCO model with base 2--fold $N$, 
the level current is taken to be the most general current $\rmK\in\Map'(T[1]N,\DD\fkm[0])$ 
obeying $\dd\rmK=0$ with components $k$ and $K$ proportional to the Heaviside and Dirac distributions
$\theta_N$ and $\delta_{\partial N}$ of $N$ and $\partial N$, respectively. 
Recall that $\theta_N\in\Map'(T[1]N,\bbR)$, 
$\delta_{\partial N}\in\Map'(T[1]N,\bbR[1])$ are defined by the relations 
$\int_{T[1]N}\varrho_N\theta_N\varphi=\int_{T[1]N}\varrho_N\varphi$, 
$\int_{T[1]N}\varrho_N\delta_{\partial N}\varphi=\int_{T[1]\partial N}\varrho_{\partial N}\varphi$
for $\varphi\in\Fun(T[1]N)$ and satisfy by Stokes' theorem the equation
$d\theta_N+\delta_{\partial N}=0$
\footnote{$\vphantom{\dot{\dot{\dot{a}}}}$ \label{foot:heavidir}
The Heaviside distribution $\theta_N$ and Dirac distribution $\delta_{\partial N}$
are singular precisely on the boundary $\partial N$ of $N$.
To make the above distributional identities to make sense, we imagine that $N$ is
extended to a larger 2--fold $N'$ by attaching an outer collar to each connected component of 
$\partial N$. It is $N'$ that rigorously corresponds to the base
2--fold of the model in the sense meant in subsect. \cref{subsec:hafsmod} and denoted by $N$ there.
For notational simplicity, we shall imagine the collars as `infinitesimally thin'   
and shall not distinguish between $N$ and $N'$}. We have 
\begin{align}
&k=\theta_N\dot\tau(\varLambda),
\label{hafsspx1}
\\
&K=\delta_{\partial N}\varLambda
\label{hafsspx1/1}
\end{align}
for some $\varLambda\in\fke$, as is immediately checked.

In the characteristic set--up, the derived TCO action $\sfS(\rmG;\Omega)$ is more usefully
expressed in terms of the components $g$, $G$ and $\omega$, $\varOmega$ of the TCO field
$\rmG$ and the background gauge field $\Omega$.
Using \ceqref{hafsspx1}, \ceqref{hafsspx1/1}, expression \ceqref{hafsmod1} furnishes 
\begin{multline}
\sfS(\rmG;\Omega)  
=\int_{T[1]N}\varrho_N
\left\langle\dot\tau(\varLambda),\mu\sdot\left(g^{-1},
\varsigma^*\varOmega+\sdot\mu\sdot(\varsigma^*\omega, G)
+dG+\tfrac{1}{2}[G, G]\right)\right\rangle
\\
-\int_{T[1]\partial N}\varrho_{\partial N}
\left\langle\Ad g^{-1}\mhfpt\left(\varsigma^*\omega+dg g^{-1}
+\dot\tau(G)\right),\varLambda\right\rangle.
\label{hafsspx2}
\end{multline}

The characteristic model's field equations can be obtained either from the derived form equation
\ceqref{hafsmod5} or by variation of the action \ceqref{hafsspx2}.
Componentwise, they take the form 
\begin{align}
&\left[\mu\sdot\left(g^{-1},\varsigma^*\varOmega+\sdot\mu\sdot(\varsigma^*\omega, G)+dG
+\tfrac{1}{2}[G, G]\right),\varLambda\right]=0\qquad \text{on $N$}, 
\vphantom{\Big]}
\label{hafsspx3}   
\\
&\dot\tau\left(\hfpt\sdot\mu\sdot\left(\Ad g^{-1}(\varsigma^*\omega+dg g^{-1}
+\dot\tau(G)),\varLambda\right)\right)=0\qquad \text{on $N$},
\vphantom{\Big]}
\label{hafsspx4}   
\\
&\sdot\mu\sdot\left(\Ad g^{-1}(\varsigma^*\omega+dg g^{-1}
+\dot\tau(G)),\varLambda\right)=0 \qquad \text{on $\partial N$}.
\vphantom{\Big]}
\label{hafsspx5}  
\end{align}
While the first method of derivation is more direct, the second one is more instructive. 
The variation $\delta\sfS$ of the action $\sfS$ displays
beside a bulk piece supported on the interior of the base 2--fold $N$ also an edge contribution
localized on the boundary $\partial N$ of $N$, that cannot be absorbed into the bulk one
by means of Stokes' theorem without generating terms containing derivatives
of the variations $\delta g$, $\delta G$ of the component
fields $g$, $G$ not allowed. This indicates that the action functional $\sfS$ is not differentiable in the sense
of refs. \ccite{Regge:1974zd,Benguria:1976in}. In a situation like this, the edge dynamics 
is added to the bulk one and the field equations split into 
bulk and edge. \ceqref{hafsspx3}, \ceqref{hafsspx4} are bulk equations yielded by
variation of $\sfS$ with respect to $g$, $G$, respectively. \ceqref{hafsspx5} is an edge 
equation obtained by varying $g$. 
There is no edge equation associated with variation with respect to $G$, because
the bulk $G$ variation generates a boundary term via Stokes' theorem that exactly cancels
that engendered by the edge $G$ variation.
Bulk and edge dynamics must be compatible: 
the edge equations must imply the restriction to the boundary of the bulk ones. 
Indeed, \ceqref{hafsspx5} clearly implies
\ceqref{hafsspx4} at the boundary. 

The characteristic model's integrability conditions can similarly be obtained either from the
derived form condition \ceqref{hafsmod6} or from the field equations \ceqref{hafsspx3}--\ceqref{hafsspx5}.
Componentwise, they read as 
\begin{align}
&\dot\tau\left(\hfpt\sdot\mu\sdot\left(\varsigma^*\phi,\mu\sdot(g,\varLambda)\right)\right)\approx 0
\qquad \text{on $N$},
\vphantom{\Big]}
\label{hafsspx6}
\\
&\sdot\mu\sdot\left(\varsigma^*\phi,\mu\sdot(g,\varLambda)\right)\approx 0
\qquad \text{on $\partial N$}.
\vphantom{\Big]}
\label{hafsspx8}  
\end{align}
A further condition yielded in this way, 
$\left[\varsigma^*\varPhi+\sdot\mu\sdot(\varsigma^*\phi, G),\mu\sdot(g,\varLambda)\right]\approx 0$ on $N$, 
is trivially satisfied by dimensional reasons. The conditions are manifestly realized if the fake
flatness requirement \ceqref{hafsmod7} is met.

We examine next the symmetries of the characteristic model starting with the level preserving 
gauge symmetry (cf. subsect. \cref{subsec:hafssym}). Since the level current is completely 
determined by the source Lie algebra datum $\varLambda$, we shall denote the level preserving
gauge transformation group $\clG_{\msM,\rmK}(N)$ as $\clG_{\msM,\varLambda}(N)$. 
From \ceqref{hafssym1}, the components $\upsilon$, $\varUpsilon$ of a 
gauge transformation $\Upsilon\in\clG_{\msM,\varLambda}(N)$ obey 
\begin{align}
&[\varUpsilon,\varLambda]=0\qquad \text{on $N$},
\vphantom{\Big]}
\label{hafsspx10}
\\
&\dot\tau(\mu\sdot(\upsilon,\varLambda)-\varLambda)=0\qquad \text{on $N$},
\vphantom{\Big]}
\label{hafsspx9}
\\
&\mu\sdot(\upsilon,\varLambda)-\varLambda=0\qquad \text{on $\partial N$}
\vphantom{\Big]}
\label{hafsspx9/1}
\end{align}
by the characteristic form \ceqref{hafsspx1}, \ceqref{hafsspx1/1} of the level current. 
\ceqref{hafsspx9/1} implies \ceqref{hafsspx9} at the base 
boundary making the bulk and edge properties compatible.
The gauge transform $\rmG^\Upsilon$ of the TCO field $\rmG$ reads in components as
\begin{align}
&g^{\upsilon,\varUpsilon}=g\upsilon,
\vphantom{\Big]}
\label{gbphafsspx1}
\\
&G^{\upsilon,\varUpsilon}=G+\mu\sdot(g,\varUpsilon).
\vphantom{\Big]}
\label{gbphafsspx2}
\end{align}

In the characteristic TCO model, the classical anomaly $\sfA(\Upsilon)$ of a 
gauge transformation $\Upsilon\in\clG_{\msM,\varLambda}(N)$ given in eq. \ceqref{hafssym5} takes the form 
\begin{equation}
\sfA(\upsilon,\varUpsilon)
=-\int_{T[1]\partial N}\varrho_{\partial N}\langle d\upsilon\upsilon^{-1},\varLambda\rangle.
\label{hafsspx11}
\end{equation}
The remarkable property of $\sfA$ is its being a pure boundary term. This makes the derived
TCO model akin to the 4--dimensional CS action \ccite{Zucchini:2021bnn}
\footnote{$\vphantom{\dot{\dot{\dot{a}}}}$ It is not difficult to understand how this relation
comes about. By comparing \ceqref{hafsmod1} and \ceqref{hafssym5}, it is
apparent that $\sfA(\upsilon,\varUpsilon)=\sfS(\upsilon,\varUpsilon;0,0)$.
On account of conditions \ceqref{hafsspx9}, \ceqref{hafsspx10}
and the invariance of the pairing, all the terms of the form  $\mu(\upsilon,\cdot\hfpt)$
and $[\varUpsilon,\cdot\hfpt]$ drop out in the expression of $\sfS(\upsilon,\varUpsilon;0,0)$
furnished by \ceqref{hafsspx2} leaving only the boundary contribution shown in \ceqref{hafsspx11}.
}. 

The special gauge symmetry has in the characteristic TCO model no particular 
features which distinguish it from the general model. We report here the
bare form component expression of the gauge transform $\rmG^A$
of the TCO field $\rmG$ by a special gauge transformation $A\in\clH_\msM(N)$, 
\begin{align}
&g^A=\tau(A)^{-1}g,
\vphantom{\Big]}
\label{gbphafsspx3}
\\
&G^A=\Ad A^{-1}(G+dAA^{-1}+\sdot\mu(\varsigma^*\omega,A))
\vphantom{\Big]}
\label{gbphafsspx4}
\end{align}
(cf. eq. \ceqref{hafssym14}). 

The gauge background gauge symmetry has too no
special features in the characteristic TCO model. 
The component
expression of the transform $\rmG^\rmU$ of the TCO field $\rmG$ under a background gauge transformation
$\rmU\in\clG_\msM(M)$ reads as  
\begin{align}
&g^{u,U}=\varsigma^*u^{-1}g,
\vphantom{\Big]}
\label{gbphafsspx5}
\\
&G^{u,U}=\mu\sdot(\varsigma^*u^{-1},G-\varsigma^*U)
\vphantom{\Big]}
\label{gbphafsspx6}
\end{align}
(cf. eq. \ceqref{hafssym9}). 

The natural question arises about whether the characteristic model is a sigma model
and, having in mind a possible relationship to derived KKS theory 
it is one over the derived coadjoint orbit $\clO_\varLambda$ of $\varLambda$ (cf. subsects. 5.1 and 5.9 of I). 
In subsect. \cref{subsec:hafssigmod}, we found that the general derived TCO model is a sigma model if
there exists a crossed submodule $\msM_\rmK$ of $\msM$ such that the level preserving gauge transformation
group $\clG_{\msM,\rmK}(N)$ turns out to equal the $\DD\msM_\rmK$--valued gauge transformation group
$\clG_{\msM_\rmK}(N)$. Inspection of eqs. \ceqref{hafsspx10}--\ceqref{hafsspx9/1} stating the
conditions which a level preserving gauge transformation $\Upsilon\in\clG_{\msM,\varLambda}(N)$ must fulfill
reveals that there are three groups relevant for settling the issue for the characteristic model:
the subgroup $\mu\ZZ\msE_\varLambda$
of $\msG$ of the elements $a\in\msG$ such that $\mu\sdot(a,\varLambda)=\varLambda$, the subgroup
$\mu\ZZ^{\,*}\!\msE_\varLambda$ of $\msG$ of the elements $a\in\msG$ such that
$\mu\sdot(a,\varLambda)=\varLambda$ mod $\ker\dot\tau$ and the subgroup $\ZZ\msE_\varLambda$ of $\msE$
of the elements $A\in\msE$ such that $\Ad A(\varLambda)=\varLambda$.
By \ceqref{hafsspx10}--\ceqref{hafsspx9/1}, a level preserving gauge transformation 
$\Upsilon\in\clG_{\msM,\varLambda}(N)$ is one such that its components
$\upsilon$, $\varUpsilon$ are valued in 
$\mu\ZZ^{\,*}\!\msE_\varLambda$, $\ZZ\msE_\varLambda$ in the base manifold interior $N$
and in $\mu\ZZ\msE_\varLambda$, $\ZZ\msE_\varLambda$ in the base manifold boundary $\partial N$. 
$\mu\ZZ^{\,*}\!\msE_\varLambda$ and $\ZZ\msE_\varLambda$  are the source and target groups of a
crossed submodule $\ZZ^{\,*}\!\msM_\varLambda$ of $\msM$. Similarly,
$\mu\ZZ\msE_\varLambda$ and $\ZZ\msE_\varLambda$  are the source and target groups of a
crossed submodule $\ZZ\msM_\varLambda$ of $\msM$, in fact the centralizer crossed submodule
of $\varLambda$ (cf. subsect. 5.1 of I). $\ZZ\msM_\varLambda$ is evidently a crossed submodule
of $\ZZ^{\,*}\!\msM_\varLambda$. 
Since $\clO_\varLambda=\DD\msM/\DD\ZZ\msM_\varLambda$, the answer to the question we posed
at the beginning of this paragraph is positive provided that $\ZZ^{\,*}\!\msM_\varLambda=\ZZ\msM_\varLambda$. 

In general, $\ZZ\msM_\varLambda$ is strictly smaller that $\ZZ^{\,*}\!\msM_\varLambda$.
A straightforward way of arranging that $\ZZ^{\,*}\!\msM_\varLambda=\ZZ\msM_\varLambda$ is by requiring
that $\ker\dot\tau=0$. A crossed module $\msM$ with this property is called quasi injective.
A restriction such as this one seems however to be exceedingly severe. The identity
$\ZZ^{\,*}\!\msM_\varLambda=\ZZ\msM_\varLambda$ can be ensured 
with less drastic limitations on $\msM$ as follows. 

%
We recall that the action $\mu\sdot$ of $\msG$ on $\fke$ leaves the subspace 
$\ker\dot\tau\subseteq\fke$ invariant. 
The crossed module $\msM$ is said to be inert on the target kernel if the
induced action $\mu\sdot$ of $\msG$ on $\ker\dot\tau$ is trivial. 
Note that every  quasi injective crossed module is inert on the target kernel, but the converse
is false in general. Therefore, target kernel inert crossed modules constitute
a broader set of crossed modules. 

Suppose that $\msM$ is a compact crossed module (cf. subsect. 5.1 of I)
inert on the target kernel.
The map $C:\mu\ZZ^{\,*}\!\msE_\varLambda\rightarrow\ker\dot\tau$ defined by
$C(a)=\mu\sdot(a,\varLambda)-\varLambda$ \linebreak 
for $a\in\mu\ZZ^{\,*}\!\msE_\varLambda$ satisfies the relation
$C(ab)=C(a)+C(b)$ for all $a,b\in\mu\ZZ^{\,*}\!\msE_\varLambda$ because of the triviality of the
$\msG$--action $\mu\sdot$ on $\ker\dot\tau$. $C$ is so a
1--cocycle of the group $\mu\ZZ^{\,*}\!\msE_\varLambda$ with coefficients in the trivial 
$\mu\ZZ^{\,*}\!\msE_\varLambda$--module $\ker\dot\tau$. From the compactness of $\mu\ZZ^{\,*}\!\msE_\varLambda$
ensuing from that of $\msG$, it follows that the 1--cocycle $C$ vanishes identically 
\footnote{$\vphantom{\dot{\dot{\dot{a}}}}$
Since $\ZZ^{\,*}\!\msE_\varLambda$ is a compact group, $C(\ZZ^{\,*}\!\msE_\varLambda)$ is a compact subset of $\ker\dot\tau$
by virtue of the continuity of $C$. The topology of $\ker\dot\tau$
can by described by means of a norm 
and every compact set of $\ker\dot\tau$ turns out to be bounded
with respect to that norm. In particular, $C(\ZZ^{\,*}\!\msE_\varLambda)$ is bounded. $C$ being a 1--cocycle entails that
for any $a\in\mu\ZZ^{\,*}\!\msE_\varLambda$ and any positive integer $n$
one has $C(a^n)=nC(a)$. This property is compatible with the boundedness 
of $C(\ZZ^{\,*}\!\msE_\varLambda)$ only if $C(a)=0$.}. 
So, $\mu\ZZ^{\,*}\!\msE_\varLambda=\mu\ZZ\msE_\varLambda$. Hence,  
$\ZZ^{\,*}\!\msM_\varLambda=\ZZ\msM_\varLambda$ as desired.

In subsect. 5.1 of I, we examined a number of Lie group crossed module mo\-dels 
to illustrate basic notions of derived KKS theory.  We can use the same models
to check whether the range of crossed modules which are inert on the target kernel
is large enough. 
The crossed module $\INN_\msG\msN=(\msN,\msG,\iota,\kappa)$
associated with a pair of a Lie group $\msG$ and a normal subgroup $\msN$ of $\msG$,
where $\iota$ is the inclusion map of $\msN$ into $\msG$ and $\kappa$ is the left
conjugation action of $\msG$ on $\msN$, is the first such model we consider. $\INN_\msG\msN$ is quasi injective and so
trivially inert to the target kernel. 
The second model we scrutinize consists in the crossed module 
$\msC(\!\xymatrix@C=1.3pc{\msQ\ar[r]^-{\pi}&\msG}\!)=(\msQ,\msG,\pi,\alpha)$
stemming from a central extension
$\!\mhfpt\xymatrix@C=1.3pc{1\ar[r]&\msC\ar[r]^-{\iota}&\msQ\ar[r]^-{\pi}&\msG\ar[r]&1}\!\mhfpt$
of Lie groups, where the action $\alpha$
is given by $\alpha(a,A)=\sigma(a)A\sigma(a)^{-1}$ for $a\in\msG$, $A\in\msQ$ with 
$\sigma:\msG\rightarrow\msQ$ a section of the projection $\pi$. 
$\msC(\!\xymatrix@C=1.3pc{\msQ\ar[r]^-{\pi}&\msG}\!)$ is not quasi injective in general,
but it is always inert on the target kernel since $\ker\pi$ is central in $\msQ$.
As final model, we examine the crossed module $\msD(\rho)=(\msV,\msG,1_\msG,\rho)$ 
of a Lie group $\msG$, a vector space $\msV$ regarded as an Abelian group, the trivial morphism $1_\msG$ of
$\msV$ into $\msG$ and a representation $\rho$ of $\msG$ in $\msV$. This crossed module is inert on the target
kernel only if the representation $\rho$ is trivial. 

As the crossed module entering as basic symmetry datum of a derived TCO model 
is always equipped with an invariant pairing, it is balanced (cf. subsect. 3.1 of I).
This, however, involves only an illusory loss of generality. 
In fact, every crossed module $\msM$ can always be trivially extended to a balanced 
crossed module $\msM^c$ (see fn. 7 of I for an explicit description of this latter).
$\msM^c$ can then be employed instead of $\msM$ without this making any difference as
far as the symmetry properties of the resulting TCO model are concerned. 
It is not difficult to show that, if $\msM$ is trivial on the target kernel, then $\msM^c$ also
is. The target kernel trivial crossed modules considered in the previous paragraph are generally non balanced.
This is however no problem, for the reasons just explained. 

In this way, the analysis of subsect. \cref{subsec:hafssigmod} and the above considerations
lead us to the conclusion that 
when the symmetry crossed module $\msM$ is compact and target kernel inert, as we assume henceforth, 
the derived characteristic TCO model field space is  
$\overline{\clF}_{\msM,\varLambda}(N)=\Map(T[1]N,\DD\msM/\DD\ZZ\msM_\varLambda)$
and that so the model is truly a sigma model over the derived coadjoint orbit 
$\clO_\varLambda=\DD\msM/\DD\ZZ\msM_\varLambda$.
In this respect, the close relationship  of the characteristic to the ordinary TCO model is quite evident. 

The quantization condition \ceqref{hafspath1} of the anomaly $\sfA(\Upsilon)$ required by the
consistency of the quantum theory (cf. subsects. \cref{subsec:hafssigmod}, \cref{subsec:hafspath}) reads as 
\begin{equation}
\int_{T[1]\partial N}\varrho_{\partial N}\langle d\upsilon\upsilon^{-1},\varLambda\rangle\in 2\pi\bbZ.
\label{hafsspx12}
\end{equation}
The following conditions on $\varLambda$ suffice to guarantee that this requirement is met.
First, the centralizer crossed module $\ZZ\msM_\varLambda$ of $\msM_\varLambda$
must be a maximal toral crossed submodule $\msJ=(\msH,\msT)$ of $\msM$
and, second, the map $\xi_\varLambda:\msT\rightarrow \msU(1)$
defined by $\xi_\varLambda(\ee^x)=\ee^{i\langle x,\varLambda\rangle}$ with $x\in\fkt$ must be a 
character of $\msT$. Such prerequisites are the same as those imposed in subsect. 5.9 of I. 
The former entails that $\varLambda$ is a regular element of $\fke$ and that the derived coadjoint orbit
of $\varLambda$ is thus $\clO_\varLambda=\DD\msM/\DD\msJ$.
The latter implies that the restriction of the mapping $x\rightarrow \langle x,\varLambda\rangle/2\pi$ to the integer
lattice $\Lambda_\msG$ of $\msT$ lies in the dual integral lattice $\Lambda_\msG{}^*$ of $\Lambda_\msG$.
Again, by another independent route, a direct contact with derived KKS theory is established. 

The canonical formulation of subsect. \cref{subsec:hafscan} can be readily adapted to the characteristic
TCO model. The canonical set--up has the virtue of revealing other important features
not immediately evident in the Lagrangian approach. 

In canonical theory, where $N=\widetilde L=\bbR^1\times L$ with either
$L=\bbS^1$ or $L=\bbI^1$, the derived TCO action \ceqref{nhafscan4} reads as 
\begin{align}
\sfS(\rmG;\Omega)    
&=\int_{T[1]\widetilde L}\varrho_{\widetilde L}
\big\langle\dot\tau(\varLambda),\mu\sdot(g^{-1},\varsigma^*{}_t\varOmega+d_tG
\label{nhafsspx1}
\\
&\hspace{3cm}
+\sdot\mu\sdot(\varsigma^*{}_t\omega,G)+\sdot\mu\sdot(\varsigma^*\omega+dgg^{-1}+\dot\tau(G),G_t))\big\rangle
\nonumber
\\
&-\int_{T[1]\partial\widetilde L}\varrho_{\partial\widetilde L}
\big\langle\Ad g^{-1}(\varsigma^*{}_t\omega+d_tgg^{-1}),\varLambda\big\rangle
\nonumber
\end{align}
after some straightforward rearrangements. The notation used in \ceqref{nhafsspx1}
is defined at the beginning of subsect.  \cref{subsec:hafscan}. 
\ceqref{nhafsspx1} shows clearly that $G_t$ is a non dynamical Lagrange multiplier field
as expected from the general analysis carried out in subsect. \cref{subsec:hafscan}. 
The constraint it enforces upon variation, corresponding to \ceqref{nhafscan5}, takes the form 
\begin{equation}
\dot\tau(\sdot\mu\sdot(\Ad g^{-1}(\varsigma^*\omega+dgg^{-1}+\dot\tau(G)),\varLambda))\approx 0.
\vphantom{\ul{\ul{\ul{g}}}}  
\label{nhafsspx2}
\end{equation}
$G_t$ is done away with by imposing the gauge fixing condition 
\begin{equation}
G_t=0
\label{nhafsspx3}
\end{equation}
equivalent to \ceqref{nhafscan6}. 

The ambient phase space of the characteristic TCO model consists of all TCO fields $\rmG$ on $L$
and so is just $\clF_\msM(L):=\Map(T[1]L,\DD\msM)$.   
The model's phase space $\overline{\clF}_{\msM,\varLambda}(L)$ is obtained
by modding out the level preserving gauge transformation action
\ceqref{gbphafsspx1}, \ceqref{gbphafsspx2} in accordance with the definition \ceqref{hafscan8}.
If $\msM$ is compact and inert on the target kernel, then 
$\overline{\clF}_{\msM,\varLambda}(L)=\Map(T[1]L,\DD\msM/\DD\ZZ\msM_\varLambda)$, 
as suited for the sigma model over
the derived coadjoint orbit $\clO_\varLambda=\DD\msM/\DD\ZZ\msM_\varLambda$
encountered in the Lagrangian analysis. 

When $N=\widetilde L$, the constituents $\rmK_t$, $\rmK$ of the
level current $\widetilde\rmK$ given in  eq. \ceqref{hafsspx1} 
obey the requirement \ceqref{khafscan1} and the relations \ceqref{khafscan2}.
We further have 
\begin{align}
&k=\theta_L\dot\tau(\varLambda),
\label{hafsspx13}
\\
&K=\delta_{\partial L}\varLambda,
\label{hafsspx13/1}
\end{align} 
where $\theta_L$ and $\delta_{\partial L}$ are the Heaviside and Dirac distributions of $L$ and
$\partial L$ satisfying $d\theta_L+\delta_{\partial L}=0$. 

For the characteristic TCO model, the presymplectic 2--form $\sfPsi$ of eq. \ceqref{hafscan2}
exhibits by virtue of the special form of the current constituent $\rmK$ of eqs. \ceqref{hafsspx13},
\ceqref{hafsspx13/1} a bulk and an edge contribution
\footnote{$\vphantom{\dot{\dot{\dot{a}}}}$ Since $\partial L$, if non empty,
consists of just two points, here and in similar relations below  
the notation $\int_{T[1]\partial L}\varrho_{\partial L}\ldots$ indicates in a somewhat overwritten
manner summation over those points with signs determined by the orientation of $L$.}
\begin{multline}
\sfPsi_\varLambda=\int_{T[1]L}\varrho_L\left\langle\dot\tau(\varLambda),
\sdot\mu\sdot\left(g^{-1}\delta g, \mu\sdot(g^{-1},\delta G)\right)\right\rangle
\\
+\frac{1}{2}\int_{T[1]\partial L}\varrho_{\partial L}
\left\langle[ g^{-1}\delta g,g^{-1}\delta g],\varLambda\right\rangle. 
\label{hafsspx15}
\end{multline}
The presymplectic 2--form $\sfPsi_\varLambda$ induces a symplectic 2--form and
so a Poisson bra\-cket structure on the phase space
$\overline{\clF}_{\msM,\varLambda}(L)$ yielded by modding out the level preserving
gauge symmetry as detailed in subsect. \cref{subsec:hafscan}. 

The symmetries of the characteristic model are described in canonical theory in a way that exactly
parallels that in which they are in the Lagrangian set--up
through the appropriate analogs of
eqs. \ceqref{hafsspx9}--\ceqref{gbphafsspx2}, \ceqref{gbphafsspx3}, \ceqref{gbphafsspx4},
\ceqref{gbphafsspx5}, \ceqref{gbphafsspx6} with the field space $\clF_\msM(N)$
replaced by the ambient phase space $\clF_\msM(L)$ and the gauge transformations groups
$\clG_\msM(N)$, $\clG_{\msM,\varLambda}(N)$, $\clH_\msM(N)$ by their phase space counterparts
$\clG_\msM(L)$, $\clG_{\msM,\varLambda}(L)$, $\clH_\msM(L)$. 

The first class Hamiltonians $\sfN(\varPi)$ of the infinitesimal gauge background preserving
gauge transformations $\varPi\in\fkh_\msM(L)$, defined in \ceqref{nhafscan7} and \ceqref{nhafscan11},
take in the characteristic model the form \vspace{-1.25mm}
\begin{multline}
\sfN(\varPi)(\rmG;\Omega)   
\\
=\int_{T[1]L}\varrho_L
\left\langle\dot\tau(\sdot\mu\sdot(\Ad g^{-1}(\varsigma^*\omega+dgg^{-1}+\dot\tau(G)),\varLambda)),
\mu\sdot(g^{-1},\varPi)\right\rangle.
\label{hafsspx17}   
\end{multline}
The weak vanishing of the $\sfN(\varPi)$, in accordance with \ceqref{nhafscan19}, singles out
the physical phase space $\overline{\clF}_{\msM,\varLambda\,{\rm phys}}(L)$ of the characteristic model
within the phase space $\overline{\clF}_{\msM,\varLambda}(L)$. 

The symplectic 2--form $\sfPsi_\varLambda$
of the phase space $\overline{\clF}_{\msM,\varLambda}(L)$ exhibits a bulk and an edge term
which have a formal structure analogous to that of the two components of the derived symplectic
structure $-i\rmB_\varLambda$ of the regular derived coadjoint orbit $\clO_\varLambda$ (cf. subsect. 5.9 
and eqs. (5.9.4), (5.9.5) of I).
This is not accidental. The relationship between the two symplectic
structures will be elucidated more formally in an appropriate transgressional framework  
in subsect. \cref{subsec:afskks} below. 

In the above canonical analysis, in particular in the last paragraph,
we did not mention the integrality condition that the source Lie algebra
datum $\varLambda$ must satisfy in the quantum theory of the characteristic model as discussed above
and that was also assumed in our treatment of a regular derived coadjoint orbit
in subsect. 5.9 of I.  
In the model's canonical theory, the quantization of $\varLambda$
is expected to stem from requiring that the symplectic 2--form $\sfPsi_\varLambda$ of the phase space 
$\overline{\clF}_{\msM,\varLambda}(L)$ equals $-i$ times the curvature of a prequantum line bundle
$\scL_\varLambda$ on $\overline{\clF}_{\msM,\varLambda}(L)$ whose properly normalized sections form
the model's prequantum Hilbert space in the spirit of geometric
quantization. Since $\overline{\clF}_{\msM,\varLambda}(L)$ is an infinite dimensional graded manifold,
this would inevitably lead us to an infinite dimensional graded geometric setting which, as we have remarked in I,
is very difficult to deal with and for this reason we want to avoid for the time being.

The problems with the geometric quantization of the characteristic derived TCO model
pointed out in the previous paragraph pair with the difficulties to work out a
full prequantization of derived KKS theory noticed in subsect. 5.8 of I.
The point is that the only viable quantization scheme of the TCO model
and the KKS theory that secretly informs it presently at our disposal,
notwithstanding its shortcomings, is the functional integral
one of subsect. \cref{subsec:hafspath}. 
Though in an indirect way and with certain underlying assumptions, this should provide the quantization of
both the characteristic TCO canonical set--up and derived KKS theory.



\subsection{\textcolor{blue}{\sffamily Canonical formulation of the TCO model and derived KKS theory}}\label{subsec:afskks}

In this subsection, we unveil the relationship between the regular case
derived KKS theory worked out in subsect. 5.9 of I 
and the canonical formulation of the characteristic derived TCO model studied in subsect. \cref{subsec:hafsspx}.  

Transgression is a procedure used in geometry and topology
for transferring cohomology classes from one space to another in the absence of a morphism
relating them. Below we shall work out a version of transgression as a chain map turning
forms of $\DD\msM$ into forms on the TCO model phase space $\clF_\msM(L)$, where the former are viewed
as functions on $T[1]\DD\msM$ and the latter as functionals on $T[1]\clF_\msM(L)$ as usual. 

In simple terms,
the transgression of a function $F\in\iFun(T[1]\DD\msM)$ involves two steps: 
$i$ generating a functional $\ev^*\!F\in\Fun(T[1]\clF_\msM(L)\,\boxplus\,T[1]L)$
by means of the pull--back of the evaluation map 
$\ev:T[1]\clF_\msM(L)\times T[1]L\rightarrow T[1]\DD\msM$ 
and $ii$ constructing an element $\sfT(F)\in\Fun(T[1]\clF_\msM(L))$ by integration
of $\ev^*\!F$ on a cycle of $L$. 
We shall examine such steps individually. 


The analysis of transgression requires setting up a suitable Cartan calculus for the space $\clF_\msM(L)\times L$
(cf. app. A.3 of I). We shall do that by separately constructing appropriate
Cartan calculi for the two Cartesian factors $\clF_\msM(L)$,
$L$ roughly proceeding along the lines followed for the space $\DD\msM$ is subsect. 5.2 of I.

With the above in mind, we have to appraise which range of
vector fields of $\clF_\msM(L)$, $L$ are the most appropriate for
contractions and Lie derivatives. 
The full vector field Lie algebras $\Vect(\clF_\msM(L))$, $\Vect(L)$ 
are too large. Working with more restricted 
algebras is enough. To this end, we reconsider the Cartan
calculus of $\DD\msM$ from a more general viewpoint. 
Let $V\in\iVect(\DD\msM)$ be an internal  vector field.
For the base coordinate $\Gamma$ of $T[1]\DD\msM$ (cf. subsect. 5.2 of I),
we have 
\begin{align}
&\Gamma^{-1}j_V\Gamma=0,
\vphantom{\Big]}
\label{nafskks3}
\\
&\Gamma^{-1}l_V\Gamma=\rmN_V(\Gamma)
\vphantom{\Big]}
\label{nafskks4}
\end{align}
on general grounds, where $j_V$, $l_V$ denote contraction and Lie derivation with respect to $V$
and $\rmN_V\in\iMap(\DD\msM,\DD\fkm[0])$ is an internal map depending on $V$.
$\rmN_V(\Gamma)$ is nothing but the coordinate expression of $V$.
For instance, if $S_{\msL\rmH}$ is the vertical vector field of left target kernel $\DD\msM_\tau$--action
of the derived Lie algebra element $\rmH\in\DD\fkm_\tau$ (cf. subsect. 5.3 of I), 
then $\rmN_{S_{\msL\rmH}}(\Gamma)=-\Ad\Gamma(\rmH)$. From (5.2.1) of I, it then follows that
the fiber coordinate $\Sigma$ of $T[1]\DD\msM$ obeys
\begin{align}
&j_V\Sigma=\rmN_V(\Gamma),
\vphantom{\Big]}
\label{nafskks5}
\\
&l_V\Sigma=\dd\rmN_V(\Gamma)+[\Sigma,\rmN_V(\Gamma)]. 
\vphantom{\Big]}
\label{nafskks6}
\end{align}
In fact, it is possible to show that, under the assumption of the validity of relations
(5.2.1) of I and \ceqref{nafskks3}, \ceqref{nafskks4} for any $V\in\iVect(\DD\msM)$,
the basic Cartan calculus relations (A.3.1)--(A.3.4) of I hold if and only if (5.2.2) of I holds,
\ceqref{nafskks5}, \ceqref{nafskks6} hold true for any $V\in\iVect(\DD\msM)$ and 
moreover the relation 
\begin{equation}
\rmN_{[V,W]}(\Gamma)=l_V\rmN_W(\Gamma)-l_W\rmN_V(\Gamma)+[\rmN_V(\Gamma),\rmN_W(\Gamma)]
\label{nafskks7}
\end{equation}
is satisfied for $V,W\in\iVect(\DD\msM)$. 

We let $\iVect_\msr(\DD\msM)$ be the set of all vector fields $V\in\iVect(\DD\msM)$ such that
$\rmN_V$ is an ordinary non internal map 
so that $\rmN_V\in\Map(\DD\msM,\DD\fkm[0])$.
From \ceqref{nafskks7}, $\iVect_\msr(\DD\msM)$ is a Lie subalgebra of $\iVect(\DD\msM)$, the restricted
subalgebra. It is shown below that 
it is possible associate with every vector field $V\in\iVect_\msr(\DD\msM)$
vector fields $\varUpsilon(V)\in\Vect(\clF_\msM(L))$, $Y(V)\in\Vect(L)$, 
called the phase space and space manifold transplants of $V$, with natural properties.
The vector fields yielded in this fashion will be the ones considered
in the Cartan calculi of $\clF_\msM(L)$, $L$ in what follows. 



We now illustrate the Cartan calculus of the phase space $\clF_\msM(L)$ and its main
properties. The action of the calculus' variational derivations on the functional algebra
$\Fun(T[1]\clF_\msM(L))$ of $T[1]\clF_\msM(L)$
is specified by that on suitable 
field coordinates of $T[1]\clF_\msM(L)$. As $\clF_\msM(L)=\Map(T[1]L,\DD\msM)$, 
$\clF_\msM(L)$ is a group manifold. By the isomorphism 
$T[1]\clF_\msM(L)\simeq\Map(T[1]L,\DD\msM)\times\Map(T[1]L,\DD\fkm)[1]$,
it is natural to use coordinates adapted to the Cartesian factors
$\Map(T[1]L,\DD\msM)$, $\Map(T[1]L,\DD\fkm)[1]$ which we may think of as base and fiber coordinates of
$T[1]\clF_\msM(L)$, hence field variables $\rmG\in\Map(T[1]L,\DD\msM)$, $\rmS\in\Map(T[1]L,\DD\fkm)[1]$.



There is a vector field $\varUpsilon(V)\in\Vect(\clF_\msM(L))$ associated
with each vector field $V\in\iVect_\msr(\DD\msM)$, 
whose coordinate expression $\rmN_{\varUpsilon(V)}\in\Map(\clF_\msM(L),\DD\fkm[0])$ is 
\begin{equation}
\rmN_{\varUpsilon(V)}(\rmG)=\rmN_V\circ\rmG.
\label{nafskks9}
\end{equation}
$\varUpsilon(V)$ is the phase space transplant of $V$. 
On account of property \ceqref{nafskks7}, the map
$\varUpsilon:\iVect_\msr(\DD\msM)\rightarrow\Vect(\clF_\msM(L))$
is a Lie algebra morphism.


The way the variational derivations of the Cartan calculus of $\clF_\msM(L)$ act
on the field coordinates is determined by the interpretation of these and consistency.
The variational differential $\delta$ acts on the field coordinate $\rmG$ according to \pagebreak 
\begin{equation}
\rmG^{-1}\delta\rmG=\rmS,
\label{nafskks1}
\end{equation}
identifying $\rmS$ as the variational Maurer--Cartan field form associated with $\rmG$. 
The action of $\delta$ on the field coordinate $\rmS$ in turn 
is mandated by the requirement of nilpotence of $\delta$ and reduces to the variational 
Maurer--Cartan equation,
\begin{equation}
\delta\rmS=-\frac{1}{2}[\rmS,\rmS].
\label{nafskks2}
\end{equation}
The variational contraction and Lie derivative along the vector fields $\varUpsilon(V)$ with 
$V\in\iVect_\msr(\DD\msM)$, $\iota_{\varUpsilon(V)}$ and $\lambda_{\varUpsilon(V)}$,  act on $\rmG$ according to 
\begin{align}
&\rmG^{-1}\iota_{\varUpsilon(V)}\rmG=0,
\vphantom{\Big]}
\label{nafskks10}
\\
&\rmG^{-1}\lambda_{\varUpsilon(V)}\rmG=\rmN_{\varUpsilon(V)}(\rmG)
\vphantom{\Big]}
\label{nafskks11}
\end{align}
on general grounds in analogy to \ceqref{nafskks3}, \ceqref{nafskks4}.
From relation \ceqref{nafskks1}, it then follows that they act on $\rmS$ as 
\begin{align}
&\iota_{\varUpsilon(V)}\rmS=\rmN_{\varUpsilon(V)}(\rmG),
\vphantom{\Big]}
\label{nafskks12}
\\
&\lambda_{\varUpsilon(V)}\rmS=\delta\rmN_{\varUpsilon(V)}(\rmG)+[\rmS,\rmN_{\varUpsilon(V)}(\rmG)]
\vphantom{\Big]}
\label{nafskks13}
\end{align}
in analogy to \ceqref{nafskks5}, \ceqref{nafskks6}. The variational derivations $\delta$, $\iota$
and $\lambda$ of $\clF_\msM(L)$ obey the basic Cartan relations
analogously to the derivations $d$, $j$ and $l$ of $\DD\msM$. 




We outline next the Cartan calculus of the space manifold $L$ illustrating its main properties. 
The action of the calculus derivations on the function algebra $\Fun(T[1]L)$ is specified 
by that on suitable coordinates on $T[1]L$. As $L=\bbS^1$ or $\bbI^1$, $T[1]L\simeq L\times\bbR[1]$. So, we can choose
the variables $r\in L$, $z\in\bbR[1]$ as natural coordinates.

It is reasonable  to assume that no action on $L$ results from any action on $\DD\msM$
via transgression. Therefore, for a vector field $V\in\iVect_\msr(\DD\msM)$ we take its
space manifold transplant $Y(V)\in\Vect(L)$ to vanish identically, $Y(V)=0$.
The map $Y:\iVect_\msr(\DD\msM)\rightarrow\Vect(L)$ is then trivially a Lie algebra morphism. 

The expressions of the derivations of the Cartan calculus of $L$ are elementary. 
The action of the de Rham differential $d$ of $L$ on $r$, $z$ reads as usual as 
\begin{equation}
dr=z,\qquad dz=0.
\label{nafskks14}
\end{equation}
The vanishing of the transplant $Y(V)$ of a vector field $V\in\iVect_\msr(\DD\msM)$ implies that 
the contraction and Lie derivative along $Y(V)$, $j_{Y(V)}$ and $l_{Y(V)}$, act trivially, 
\begin{equation}
j_{Y(V)}r=0, \qquad l_{Y(V)}r=0, \qquad j_{Y(V)}z=0, \qquad l_{Y(V)}z=0. 
\label{nafskks15}
\end{equation}

The formulation of the Cartan calculi of the phase space $\clF_\msM(L)$ and space manifold $L$ 
furnishes us readily that of the product space $\clF_\msM(L)\times L$. 
In fact, the shifted tangent bundle $T[1](\clF_\msM(L)\times L)$ of $\clF_\msM(L)\times L$
is isomorphic the external direct sum $T[1]\clF_\msM(L)\,\boxplus\,T[1]L$ (cf. fn. \cref{foot:boxplus})
which turns out to be just the product $T[1]\clF_\msM(L)\times T[1]L$ as a manifold.
The transplant of a vector field $V\in\iVect_\msr(\DD\msM)$
in $\clF_\msM(L)\times L$ is the vector field $\varUpsilon(V)+Y(V)$ of $\clF_\msM(L)\times L$
\linebreak with the transplants $\varUpsilon(V)$, $Y(V)$ of $V$ 
in $\clF_\msM(L)$, $L$ as components. The differential and the contraction and Lie derivative along 
$\varUpsilon(V)+Y(V)$ of $\clF_\msM(L)\times L$ \linebreak are given by $\delta+d$ and $\iota_{\varUpsilon(V)}+j_{Y(V)}$, 
$\lambda_{\varUpsilon(V)}+l_{Y(V)}$ in terms of the differentials 
 \linebreak $\delta$, $d$ and the contractions and Lie derivatives $\iota_{\varUpsilon(V)}$, $j_{Y(V)}$, 
$\lambda_{\varUpsilon(V)}$, $l_{Y(V)}$ along  \linebreak $\varUpsilon(V)$, $Y(V)$ of $\clF_\msM(L)$, $L$, respectively.
With this framework available, we can now proceed to
working out the relevant transgression map following the two steps
outlined at the beginning of the present subsection.



The first step of the construction of the transgression map 
consists in writing down the expressions of the evaluation map
$\ev:T[1]\clF_\msM(L)\times T[1]L\rightarrow T[1]\DD\msM$ of $T[1]\clF_\msM(L)$ and its pull-back. 
The map $\ev$ is given explicitly by 
\begin{equation}
\ev(\rmG,\rmS,r,z)=(\rmG(r,z),\rmS(r,z)+\rmG^{-1}\dd\rmG(r,z)).
\label{nafskks16}
\end{equation}
The reason why the term  $\rmG^{-1}\dd\rmG$ is added to $\rmS$ in the right hand side will be explained 
momentarily.

The pull--back $\ev^*$ of $\ev$ generates functionals of $T[1]\clF_\msM(L)\,\boxplus\,T[1]L$
from functions of $T[1]\DD\msM$. Since our goal is obtaining ordinary non internal functionals
of the former space as a result, we have to restrict the range of functions of the latter space
on which we act with $\ev^*$. To this end, it is enough to limit ourselves to the restricted subalgebra
$\iFun_\msr(T[1]\DD\msM)$ formed 
by the functions $F\in\iFun(T[1]\DD\msM)$ of the form
$F=E_F(\Gamma,\Sigma)$, where $E_F\in\Fun(\DD\msM\times\DD\fkm[1])$ is an ordinary function.
The evaluation map pull--back map obtained in this way is a degree $0$ algebra morphism
$\ev^*:\iFun_\msr(T[1]\DD\msM)\rightarrow\Fun(T[1]\clF_\msM(L)\,\boxplus\,T[1]L)$. Explicitly, 
by \ceqref{nafskks16}, for any $F\in\iFun_\msr(T[1]\DD\msM)$ we have 
\begin{equation}
\ev^*\mhfpt F(\rmG,\rmS,r,z)
=E_F(\rmG(r,z),\rmS(r,z)+\rmG^{-1}\dd\rmG(r,z)).
\label{nafskks17}
\end{equation}
Using relations \ceqref{nafskks9}--\ceqref{nafskks13} and \ceqref{nafskks14}, \ceqref{nafskks15},
it is straightforward to verify that the pull--back map $\ev^*$ has the expected properties.  
In particular, 
\begin{equation}
\ev^*\mhfpt dF=(\delta+d)\ev^*\mhfpt F.
\label{nafskks18}
\end{equation}
The validity of this important relation rests crucially on the addition of the key term $\rmG^{-1}\dd\rmG$
to $\rmS$ in the expression of the evaluation map, eq. \ceqref{nafskks16}. 
For any vector field $V\in\iVect_\msr(\DD\msM)$, we have analogously 
\begin{align}
&\ev^*\mhfpt j_VF=\iota_{\varUpsilon(V)}\ev^*\mhfpt F,
\vphantom{\Big]}
\label{nafskks19}
\\
&\ev^*\mhfpt l_VF=\lambda_{\varUpsilon(V)}\ev^*\mhfpt F.
\vphantom{\Big]}
\label{nafskks20}
\end{align}

The second step of the construction of the transgression map consists in the integration on a 
cycle of $L$ of the functionals of $\Fun(T[1]\clF_\msM(L)\boxplus T[1]L)$
yielded \linebreak by operating with the pull--back map $\ev^*$ on the functions of
$\iFun_\msr(T[1]\DD\msM)$. This yields functionals
of $\Fun(T[1]\clF_\msM(L))$ as a result. 
The cycle can be conveniently encoded by a 
current $C\in\Fun'(T[1]L)$ obeying $dC=0$.
The transgression of a function $F\in\iFun_\msr(T[1]\DD\msM)$ 
resulting from combining the two steps described above has therefore the form
\begin{equation}
\sfT(F)=\int_{T[1]L}\varrho_L\,C\ev^*\!F. 
\label{nafskks21}
\end{equation}

We assume below for simplicity that the current $C$ is homogeneous. Since $L$ is 1--dimensional,
its degree $p$ can take only the values $0$, $1$. Transgression then 
furnishes a degree $p-1$ linear map $\sfT:\iFun_\msr(T[1]\DD\msM)\rightarrow\Fun(T[1]\clF_\msM(L))$. 
By \ceqref{nafskks18} and the fact that $dC=0$, $\sfT$ is a chain map,
\begin{equation}
\sfT(dF)=-(-1)^p\delta\sfT(F).
\label{nafskks22}
\end{equation}
For a vector field $V\in\iVect_\msr(\DD\msM)$, we have similarly \pagebreak 
\begin{align}
&\sfT(j_VF)=-(-1)^p\iota_{\varUpsilon(V)}\sfT(F),
\vphantom{\Big]}
\label{nafskks23}
\\
&\sfT(l_VF)=\lambda_{\varUpsilon(V)}\sfT(F).
\vphantom{\Big]}
\label{nafskks24}
\end{align}
by \ceqref{nafskks19}, \ceqref{nafskks20} and the fact that $j_{Y(V)}C=0$ and 
$l_{Y(V)}C=0$ by virtue of \ceqref{nafskks15}
\footnote{$\vphantom{\dot{\dot{\dot{a}}}}$  
The extra $-1$ factor appearing in \ceqref{nafskks22}, \ceqref{nafskks23} is due to 
the Berezin measure $\varrho_L$ having degree $-1$.}.

The transgression procedure illustrated above admits a number of  generalizations.
We consider in what follows the extension of transgression to $\INN\bbR$--valued
maps, where $\INN\bbR=(\bbR,\bbR,\id_\bbR,o_\bbR)$ is the inner derivation crossed module of the Abelian Lie algebra
$\bbR$ (cf. subsect. 3.1 of I), where $o_\bbR$ is the trivial action. 

A map $\rmF\in\iMap(T[1]\DD\msM,\SD\INN\bbR)$
(cf. fn. 4 of I for notation) is characterized by its components
$f,F\in\iMap(T[1]\DD\msM,\rmS\hfpt\bbR)\simeq\iFun(T[1]\DD\msM)$. We let $\iMap_\msr(T[1]\DD\msM,\SD\INN\bbR)$
be the vector subspace of $\iMap(T[1]\DD\msM,\SD\INN\bbR)$ constituted by all the maps 
$\rmF\in \iMap(T[1]\DD\msM,\SD\INN\bbR)$ with components %
$f,F\in\iFun_\msr(T[1]\DD\msM)$.
The evaluation map pull--back is the linear mapping
$\ev^*:\iMap_\msr(T[1]\DD\msM,\SD\INN\bbR)$ $\rightarrow\Map(T[1]\clF_\msM(L)\,\boxplus\,T[1]L,\SD\INN\bbR)$
induced by 
the previously defined pull--back $\ev^*$ map acting componentwise. 

Let $\rmC\in\Map'(T[1]L,\INN\bbR)$ be a derived current satisfying $\dd\rmC=0$,
where here $\dd$ denotes the derived differential. $\rmC$
may be regarded as a derived cycle of $L$. 
The transgression of a map $\rmF\in\iMap_\msr(T[1]\DD\msM,\SD\INN\bbR)$ reads as 
\begin{equation}
\sfT(\rmF)=\int_{T[1]L}\varrho_L\,(\rmC,\ev^*\!\rmF)_\bbR.
\label{nafskks25}
\end{equation}
In this expression, $(\cdot,\cdot)_\bbR$ is the pairing of $\iMap(T[1]L,\SD\INN\bbR)$
associated according to (3.3.15) of I
with the canonical invariant pairing $\langle\cdot,\cdot\rangle_\bbR$ of $\INN\bbR$ given 
by $\langle x,X\rangle_\bbR=xX$ with $x\in\bbR$, $X\in\bbR$. In this way,
a degree $0$ linear transgression map $\sfT:\iMap_\msr(T[1]\DD\msM,\SD\INN\bbR)\rightarrow\Fun(T[1]\clF_\msM(L))$ 
is defined. 
$\sfT$ obeys relations analogous to \ceqref{nafskks22}--\ceqref{nafskks24}. 
Indeed, using \ceqref{nafskks25} and proceeding as in the proof of those relations, one can verify that  
\begin{equation}
\sfT(\dd\rmF)=-\delta\sfT(\rmF) 
\label{nafskks26}
\end{equation}
\vfil\eject\noindent
and that for $V\in\iVect_\msr(\DD\msM)$ 
\begin{align}
&\sfT(j_V\rmF)=-\iota_{\varUpsilon(V)}\sfT(\rmF),
\vphantom{\Big]}
\label{nafskks27}
\\
&\sfT(l_V\rmF)=\lambda_{\varUpsilon(V)}\sfT(\rmF).
\vphantom{\Big]}
\label{nafskks28}
\end{align}

We now come to the issue that ultimately motivates the elaborate transgression theory developed up to this point 
in this subsection: using transgression to generate a Poisson structure of the derived TCO model's phase space
$\clF_\msM(L)$ from the Poisson structure of the homogeneous space $\DD\msM/\DD\msJ$ associated with the curvature
$\rmB$ of a connection $\rmA$ of the unitary derived line bundle $\clL_\beta$ with $\msJ$ a maximal toral crossed submodule
of $\msM$ and $\beta$ an assigned character of $\msJ$ in the constructive derived framework of subsect.
5.6 of I.



The inner derivation crossed module $\INN\fku(1)$ of the Lie algebra $\fku(1)$
is essentially the same as the crossed module $\INN\bbR$ by the identity $\fku(1)=i\bbR$.
Since 
$\iMap(T[1]\DD\msM,\SD\INN\fku(1))=i\iMap(T[1]\DD\msM,\SD\INN\bbR)$ as vector spaces,
a restricted space
$\iMap_\msr(T[1]\DD\msM,\SD\INN\fku(1))=i\iMap_\msr(T[1]\DD\msM,\SD\INN\bbR)$
and a transgression map 
$\sfT:\iMap_\msr(T[1]\DD\msM,\SD\INN\fku(1))\rightarrow i\Fun(T[1]\clF_\msM(L))$
can be defined 
with the same properties as before.

A connection $\rmA$ of $\clL_\beta$ is restricted if $\rmA\in\iMap_\msr(T[1]\DD\msM,\DD\INN\fku(1)[1])$.
On account of relation (5.5.16) of I, the curvature $\rmB$ of $\rmA$ is then 
restricted as well, i.e. $\rmB\in\iMap_\msr(T[1]\DD\msM,\DD\INN\fku(1)[2])$. 
The transgressions $\sfT(\rmA)$, $\sfT(\rmB)$ of $\rmA$, $\rmB$
are so degree $1$, $2$ elements of $i\Fun(T[1]\clF_\msM(L))$, respectively. 
Further,  
\begin{equation}
\sfT(\rmB)=-\delta\sfT(\rmA)
\label{nafskks30}
\end{equation}
by \ceqref{nafskks26} owing to (5.5.16) of I and \hphantom{xxxxxxxxxxxx}
\begin{equation}
\delta\sfT(\rmB)=0
\label{nafskks31}
\end{equation}
owing to the Bianchi identity (5.5.17) of I. 
By relations \ceqref{nafskks30}, \ceqref{nafskks31},
$-i\sfT(\rmB)$ can be regarded as a presymplectic 
form on the derived model's TCO phase space 
$\clF_\msM(L)$ with presymplectic potential $-i\sfT(\rmA)$.
A Poisson bracket $\{\cdot,\cdot\}_{\sfT(\rmA)}$ is therefore available
for the algebra $\Fun_\rmA(\clF_\msM(L))$ of Hamiltonian phase space functionals on
general grounds.



In subsect. 5.6 of I, we studied the space $\iDFnc_\rmA(\DD\msM)$ of Hamiltonian
derived functions of $\DD\msM$. As $\iDFnc_\rmA(\DD\msM)\subset
\iDFnc(\DD\msM)=\iMap(T[1]\DD\msM,\DD\bbR)$ 
and $\iMap(T[1]\DD\msM,\DD\bbR)=\iMap(T[1]\DD\msM,\DD\INN\bbR)$ as vector spaces,
it is natural to limit our set--up to the space of restricted derived Ha\-miltonian
functions $\iDFnc_{\msr\rmA}(\DD\msM)$ $=\iDFnc_\msA(\DD\msM)\cap\iMap_\msr(T[1]\DD\msM,\DD\INN\bbR)$.
If $\rmF\in\iDFnc_{\msr\rmA}(\DD\msM)$ is a restricted Hamiltonian function, 
its Hamiltonian vector field
$P_\rmF\in\iVect_\msr(\DD\msM)$ is also restricted. By (5.6.5) of I and 
\ceqref{nafskks26}, \ceqref{nafskks27}, 
\begin{align}
\delta\sfT(\rmF)-i\iota_{\varUpsilon(P_\rmF)}\sfT(\rmB)=0.
\label{nafskks32}
\end{align}
Hence, the functional $\sfT(\rmF)$ is Hamiltonian too.

In this fashion, a linear map $\sfT:\iDFnc_{\msr\rmA}(\DD\msM)\rightarrow\Fun_\rmA(\clF_\msM(L))$
is established. $\sfT$ has the distinguished property that 
\begin{equation}
\{\sfT(\rmF),\sfT(\rmH)\}_{\sfT(\rmA)}=\sfT(\{\rmF,\rmH\}_\rmA)
\label{nafskks33}
\end{equation}
for $\rmF,\rmH\in\iDFnc_{\msr\rmA}(\DD\msM)$, as is straightforwardly verified.
In spite of its formal appearance, relation \ceqref{nafskks32} cannot be interpreted
as indicating that $\sfT$ is a Poisson map. In fact, 
while $\Fun_\rmA(\clF_\msM(L))$ is a functional algebra
and $\{\cdot,\cdot\}_{\sfT(\rmA)}$ is a genuine Poisson bracket structure on it, $\iDFnc_\rmA(\DD\msM)$
is a mere vector space and $\{\cdot,\cdot\}_\rmA$ is only a twisted Lie bracket structure thereon 
reducing to a genuine Poisson structure only upon restriction to
the short algebra $\iDFnc_{\varpi\rmA}(\DD\msM)$ (cf. subsect. 5.6 of I). The twisting
of the Lie bracket $\{\cdot,\cdot\}_\rmA$ is compatible with the Jacobi property of $\{\cdot,\cdot\}_{\sfT(\rmA)}$
since for any function triple $\rmF,\rmH,\rmK\in\iDFnc_{\msr\rmA}(\DD\msM)$
\begin{equation}
\sfT(\langle\rmF,\rmH,\rmK\rangle_\rmA)
=i\delta\iota_{\varUpsilon(P_\rmK)}\iota_{\varUpsilon(P_\rmH)}\iota_{\varUpsilon(P_\rmF)}\sfT(\rmB)=0
\label{nafskks34}
\end{equation}
by grading reasons (cf. eqs. (5.6.7), (5.6.8) of I). 

We can now establish a relationship between the regular case
derived KKS theory developed in subsect. 5.9 of I
and the canonical formulation of the characteristic derived TCO model
worked out in subsect. \cref{subsec:hafsspx}. 
We shall do that using the transgression map $\sfT$ of eq.  \ceqref{nafskks25} 
in the special case where the derived current $\rmC$ has components 
\begin{equation}
c=\theta_L, \qquad C=\delta_{\partial L}  
\label{nafskks35}
\end{equation}
corresponding to the distributional factors of the characteristic model's level current
in the canonical formulation (cf. eqs. \ceqref{hafsspx13}, \ceqref{hafsspx13/1}). 
By a straightforward computation, it can be verified that the presymplectic form $\sfPsi_\varLambda$  
of the characteristic model's ambient phase space $\clF_\msM(L)$ 
given by eq. \ceqref{hafsspx15} is precisely the presymplectic form $-i\sfT(B_\varLambda)$
yielded by transgression of the curvature $\rmB_\varLambda$ of the unitary 
connection $\rmA_\varLambda$ of the derived line bundle $\clL_\varLambda$ 
given componentwise by eqs. (5.9.2), (5.9.3) and (5.9.4), (5.9.5) of I, 
\begin{equation}
\sfPsi_\varLambda=-i\sfT(\rmB_\varLambda). 
\label{nafskks37}
\end{equation}
The observation made at the end of subsect. \cref{subsec:hafsspx} that $\sfPsi_\varLambda$ 
exhibits a bulk and an edge term with a formal structure analogous to that of the
two components of the derived symplectic structure $-i\rmB_\varLambda$ of the regular
orbit $\clO_\varLambda$ now takes a more precise meaning in the light of the above
transgressional analysis and add new evidence for the close relationship between 
the characteristic model and KKS theory.



\subsection{\textcolor{blue}{\sffamily Conclusions}}\label{subsec:dtcosend}

In this section, we have studied in detail a 2-dimensional derived TCO model, its symmetries
and its sigma model interpretation in both the Lagrangian and canonical perspective and provided
significant evidence to support
the claim the model furnishes the partition function realization of an underlying Wilson surface
upon quantization, which was the ultimate motivation of our endeavour.
Most importantly, we have shown that the characteristic version of
the model is intimately related to derived KKS theory and may yield important clues on the
eventual geometric quantization of this latter, an open problem. The unifying element
of the multiple constructions we have carried out is the derived geometric framework
of higher gauge theory, whose basicness the present work highlights conclusively.

\vfil\eject

\vfil\eject

\markright{\textcolor{blue}{\sffamily Acknowledgements}}

\noindent
\textcolor{blue}{\sffamily Acknowledgements.}
The author thanks E. Meinrenken for correspondence. 
The author acknowledges financial support from the grant GAST of INFN Research Agency
under the provisions of the agreement between Bologna University and INFN.

\vfil\eject

\noindent
\section*{\textcolor{blue}{\sffamily References}}

\end{document}